\documentclass[apjl]{emulateapj}
\usepackage{graphicx}

\shorttitle{COMPLETE Shells in Perseus}
\shortauthors{Arce, et al.}

\begin{document}


\title{A Bubbling Nearby Molecular Cloud: {\it COMPLETE} Shells in Perseus}


\author{H\'{e}ctor G. Arce}
\affil{Department of Astronomy, Yale University, P.O. Box 208101, New Haven CT 06520}
\email{hector.arce@yale.edu}

\author{Michelle A. Borkin}
\affil{School of Engineering and Applied Sciences, Harvard University, 29 Oxford Street, Cambridge MA 02138}
\email{michelle\_borkin@harvard.edu}

\author{Alyssa A. Goodman}
\affil{Harvard-Smithsonian Center for Astrophysics, 60 Garden Street, Cambridge MA 02138}
\email{agoodman@cfa.harvard.edu}

\author{Jaime E. Pineda}
\affil{ESO, Karl Schwarzschild Str. 2, 85748 Garching bei Munchen, Germany  \& \\ 
Jodrell Bank Centre for Astrophysics, School of Physics and Astronomy, University of Manchester, Manchester, M13 9PL, UK}
\email{jaime.pineda@manchester.ac.uk}

\and

\author{Christopher N. Beaumont\footnote{SAO predoctoral fellow. Current address and email: 
Harvard-Smithsonian Center for Astrophysics, 60 Garden Street, Cambridge MA 02138. cbeaumont@cfa.harvard.edu}}
\affil{Institute for Astronomy, University of Hawaii at Manoa, 2680 Woodlawn Drive, Honolulu HI 96822}
\email{beaumont@ifa.hawaii.edu}

\begin{abstract}

We present a study on the shells (and bubbles) in the Perseus molecular cloud  using the COMPLETE survey large-scale $^{12}$CO(1-0) and $^{13}$CO(1-0) maps.
The twelve shells reported here are spread throughout most of the Perseus cloud and have circular or arc-like morphologies with a range in radius of about 0.1 to 3 pc.
 Most of them have not been detected before most likely as maps of the region lacked the coverage {\it and} resolution needed to distinguish them.
 The majority of the shells are coincident with infrared nebulosity of similar shape and have a candidate powering source near the center. We
suggest they  are formed by the interaction of spherical or very wide-angle winds powered by young stars inside or near the Perseus 
molecular cloud 
---a cloud that is commonly considered to be mostly forming low-mass stars.
Two of the twelve shells are powered by high-mass stars close to the cloud, while the others appear to be powered by low or intermediate mass stars
in the cloud.  We argue that winds with a mass loss rate of about $10^{-8}$ to $10^{-6}$ M$_{\sun}$~yr$^{-1}$
are required to produce the observed shells. 
Our estimates indicate that the  energy input rate from these stellar winds is similar to the turbulence dissipation rate. We conclude that in Perseus 
the total energy input from both collimated protostellar outflows and powerful spherical winds from young stars is sufficient to maintain the 
turbulence in the molecular cloud.    Large scale molecular line and IR continuum maps 
of a sample of clouds will help determine the frequency of this phenomenon in other star forming regions.




\end{abstract}

\keywords{ISM: bubbles --- stars: winds, outflows --- stars: formation ---   stars: preÐmain-sequence --- 
 turbulence --- ISM: individual (Perseus)}

\section{Introduction}
\label{sec:intro}

Stars go through a mass loss stage during the early stages of their life. The properties of the wind driven by the forming star will depend, in part, 
on the mass and evolutionary stage of the driving source. 
In general, young embedded protostars drive powerful collimated (jet-like) outflows, and as protostars evolve the mass loss rate decreases and the outflow becomes less collimated \citep{Bontemps+96,AS06}.  During the late stages of star formation, low and high mass pre-main sequence stars (T-Tauri and Herbig Ae/Be stars) drive
wide-angle winds, that in some cases  coexist with a collimated jet-like wind component.  Regardless of their degree of collimation, these supersonic winds  can entrain and accelerate the surrounding gas, thereby injecting momentum and energy into the surrounding gaseous environment.  

Early on in the study of winds from young stars it was realized that they have the potential to significantly affect
the dynamics and structure of their parent cloud. \citet{NS80} proposed a model in which the collision between
bubbles formed by the interaction of
spherical winds from low-mass  pre-main sequence stars and the surrounding quiescent gas would dominate the dynamics of molecular clouds. 
Although the idea that winds from young stars can have a significant impact 
on the cloud is still relevant \citep{Arce+07}, 
the  bubble model of \citet{NS80} has lost its original appeal as 
T Tauri stars are rarely observed to have a major impact on the surrounding molecular gas. Instead, it is the
  bi-polar collimated winds from younger embedded protostars that  have been commonly observed to have 
  a significant effect on their surroundings through  well-known molecular outflows.

Parsec-scale circular cavities (typically referred to as bubbles) are regularly found in regions of high-mass star formation, and are
 likely created by spherical winds from high-mass stars
 \citep[e.g.,][]{Churchwell+06,Churchwell+07,BW10,Deharveng+10}. High-mass stars  evolve much faster than low-mass stars and reach the main sequence stage while they are 
still  accreting material and are embedded in their natal cloud. During the main sequence phase, high-mass stars can drive spherical winds with mass loss rates of $\sim 10^{-6}$ M$_{\sun}$~yr$^{-1}$ that can easily
create the observed bubbles. The conventional thought has been that in regions  where predominantly low and intermediate mass stars form
 such bubbles  do not exist, 
as the spherical winds from pre-main sequence stars are thought to be too weak to produce them. In this study we hope to cast away this notion.

Here we present a study of circular shells detected in CO line maps of the Perseus molecular cloud complex, a cloud generally thought 
to predominantly form low-mass stars. 
We find twelve shells in the molecular gas that are very likely produced by powerful spherical (or quasi-spherical) winds from young stars. Ten of these are from presumed low or intermediate mass stars (thought to be late B or of later spectral type) within the cloud and two are from high-mass stars close to the cloud. We find that these shells have the potential to have a significant impact on the cloud, as they have enough energy to drive the turbulence in Perseus and to disperse a small fraction of the cloud's gas. Yet, the few shells that we find are too sparse to drive the shell-shell collisions and produce  the clumps predicted by the original T Tauri  bubble model of  \citet{NS80}.

The Perseus molecular cloud complex is a chain of clouds with a total mass of about $7 \times 10^3$ M$_{\sun}$, and encompasses a total area of about 70~pc$^{2}$ \citep{Evans+09, Arce+10}. There is a span of distance estimates to Perseus that range from 230~pc to 350~pc, in part due to different parts of the cloud being at slightly different distances along the line of sight and also due to the  uncertainties in distance measurements
(see Enoch et al.~2006, and references therein for discussion).
To simplify our calculations, we adopt a single fiducial distance of  250$\pm$50~pc,   
similar to other recent cloud-wide studies of Perseus. 
The Perseus cloud complex contains two rich protostellar clusters,  IC~348 and NGC~1333, and a number of other regions that are mostly forming low-mass stars, 
where most young stars associated with the cloud have spectral type of A or later. 
There are however, three known B stars that reside in the cloud: HD~281159, a B5 star in the IC~348 cluster; BD+30 549, a B8 star in NGC~1333; and SVS 3, a B5 star in NGC~1333. 
As we will show in this work (and others have also suggested) there are at least two nearby high-mass stars that interact with the cloud through their winds. 

The entire Perseus region was first surveyed in $^{12}$CO by \citet{Sargent79}, and since then it has been mapped in CO at different angular resolutions (all with beams $>1.3\arcmin$) by a number of authors \citep[e.g.,][]{BC86,UT87,Padoan+99,Sun+06}. The CO maps used in this study (see below) cover approximately the same area as previous studies of the Perseus cloud, but at a significantly higher angular resolution. 
It is possible that many of the shells discussed here were not previously detected due to the low angular resolution of previous maps.
Recently, the Perseus molecular cloud complex was also observed in its entirety in the mid- and far-infrared as part of the 
``From Molecular Cores to Planet-forming Disks'' (a.k.a., c2d) {\it {\it Spitzer} } Legacy Project \citep{Jorgensen+06,Rebull+07, Evans+09}. 
These data have an angular resolution that is significantly less than $1 \arcmin$, and many of the shells detected in the molecular line maps are also observed as circular nebulosities
in the c2d infrared images.

\section{Data}
\label{datasection}

 We use the $^{12}$CO(1-0) and $^{13}$CO(1-0) line maps  of Perseus, collected as part of the COMPLETE (COordinated Molecular Probe Line Extinction Thermal Emission) Survey of Star Forming Regions\footnote{see http://www.cfa.harvard.edu/COMPLETE}, described in detail in \citet{Ridge+06a}.
 The $^{12}$CO J=1-0 (115.271 GHz) and the $^{13}$CO J=1-0 (110.201 GHz) line maps were  observed between 2002 and 2005 using the 14-meter Five College Radio Astronomy Observatory (FCRAO) telescope with the SEQUOIA 32-element focal plane array.  
 The dual-IF digital correlator allowed the simultaneous observation of $^{12}$CO and $^{13}$CO, and a total bandwidth of 25 MHz with 1024 channels in
each IF, providing a velocity resolution of approximately 0.064~km~s$^{-1}$ and 0.066~km~s$^{-1}$ in the $^{12}$CO and $^{13}$CO maps, respectively.
The telescope beam at these frequencies
is about $46\arcsec$, and the resulting maps are Nyquist-sampled with $23\arcsec$ pixels. 
 All figures shown here present the  $^{12}$CO and $^{13}$CO in the  $T_A^*$ scale. 
 To convert to the main beam antenna temperature scale, the spectra at 115 and 110 GHz need to be divided by 0.45 and 0.49, respectively.
 The  median $1-\sigma$ rms per channel in
the $^{12}$CO and $^{13}$CO  maps are approximately 0.23 and 0.11 K, respectively, in the $T_A^*$ scale.

\section{Shell and Source Identification}
\label{shellid}
\begin{figure}
\plotone{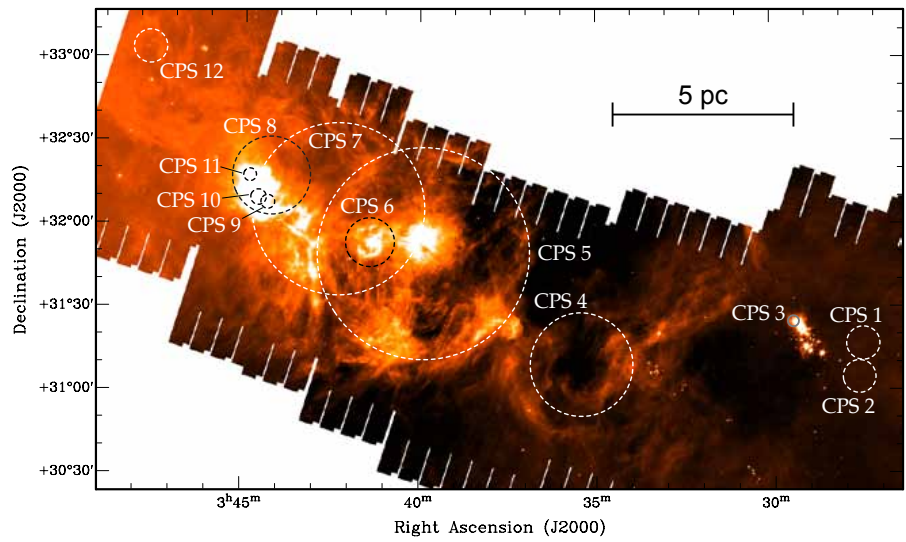}
\caption{Shells in Perseus.  The position of each CPS is shown as a dashed circle overlaid on the c2d {\it Spitzer}  MIPS 1 ($24 \micron$) image (from Rebull et al.~2007).
\label{allmips}}
\end{figure}

\begin{figure}
\plotone{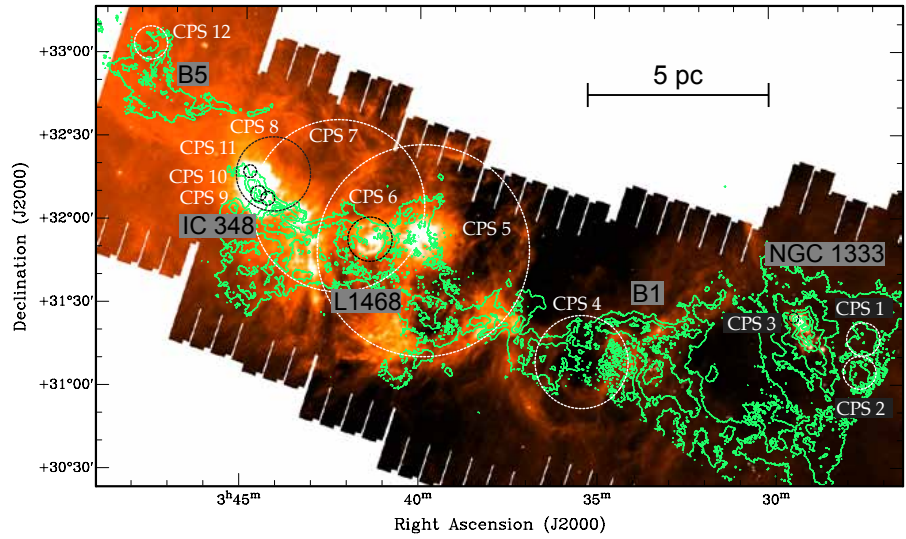}
\caption{Shells in the Perseus molecular cloud.  Contours show the integrated intensity map of the $^{13}$CO(1-0) emission. 
Starting contour and contour steps are both 3 K km s$^{-1}$.  These are overlaid on 
the c2d {\it Spitzer}  MIPS 1  image. Shells are shown and labeled as in Figure~\ref{allmips}. The names of different clouds in Perseus are labeled in black. 
\label{all13co}}
\end{figure}

In the molecular line maps, the shells in Perseus are seen as bright 
rings or arc structures surrounding  regions of less (or no) emission and many are coincident with infrared nebulosity seen in the MIPS 24~\micron \/ c2d map.   
These shells, which we denote as COMPLETE Perseus Shell (CPS) 1 through 12, are listed in Table~\ref{shellstab}. Their positions in the cloud are shown 
with respect to the 24~\micron \/ emission and the $^{13}$CO(1-0) integrated intensity emission of the Perseus region in Figures~\ref{allmips} and \ref{all13co}. These figures serve as
general finding charts that show the overall distribution of the shells in the Perseus cloud complex. More detailed figures that show the emission and velocity structure of each CPS are shown below.

\begin{figure}
\epsscale{1.0}
\plotone{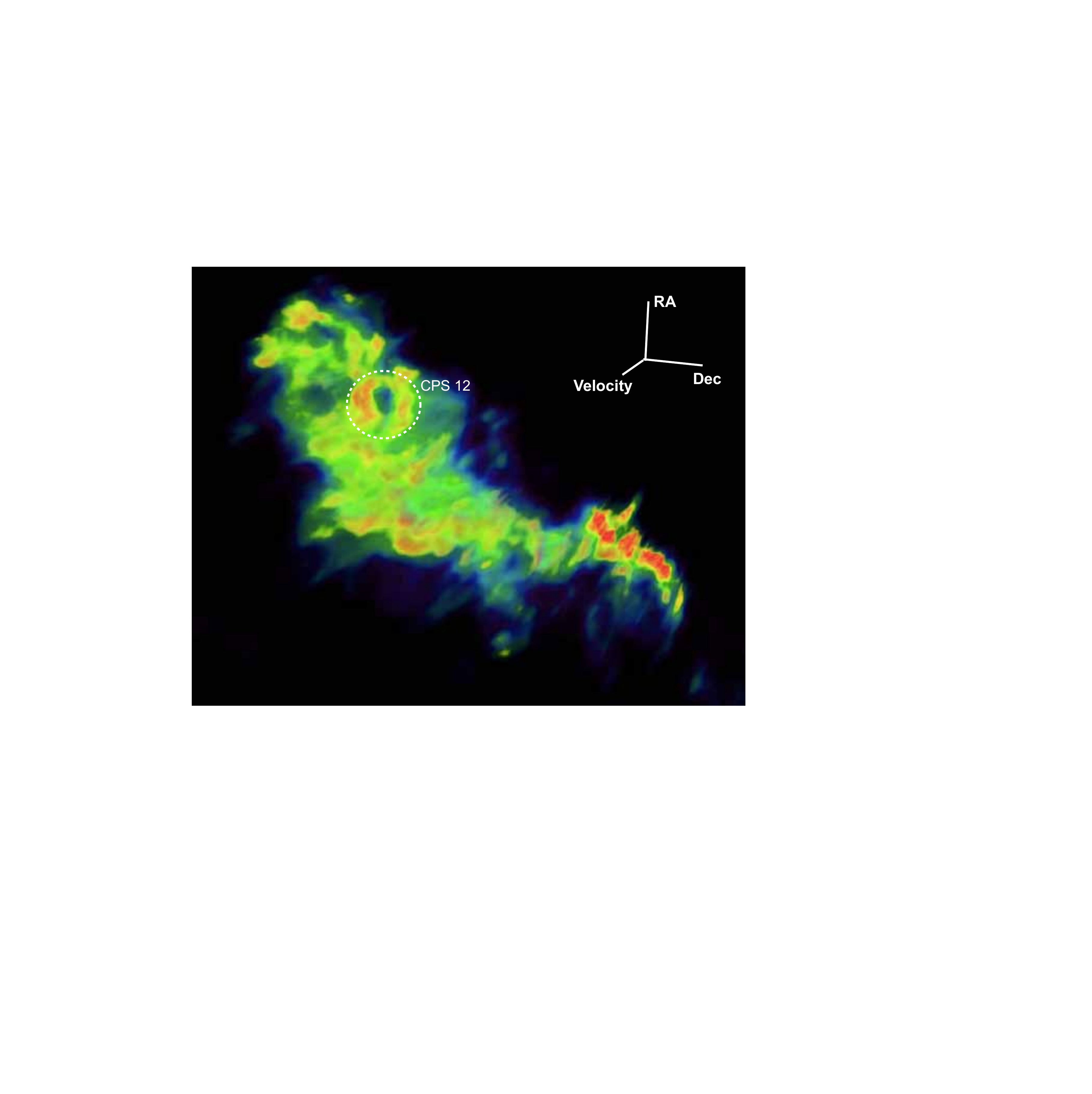}
\caption{Three-dimensional volume rendering of the $^{13}$CO(1-0) data around the B5 dark cloud, at the easternmost edge 
of the Perseus molecular cloud complex. Both the color (i.e., rainbow color map) and transparency of each voxel are mapped to the intensity of each pixel at each channel. Thus the brightest regions in the channel maps correspond to solid red 3D regions, and the dim emission in the channel maps correspond to transparent blue 3D regions.  
Visible in the upper left portion of the image is CPS 12. This kind of rendering was 
used in \citet{Arce+10} to search for high-velocity features. 
\label{3dfig}}
\end{figure}

The shells in Perseus were found through different methods. 
Some were found searching for high-veolcity features with a 3D visualization technique (see Figure~\ref{3dfig}), as part of the outflow study conducted by \citet{Arce+10}. Others were found by 
 comparing the molecular gas distribution with the c2d  IRAC and MIPS images in regions close to bright, circular and arc-like, nebulosities. A few were detected by studying how the morphology 
 of the gas changed with velocity in our molecular line data cubes (see below for more detail). 
 The sample of shells listed here is most likely incomplete, as we probably fail to detect small 
shells with radius of less than $1.9\arcmin$ ($\sim 0.14$ pc)  since it is hard to identify a shell structure with a diameter that extends less than 5 pixels in our map.  

The regions associated with the CO shells were easily  identified as circular (or arc) structures brighter than the surrounding molecular gas.
We used the CO  maps to obtain an estimate of the radius, 
thickness and center of each shell. We created CO integrated intensity maps over the velocity range in which the shell was clearly
detected (shown in Figures~\ref{cps1fig2} to \ref{cps12fig2}).  The center of each CPS  was first estimated by visually fitting a circle to the CO integrated intensity map of the shell. Using circles at  different positions and with different radii, we estimate the error in the center position to be about $\pm 45\arcsec$ for the shells that have a structure that delineates more than half a circle's circumference (i.e., CPS 1, 2, 3, 6, 9, 10, 11, 12). In shells that show only an arc in CO, but exhibit an almost complete circular structure in the IR maps (i.e., CPS 4, 5) we used the
 24~\micron \/ map to estimate the center position of the CPS (which gives similar error as using the CO maps).  The position and size for shells that show arcs that only trace less than half a circle's circumference are less constrained. For these shells (CPS 7, 8) the errors in the center position estimate are larger by a factor of four. 

\begin{figure}
\epsscale{1.0}
\plotone{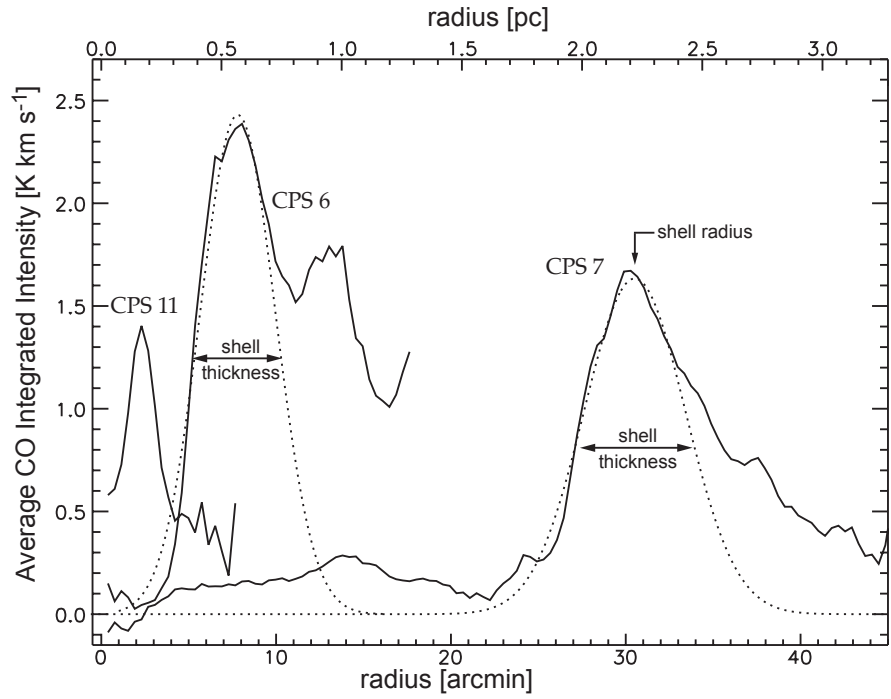}
\caption{Azimuthally averaged radial intensity profiles of three shells. The solid curves show the intensity profiles (from the integrated intensity map of each shell). A gaussian fit to the intensity profile  was performed on each shell to estimate its radius and thickness. As an example, we show the  fit to the radial profile of CPS 6 and CPS 7 (dotted line).
\label{shellprofiles}}
\end{figure}

We obtained the shell radius and thickness by fitting a gaussian to the 
azimuthally averaged profile of the CO integrated intensity map of each CPS (as shown in Figure~\ref{shellprofiles}). The position of the profile's peak gives the shell radius and the full width at half maximum (FWHM) gives the thickness of the shell. 
We used the shell radius and thickness to define the area of the annulus used to estimate the mass, hereafter ``mass region'' (see \S~\ref{mass} for more detail). 
Although the same procedure was used to obtain the mass region for each CPS,  for some shells we had to constrain it to a smaller area to avoid including emission from unrelated structures (i.e., clouds, filaments or other shells)  in the  mass estimates (see discussion for each CPS below).

\begin{figure}
\epsscale{1.0}
\plotone{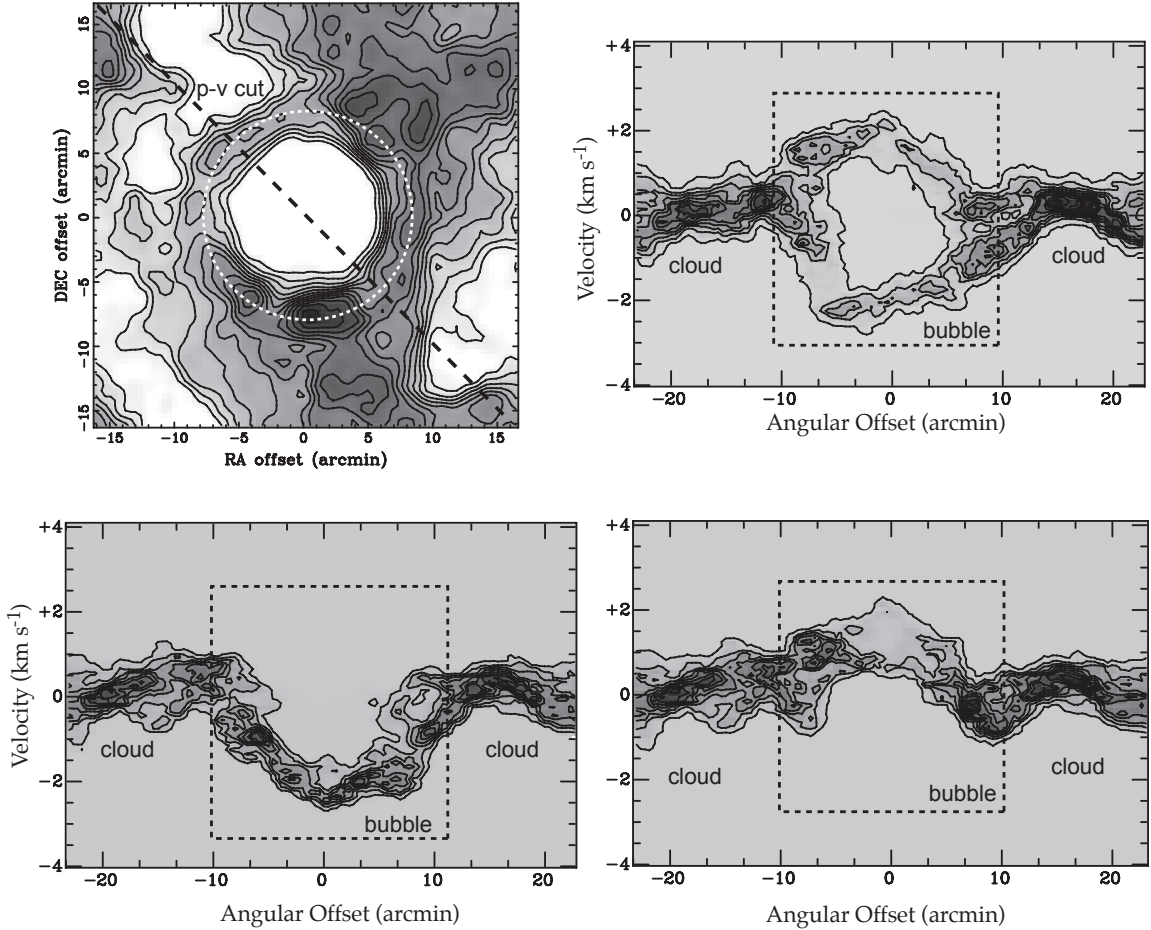}
\caption{Model of expanding bubble inside a turbulent medium. ({\it a}) Integrated intensity map (from  -1 to 0~km~s$^{-1}$)   of model expanding bubble. In this model, the bubble center coincides with the center of the cloud that is at a distance of 250 pc, at a $V_{LSR} = 0$~km~s$^{-1}$, and it is 1.3 pc thick. The cloud velocity FWHM is 1.2~km~s$^{-1}$ and the spectral index of the turbulent velocity field ($\beta$) is 2.0. The bubble radius, thickness, and expansion velocity are  0.6 pc, 0.25 pc, and 2~km s$^{-1}$ (see \S~\ref{shellid} for a description of the model).  This bubble is fully immersed in the cloud. ({\it b}) $p-v$ diagram of bubble along the direction shown in {\it a}. ({\it c}) $p-v$ diagram of bubble displaced towards the back of the cloud. In this model all parameters are the same as the bubble shown in ({\it a}), except the cloud thickness (here is 0.6 pc) and the center of the bubble is offset (towards the back) with respect to the cloud's center by  0.4 pc.  ({\it d}) $p-v$ diagram of bubble displaced in position and velocity with respect to the cloud's center. In this model the bubble's center is displaced 0.3 pc towards the front of the cloud and the bubble's central velocity is displaced 0.5~km s$^{-1}$ towards blue velocities with respect to the cloud's central velocity.
\label{shellmodel}}
\end{figure}

In order to exclude ring structures that might simply be due to random  
patterns in the cloud (e.g., due to turbulence or overlapping filaments), we only considered circular or arc-like structures
 that have a velocity distribution corresponding to that of an expanding bubble or thick ring.   
 For an expanding bubble fully immersed in a cloud and detectable at {\it all} velocities, the radius of the ring seen on the plane of the sky will increase as the velocity increases from the bluest velocity where the shell is detected (i.e., the front cap of the bubble) up to the velocity of the bubble source (i.e., the bubble's central LSR velocity), and will then decrease with increasing velocity up to the reddest velocity where the shell is detected (i.e., the back cap of the shell). A model of an expanding bubble inside a turbulent cloud with uniform density\footnote{This model has 8 parameters: $r_b$, $dr_b$, $p_{offset}$, $v_{offset}$, $dr_{cloud}$, $V_{\mathrm{exp}}$, $v_{rms}$, and $\beta$. The expanding bubble is modeled following \citet{CP05}. The parameters $r_b$, $dr_b$, and $V_{\mathrm{exp}}$ describe the radius, thickness and expansion velocity of the bubble, respectively, and  $p_{offset}$ and $v_{offset}$ are the offset in the bubble's central position and velocity with respect to the cloud's center and velocity.  The ambient cloud is modeled as a face-on slab of thickness $dr_{cloud}$ with uniform density. All of the material evacuated from the bubble interior is redistributed uniformly throughout the shell. The velocity field of the ambient medium has a dispersion of $v_{rms}$. Its spatial structure is fractal-like, with a power-spectrum of $P(k) \sim  k^{-\beta}$ \citep{Stutzki+98}.   These parameters are used to generate three-dimensional density and velocity fields, which are then transformed into position-position-velocity coordinates to simulate observations. In  Figure~\ref{shellmodel} the integrated intensity greyscale  is proportional to the gas column density.} 
 is shown in Figure~\ref{shellmodel}. In Figure~\ref{shellmodel}a, the integrated intensity map (over a limited velocity range close to the cloud velocity) shows a clumpy circular structure\footnote{In this model, which uses a uniform density cloud, the clumpy structure of the shell is solely due to the cloud's turbulence. We would expect that a shell in a real turbulent molecular cloud with non-uniform density would be as (or more) clumpy than the model shell presented here.}. 
 The corresponding position-velocity ($p-v$) diagram that cuts through the center of the bubble is shown in  Figure~\ref{shellmodel}b. Here the bubble is identified as 
 a circular or ring-like structure where the highest velocities (along the line of sight) are seen at the center of the  bubble, 
 and the gas with positive (negative) velocities correspond to the back (front) hemisphere of the bubble.
 If molecular gas does not surround the entire bubble, then the gas displaced by the bubble (i.e., the shell) will be observed over a more limited range of velocities. 
 For example, if the center of the bubble is displaced towards the back of the cloud, the gas will only show the front hemisphere of the bubble. In this case the $p-v$ will resemble  
 Figure~\ref{shellmodel}c, where the bubble appears as a ``V'' or ``U'' structure. Likewise, for a bubble with the center displaced toward the front  of a thin cloud  only the back hemisphere will be observed. This will be seen as an upside down ``V'' or ``U'' structure in the $p-v$ diagram, as in Figure~\ref{shellmodel}d. 
 The shells in Perseus were not detected at all velocities (similar to those shown in Figures~\ref{shellmodel}c and \ref{shellmodel}d)
  which indicates that most (if not all) of the bubbles (or winds) responsible for the observed shells are  not fully immersed in the cloud (see discussion in \S~\ref{geometry}).
We thus considered shells that exhibit a detectable change in the radius with velocity
consistent with the expected kinematics of an expanding bubble that may or may not be fully surrounded by molecular gas.   

 We searched for a candidate source for each COMPLETE Perseus Shell using SIMBAD, and the 2MASS and {\it Spitzer}  c2d catalogs. We used these catalogs to look for young (pre-main sequence) stars associated  with the cloud, close to the center of each shell. We expect that young stars (with age $\sim 1-2$ Myr or less) will be the most likely drivers of winds with enough mass loss rate to drive the observed shells (see discussion in \S~\ref{origins}). The velocity of young (low-mass) stars with respect to their ambient cloud is expected to be relatively low, between 0.1 and  0.2~km~s$^{-1}$ \citep{Hartmann02, Jorgensen+07}. Thus, at the distance of Perseus, we expect sources that are about 1 Myr of age to be within  3\arcmin \/ from the center of the circular cavity delineated by the shell. 
Using the c2d survey, \citet{Jorgensen+06} show that the density  of YSO candidates (YSOc) in Perseus outside the IC348 and NGC1333 clusters is 43.3 YSOc deg$^{-1}$. If we assume that 
the c2d YSOc catalog lists 90\%  of the total PMS in the cloud (i.e., the catalog is $\sim$ 10\% incomplete, see Evans et al. 2009), 
the density should then be closer to 48 young stars per square degree. Hence,  there is only a 30\% probability of finding a pre-main sequence star by chance within a circular area with a radius of 3\arcmin, for shells outside the central region of the IC 348 and NGC 1333 cluster.  
For small shells inside these clusters (i.e., CPS 3, 9, 10, 11), we search for possible sources within 1-1.5\arcmin \/ radius (depending on the shell size). In these two clusters the surface density of YSOc is relatively large and a few to several tens of pre-main sequence stars may lie within a radius of 1 to 1.5\arcmin. However, for all these small shells  we can    
 identify the most likely  candidate source from the list of stars in the region.
We find a realistic candidate source for nine of the twelve shells, which we list in Table~\ref{shellstab}.  Their properties are given in  
 Table~\ref{sourceprop}.  
 
 The ring-like (or arc-like) morphology of the Perseus shells, their velocity-dependent structure, and  in (most cases) the clear association with a source or an IR nebulosity with a similar morphology indicate that most, if not all, of these shells likely trace the region where (very wide angle or spherical) winds from young stars interact with the molecular gas (see \S~\ref{origins}).
It is clear that not all shells show the same kind of evidence supporting their presumed wind-driven nature. We quantify this by assigning a 
 ``confidence level'' to each CPS based on the pieces of evidence that support the hypothesis that these structures are wind-driven, as opposed to, 
 for example, being chance superposition of cloud filaments or random structures formed by the cloud's turbulence. The levels range from 2 to 5, where the higher the grade the more supporting evidence exist of that particular shell. The grade is assigned by  adding the total number of pieces of evidence for each shell, as shown in Table~\ref{shellgrade}: 1) 
 change in shell radius with velocity; 2) coincidence with IR nebulosity of similar shape and size;
  3) CO or IR emission that clearly delineates more than half a circle;  4) a $p-v$ diagram that is consistent with that of an expanding bubble; 
  and 4) a realistic candidate source near the shell center.
We note that even shells that do not show a clear $p-v$ diagram consistent with that of an expanding bubble show 
 evidence of changing radius with velocity, but their $p-v$ diagrams are contaminated by nearby shells (as is the case for CPS 7) or are just limited by the fact that the CO shell does not trace a full circle (as in the case of CPS 4, 5, 7 and 8).
We believe that all the shells discussed here are likely to be wind-driven structures, even shells with a confidence level of 2. A lower grade simply means that there is less evidence in the current data to support the shell authenticity compared to shells with a higher grade. 



Below we give a description for each CPS, the area and velocities associated with the CO shell, and discuss their candidate powering source.
  For each CPS we present two figures. In one we present 
channel maps of the region surrounding the shell. In the other figure we show, on the left panel,
an integrated intensity contour map of the shell (in most cases overlaid on the MIPS 24~\micron \/ image of the region), and on the right panel we present a $p-v$ diagram of the shell.
We use information from the integrated intensity map and the $p-v$ diagram to fit a simple spherical expanding bubble model  \citep[based on][]{CP05}
to the shell structure in the channel maps. 
The model consists of three parameters: bubble radius, expansion velocity, and central LSR velocity. The radius is set by our estimate of the CPS radius 
(shown in Table~\ref{shellstab}). We use the $p-v$ diagram of each source to obtain a first guess of the central velocity 
(e.g., the upper  part of the ``U'' structure in Figure \ref{shellmodel}c) and the expansion velocity of the bubble (e.g., the difference between the central velocity and the lowest point of the ``U'' structure in Figure~\ref{shellmodel}c). We then fit by eye the expanding bubble model to the channel maps by slightly changing our first guess to the expansion velocity and central velocity, while maintaining the bubble radius constant. The uncertainty in the estimate of the bubble central LSR velocity is  approximately $\pm 0.5$ km~s$^{-1}$. In most of the shells the expansion velocity is not well constrained as the CO shell that traces the expanding bubble is observed over a limited range of
velocities. Hence, the derived expansion velocity is a lower limit.

\subsection{CPS 1}
\label{cps1sec}
\begin{figure}
\epsscale{1.0}
\plotone{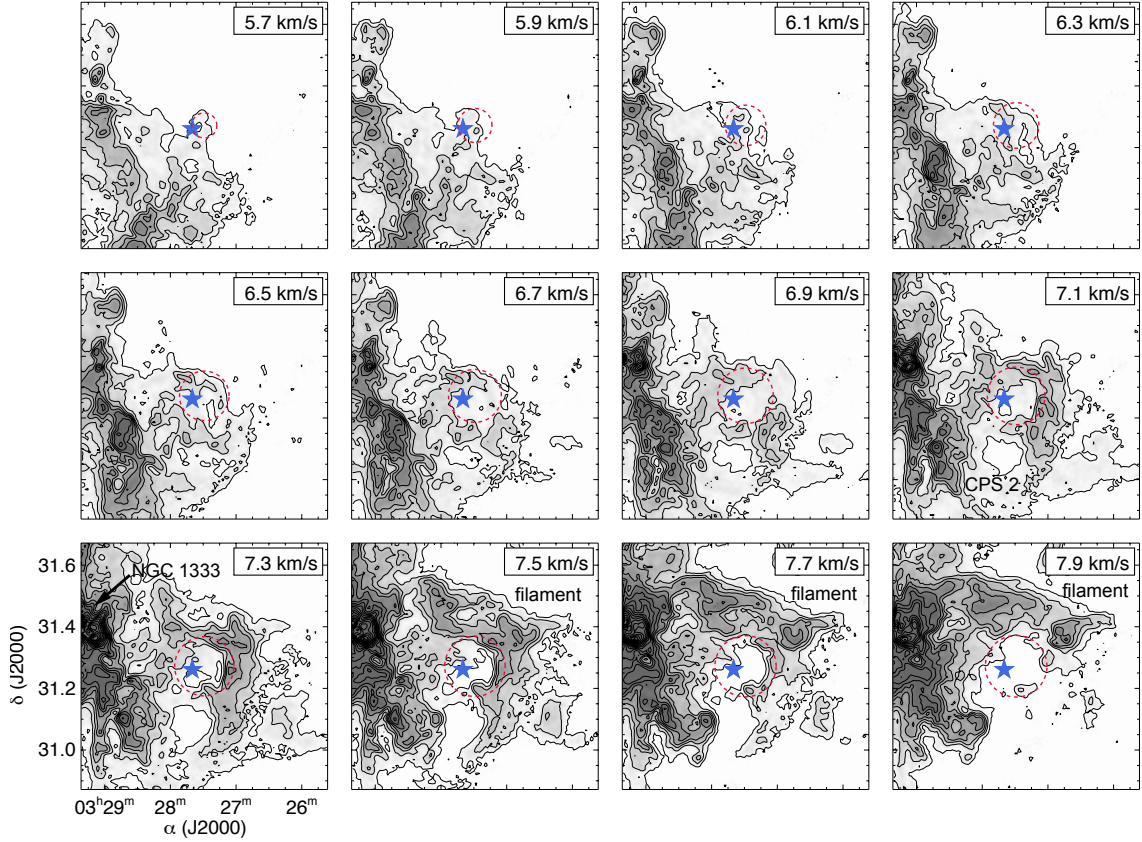}
\caption{Channel maps of $^{13}$CO emission near CPS 1. 
 The number on the upper right corner of each panel indicates the central $V_{LSR}$  of the channel map. 
 Starting contour and contour steps are both 0.5 K.
The filled star symbol shows the position of the candidate driving source BD+30 543. Dashed circles shows the expected extent, 
at different radial velocities, of an expanding bubble with a radius, $V_{\mathrm{exp}}$ and central LSR velocity of 6\arcmin, 2.5~km~s$^{-1}$ and
 8.0~km~s$^{-1}$, respectively (using the model discussed in \S~\ref{shellid}). The location of an unrelated filament, CPS 2 and the central part of
 NGC 1333 are also shown.
\label{cps1fig1}}
\end{figure}

\begin{figure}
\epsscale{1.0}
\plotone{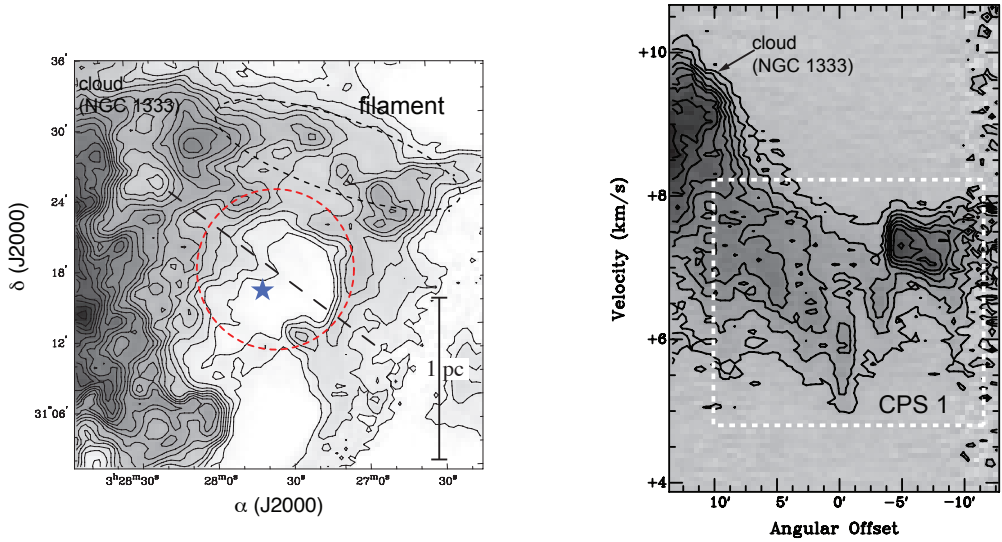}
\caption{Integrated intensity map and  $p-v$ diagram of $^{13}$CO emission near CPS 1.
The left panel shows the $^{13}$CO(1-0) integrated intensity (for  $7.2 < V_{LSR} < 8.6$~km~s$^{-1}$) depicting CPS 1. 
Starting contour and contour steps are both 0.5 K km s$^{-1}$.  The dashed circle shows the approximate extent of CPS 1. 
The filled star symbol shows the position of the candidate driving source BD+30 543.
The right panel shows the $p-v$ diagram along the cut shown by the diagonal dashed line in the integrated intensity map. 
Positions northeast (southwest) of the center of CPS 1 are shown as positive (negative) offsets. 
The approximate extent of the $^{13}$CO emission associated with CPS 1 in the $p-v$ diagram is shown as a dashed white rectangle.
\label{cps1fig2}}
\end{figure}

\begin{figure}
\epsscale{1.0}
\plotone{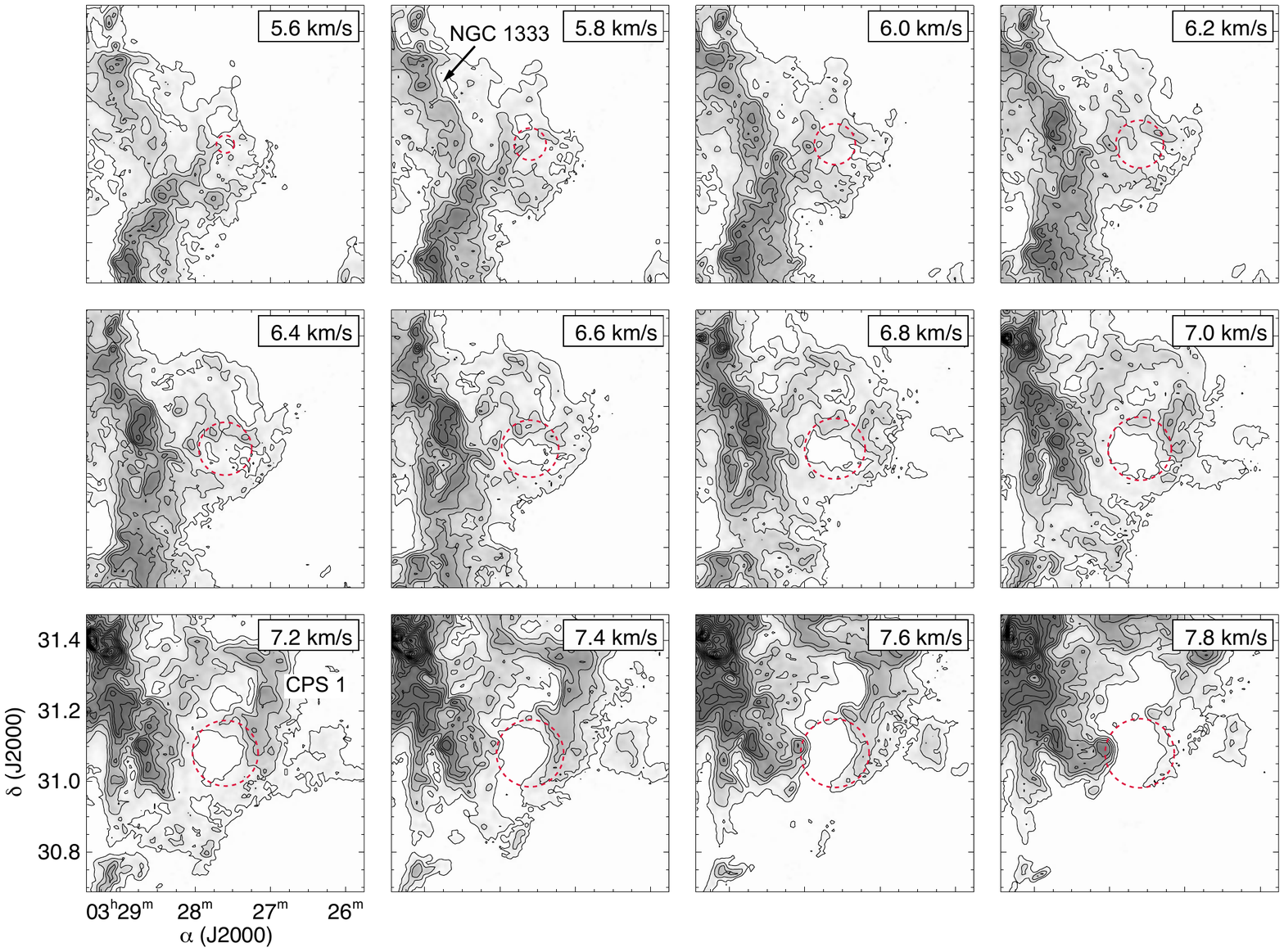}
\caption{Channel maps of $^{13}$CO emission near CPS 2. 
 The number on the upper right corner of each panel indicates the central $V_{LSR}$  of the channel map. 
 Starting contour and contour steps are 0.6 and 0.5 K, respectively.
 Dashed circles shows the expected extent,  
at different radial velocities, of an expanding bubble with a radius, $V_{\mathrm{exp}}$ and central LSR velocity of 5.9\arcmin, 2.5~km~s$^{-1}$ and
8.0~km~s$^{-1}$, respectively (using the model discussed in \S~\ref{shellid}). The location of  CPS 1 and the central part of
 NGC 1333 are also shown.
 \label{cps2fig1}}
\end{figure}

\begin{figure}
\epsscale{1.0}
\plotone{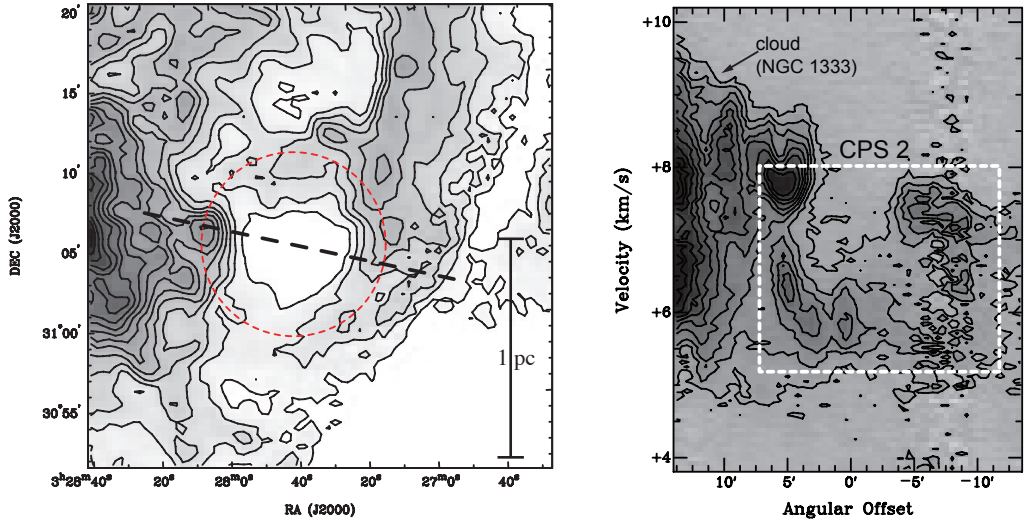}
\caption{Integrated intensity map and  $p-v$ diagram of $^{13}$CO emission near CPS 2.
The left panel shows the $^{13}$CO(1-0) integrated intensity (for  $6.6 < V_{LSR} < 8.0$~km~s$^{-1}$) depicting CPS 2. Starting contour and contour steps are both 0.3 K km s$^{-1}$.  The dashed circle shows the approximate extent of CPS 2. 
The right panel shows the $p-v$ diagram along the cut shown by the diagonal dashed line in the integrated intensity map. 
Positions northeast (southwest) of the center of CPS 2 are shown as positive (negative) offsets. 
The approximate extent of the $^{13}$CO emission associated with CPS 2 in the $p-v$ diagram is shown as a dashed white rectangle.
\label{cps2fig2}}
\end{figure}

\begin{figure}
\epsscale{1.0}
\plotone{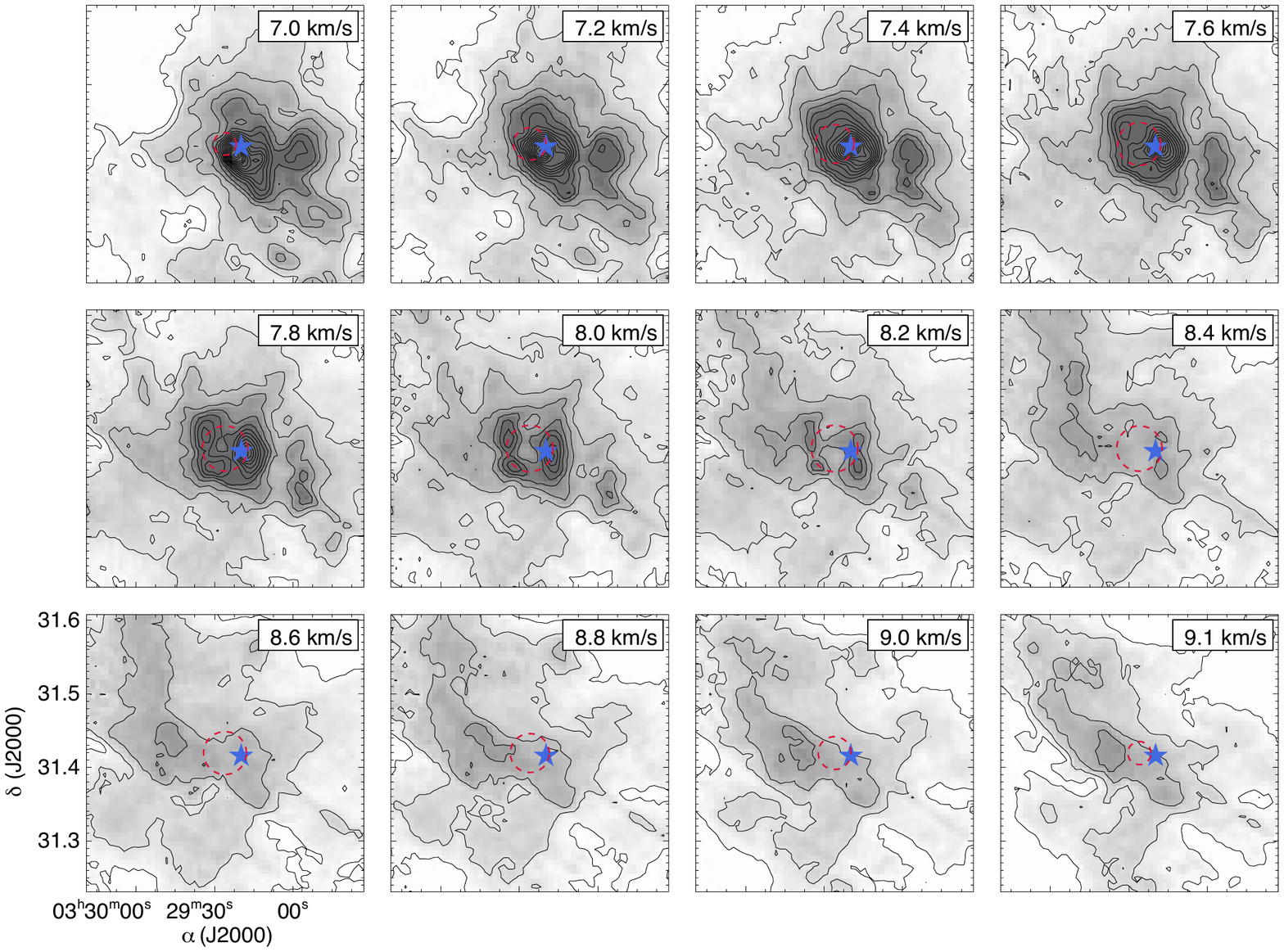}
\caption{Channel maps of $^{12}$CO emission near CPS 3. 
 The number on the upper right corner of each panel indicates the central $V_{LSR}$ of the channel map. 
 Starting contour and contour steps are 4 and 1 K, respectively.
 Dashed circles shows the expected extent, 
at different radial velocities, of an expanding bubble with a radius, $V_{\mathrm{exp}}$ and central LSR velocity of 1.9\arcmin, 1.2~km~s$^{-1}$ and
 8.1~km~s$^{-1}$, respectively (using the model discussed in \S~\ref{shellid}).
 The filled star symbol shows the position of the candidate driving source BD+30 549.
 \label{cps3fig1}}
\end{figure}

CPS 1 is located about $17\arcmin$ west of the central part of the NGC 1333 cluster and it is only distinguishable in the $^{13}$CO \/ map
(see Figure~\ref{all13co}). 
It was first detected by 
looking at the molecular line channel maps. 
The shell structure is most clearly seen at 
$7.1 < V_{LSR} < 7.7$~km~s$^{-1}$, where it shows an approximately circular morphology with $\sim 3\arcmin$  (0.2 pc) thick walls  
 (see Figure~\ref{cps1fig1}). Yet, in the channel maps and $p-v$ plot 
 emission that might be associated with the expanding shell can be seen over a wider range of velocities,  
 from $V_{LSR} \sim 5.5$ km~s$^{-1}$ to  $\sim 8.0$ km~s$^{-1}$ (see Figure~\ref{cps1fig1} and Figure~\ref{cps1fig2}).    
 The morphology of the $^{13}$CO emission within this range of velocities  
 exhibits a structure consistent with an expanding bubble with only the near-side hemisphere observable;
 the shell radius increases with increasing $V_{LSR}$ and the $p-v$ diagram shows a ``U'' or ``V'' shape (similar to the model shown in Figure~\ref{shellmodel}c). 
 From the $p-v$ diagram and the channel maps we estimate an expansion velocity of $\sim 2.5$~km~s$^{-1}$, and a central velocity at 
$V_{LSR} = 8.0$ km~s$^{-1}$.
To the north of CPS 1 there is a filament, seen in Figure~\ref{cps1fig1} at $V_{LSR} > 7.5$ km~s$^{-1}$, which is not related to CPS 1. 
To the east of CPS 1 lies  gas associated with the central region of the NGC 1333 cluster, most prominent at $V_{LSR} = 7.9$ km~s$^{-1}$.
Unlike most of the other shells, CPS 1 is not easily discernible in $^{12}$CO, as for this region (and velocities) the $^{12}$CO(1-0) emission  
practically fills the ring structure seen in $^{13}$CO. Neither the IRAC or MIPS images show any evidence of a ring-like structure in this area. 

SIMBAD lists only two stars projected within a $3\arcmin$ radius of the center of CPS 1.
Out of these, a likely source is the source SSTc2d J032740.5+311540 (also known as 2MASS J03274053+3115392).
This source is classified as a star with an IR excess starting at 24~\micron. This is very likely a young star, but it was
not classified as a YSO candidate as this area of the sky 
was not covered by the c2d IRAC maps and IRAC detection is needed for a positive YSOc classification \citep{Evans+09}.
It is  very likely that this c2d source is BD+30 543, which has been classified as   
an F2V star and has been estimated to be at the same distance (220~pc) as NGC 1333 \citep{Cernis90}.
There is no bright star at the (low-accuracy) position quoted in SIMBAD for BD+30 543, and the closest bright c2d and 2MASS point source is  SSTc2d J032740.5+311540.


\subsection{CPS 2}
\label{cps2sec}
This shell is located just south of CPS 1 in the western edge of the NGC~1333 region. Similar to CPS 1, the circular structure is only clearly discernible in the $^{13}$CO \/  map; 
it is barely seen in the $^{12}$CO \/ map, and there is no IR nebulosity in the {\it Spitzer}  images associated with this shell.
CPS 2  was first recognized when we studied the distribution of the $^{13}$CO(1-0) emission as a function of velocity.
The roundish structure is most clearly seen in at 
$V_{LSR}$ between 6.8 and 7.2~km~s$^{-1}$,
 but arcs that appear to be part of the expanding shell can be seen at LSR velocities ranging from about 5.8 to 8.0~km~s$^{-1}$ (see Figure~\ref{cps2fig1}). 
The channel maps and $p-v$ diagram of this shell are consistent with that of an expanding bubble with only the front (blueshifted) hemisphere. Similar to CPS 1, the $p-v$ diagram
shows a ``U'' shape with a minimum LSR velocity of $\sim 5.5$ km~s$^{-1}$ and top velocity of $\sim 8.0$ km~s$^{-1}$ (Figure~\ref{cps2fig2}). From the $p-v$ diagram and the channel maps we estimate   
an expansion velocity of 2.5~km~s$^{-1}$ and the central velocity to be at $V_{LSR}$ = 8.0 km~s$^{-1}$.  
The mass region for this source  was constrained to avoid 
including emission associated with CPS 1.

It is not clear which star powers CPS 2.
SIMBAD shows only one star within $4.8\arcmin$ of the presumed shell center: VSS IX-51. This star has a very low measured optical polarization of  $0.78 \pm 0.37$, which could suggest that this is a foreground star \citep{VSS76}. This area of the sky was not covered by the c2d IRAC maps and thus no c2d YSO candidate lies within the circular structure. There are, however, two point sources with significant 
$24\micron$ emission (SSTc2d J032739.1+310516 and J032742.1+310529) and nineteen 2MASS sources with reliable photometry (from the 2MASS point source catalog) that lie within
$2.5\arcmin$ of the center. None of these stars are relatively more likely to be a candidate shell-powering source; any of these could be the candidate driving source of CPS 2.

\subsection{CPS 3}
\label{cps3sec}
\begin{figure}
\epsscale{1.0}
\plotone{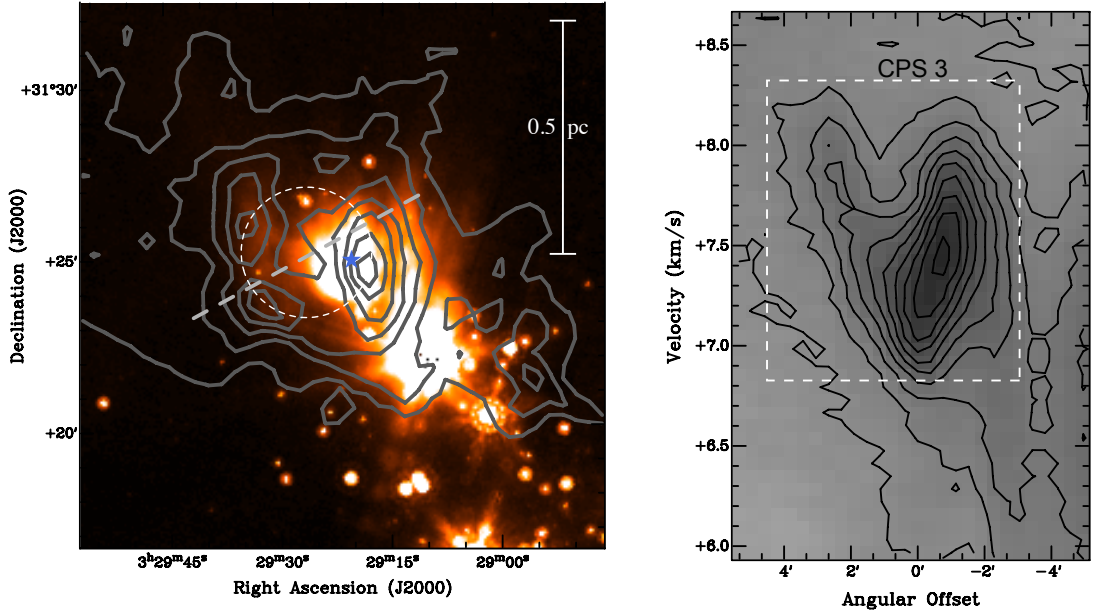}
\caption{ Integrated intensity map and  $p-v$ diagram of $^{12}$CO emission near CPS 3.
The left panel shows the $^{12}$CO(1-0) integrated intensity contours (for  $7.75 < V_{LSR} < 8.25$~km~s$^{-1}$) depicting CPS 3, overlaid on the  MIPS 24 $\micron$ \/ map. Starting contour and contour steps are 3.4 and 0.58 K km s$^{-1}$, respectively.
 The dashed circle shows the approximate extent of CPS 3.  
  The filled star symbol shows the position of the candidate driving source BD+30 549.
The right panel shows the $p-v$ diagram along the cut shown by the diagonal dashed line in the integrated intensity map. 
Positions southeast (northwest) of the center of CPS 3 are shown as positive (negative) offsets. 
The approximate extent of the CO emission associated with CPS 3 in the $p-v$ diagram is shown as a dashed white rectangle.
\label{cps3fig2}}
\end{figure}

BD+30 549 is the B8 star illuminating the NGC~1333 optical nebula. 
 The shell associated with this source was discovered by examining the $^{12}$CO channel maps, where 
 we detect a bright structure that has the characteristics of a small expanding bubble.
The $^{12}$CO \/ data show strong emission in a region of about $2\arcmin$ in size  near the position of BD+30 549, at $V_{LSR} \sim 7.2$~km~s$^{-1}$, 
with a peak intensity that is more than three times the average intensity of the surrounding area (see Figure~\ref{cps3fig1}). 
As the LSR velocity increases, 
the   morphology of the high intensity emission changes to a ring-like structure that slightly increases in size, 
consistent with the velocity structure expected for an
expanding bubble with its front cap seen at $V_{LSR} \sim 7$~km~s$^{-1}$ .
The ring structure is discernible  up to $V_{LSR} \sim 8.2$~km~s$^{-1}$, where the shell walls show
a peak intensity that is about a factor of approximately 2 higher than the surrounding emission. 
At larger $V_{LSR}$,  CO  cloud emission 
is present in this region, but  shell walls are not discernible. 
Our simple expanding bubble model indicates that there should be  a ring like structure for $V_{LSR} > 8.4$~km~s$^{-1}$.
This might not be detected clearly in our data due to our relatively large beam (compared to the size of the shell) and intervening (optically thick) CO emission. 
Higher angular resolution observation of an optically thin line tracer is needed to fully understand the kinematics of this bubble. 
The $p-v$ diagram
shows a clear ``V'' shape with a minimum and maximum LSR velocity at  $\sim 6.8$ and 8.2~km~s$^{-1}$, respectively (Figures~\ref{cps3fig2}).
From the channel maps and $p-v$ diagram we estimate  an expansion velocity of 1.2~km~s$^{-1}$ and the central velocity to be at $V_{LSR}$ = 8.1 km~s$^{-1}$. 
 
The main beam temperature ($T_{mb}$) at the front cap (at $V_{LSR} \sim 7.2$~km~s$^{-1}$)
 is about 45 K (the highest in the entire Persues $^{12}$CO \/ map), 
and in the ring structure (at $V_{LSR} \sim 7.9$~km~s$^{-1}$) the peak $T_{mb}$ is about 30 K.
Assuming the $^{12}$CO  is optically thick 
we can use  
\begin{equation}
T_{ex} = \frac{5.53}{\mathrm{ln}[1 + 5.53/(T_{peak}+0.82)]}
\end{equation}
to obtain an estimate of the excitation temperature ($T_{ex}$), where $T_{peak}$ is the peak intensity 
corrected for main beam
efficiency (Rohlfs \& Wilson 1996).
We find that the excitation temperature in these regions is about 50 and 35 K, 
similar to that found in the molecular gas of high-mass star forming regions. 
It is very likely that such high temperature is due to the strong radiation from BD+30 549.

The candidate source of this expanding bubble lies about $80\arcsec$ to the west of the presumed center of the shell. The bubble's asymmetry, 
with respect to the position of BD+30 549, could be due to a slightly higher density in the gas immediately  west of the bubble, 
where there is strong $^{13}$CO \/ emission. 
There is prominent nebular emission at optical and IR wavelengths associated with this source that  fills the region enclosed by the CO shell (see Figure~\ref{cps3fig2}).


\subsection{CPS 4}
\label{cps4sec}
\begin{figure}
\epsscale{1.0}
\plotone{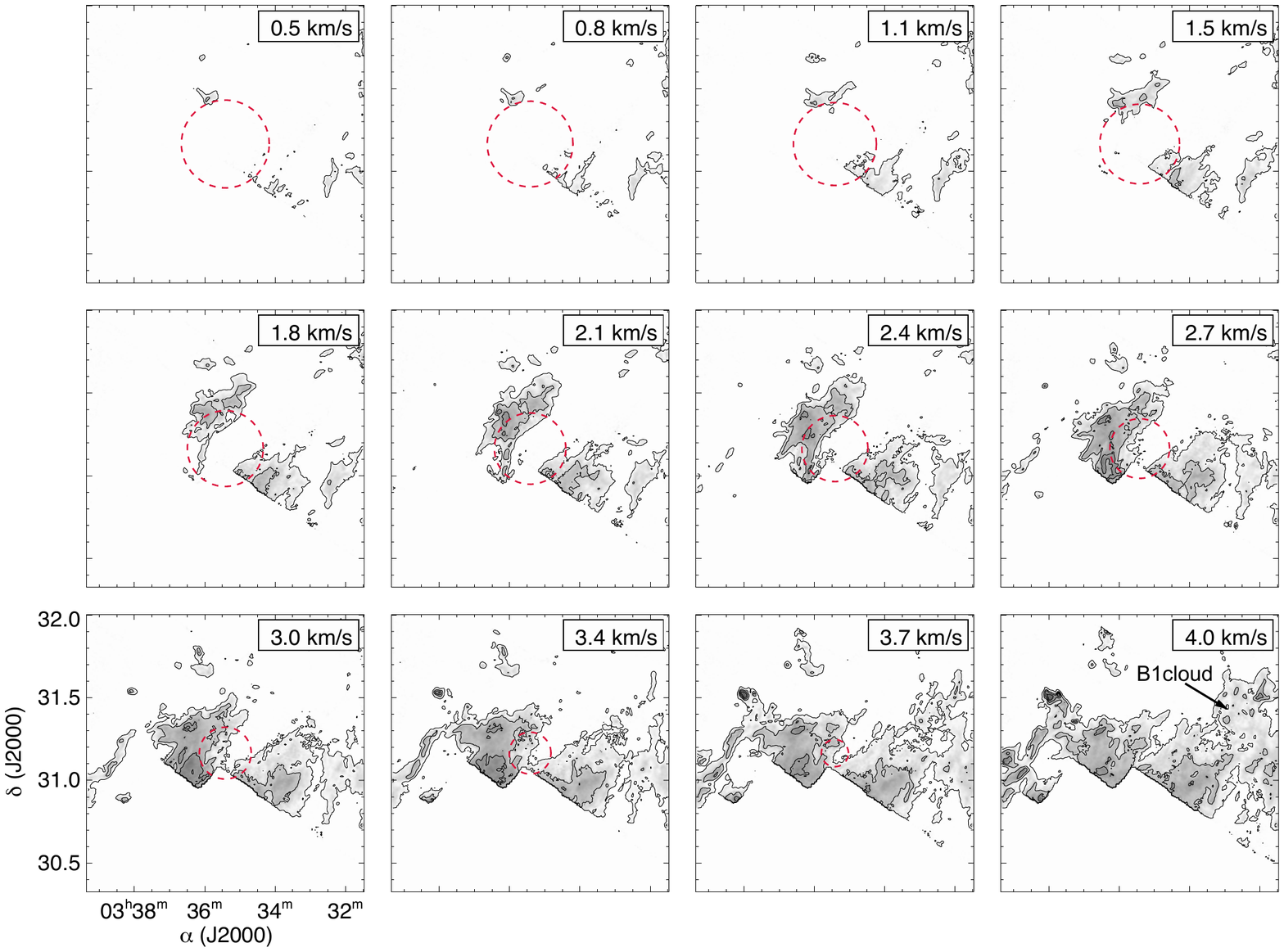}
\caption{Channel maps of $^{12}$CO emission near CPS 4. 
 The number on the upper right corner of each panel indicates the central $V_{LSR}$ of the channel map. 
  Starting contour and contour steps are 0.75 and 1.5 K, respectively.
 Dashed circles shows the expected extent, 
at different radial velocities, of an expanding bubble with a radius, $V_{\mathrm{exp}}$ and central LSR velocity of 16.8\arcmin, 5~km~s$^{-1}$ and
-1.1~km~s$^{-1}$, respectively (using the model discussed in \S~\ref{shellid}).
The location of  the B1 cloud is shown.
\label{cps4fig1}}
\end{figure}

\begin{figure}
\epsscale{1.0}
\plotone{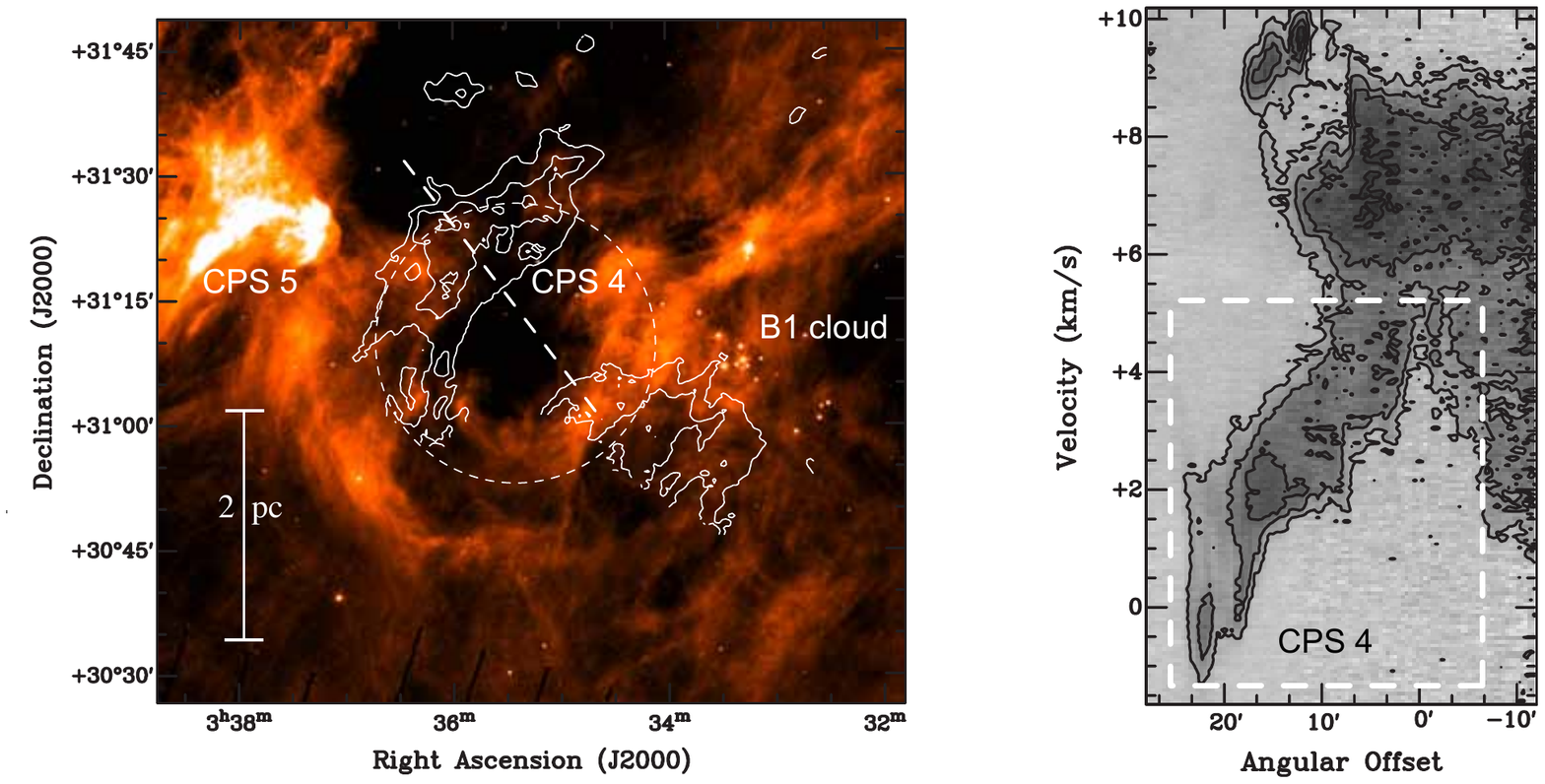}
\caption{ Integrated intensity map and  $p-v$ diagram of $^{12}$CO emission near CPS 4.
The left panel shows the $^{12}$CO(1-0) integrated intensity contours (for  $1.1 < V_{LSR} < 2.7$~km~s$^{-1}$) depicting CPS 4, overlaid 
on the  MIPS 24 $\micron$ \/ map. Starting contour and contour steps are both 1.5 K km s$^{-1}$.
 The  dashed circle shows the approximate extent of the circular IR nebulosity associated with CPS 4. 
The right panel shows the $p-v$ diagram along the cut shown by the diagonal dashed line in the integrated intensity map. 
Positions northeast (southwest) of the center of CPS 4 are shown as positive (negative) offsets. 
The approximate extent of the CO emission associated with CPS 4 in the $p-v$ diagram is shown as a dashed white rectangle.
\label{cps4fig2}}
\end{figure}

The presumed center of this shell lies about $27\arcmin$ east of the group of protostars in the B1 cloud (see Figure~\ref{all13co}), and it is clearly noticeable as a circular structure of about $26\arcmin$ in diameter in the {\it Spitzer}  IRAC 5.8 and 8.0~$\micron$ \/ and MIPS 24 and 70~$\micron$ \/ maps \citep{Jorgensen+06,Rebull+07}. The existence of the molecular gas shell was first noticed when we conducted a search for high velocity $^{12}$CO using the 3D rendering technique described in \citet{Arce+10},  and found high-velocity gas close to the IR nebulosity.
Unlike the c2d IR images, the 
 molecular line emission does not show a clear ring structure.   
 In this region, the gas associated with CPS4  is only clearly discernible in the northern and eastern part of the structure. In the west the shell structure is confused with
  gas associated with the B1 cloud. The southern part of CPS 4 is not covered by the molecular line maps (see Figures~\ref{cps4fig1} and \ref{cps4fig2}).   
$^{12}$CO \/ emission presumably associated with CPS 4 is first seen in at $V_{LSR} \sim -1$~km~s$^{-1}$ in a small area in the northern edge of the circular IR nebulosity, and as the LSR velocity increases more emission is seen along both the northern and eastern parts of the nebulosity (see Figure~\ref{cps4fig1}). At $V_{LSR} >  3.0$~km~s$^{-1}$ gas associated with the B1 cloud  ``contaminate'' the region and it is hard to discern the  emission associated with CPS 4. 
The $p-v$ diagram shows a clear velocity gradient in the CO arc defined by CPS 4, with a minimum and maximum LSR velocity at  $\sim -1$ and 4~km~s$^{-1}$, respectively (Figures~\ref{cps4fig2}). The CO velocity structure of the arc is 
consistent with it being part of the backside (redshifted) hemisphere of an expanding bubble. 
Using the channel maps and $p-v$ diagram we determine that a reasonable fit to the data is an expanding bubble with 
an expansion velocity of 5.0~km~s$^{-1}$ and a central velocity at $V_{LSR}$ = -1.1 km~s$^{-1}$.


A SIMBAD search shows  HD 22179  as the only star within 
a $3.8 \arcmin$  radius of the estimated center of CPS 4   that could be a candidate  driving source of this  shell. 
HD 22197 is a relatively young star,
with a reported age of  30 to 100 Myr \citep{Hillenbrand+08, Roccatagliata+09}, is a relatively bright X-ray source \citep[1RXS J033530.2+311336,][]{Voges+99}, and exhibits mid-IR excess emission associated with a debris disk   ---all properties that make it a suitable candidate source for CPS 4. 
However,  the estimated distance to this source is $100 \pm 20$~pc \citep{Hillenbrand+08, Roccatagliata+09}, significantly smaller than the distance to the Perseus dark cloud.  
Although different sources give slightly different spectral types for HD 22179, 
all are within a few sub-classes and all agree that HD 22197 is an early G-type star 
(G0 by SIMBAD; G3V by Guillout et al.~2009; G5IV by Li \& Hu 1998).  Using the photometry of this source (reported in SIMBAD) and its spectral type provides an estimate of the distance consistent with that given by  \citet{Hillenbrand+08}. In addition, from its kinematics \citet{Guillout+09} estimated that 
 HD 22197 has a 30\% probability of being part of the IC 2391 association and a 50\% probability of being a member of the Pleiades, two clusters that are significantly closer to the Sun than 
 the Perseus molecular cloud. It is therefore very likely that HD 22197 is not associated with the Perseus cloud. 
 
 We searched for other possible sources for CPS 4 in the c2d catalog, but it 
 shows no YSO candidate within a radius of $3\arcmin$ from the center of the shell. There are, however,    
  more than  1154 point sources in this area listed in the c2d point source catalog. 
 Near the center of the shell the IRAC1 image shows a group of stars clustered around
a relatively bright one, SSTc2d J033525.4+310925 (at about $36\arcsec$ from the center of CPS4) classified as ``star'' with  the  slope of the SED between 
between 2 and 20 $\micron$ (or $\alpha$)
equal to -2.55, and near to it (at approximately $1\arcmin$ away from the center of CPS 4) lies  SSTc2d J033526.2+310902, a source that is classified as ``rising'', with  $\alpha = 0.77$. These two seem the most likely candidates from the long list of c2d sources in this region.  However, there are no distances to these sources and no way to determine if any of these are associated with (or are close to) the Perseus cloud.  Note that the fact that both the central velocity of the shell is $V_{LSR} \sim -1$~km~s$^{-1}$, significantly different from the $V_{LSR}$ of the cloud in this region ($\sim 7$~km~s$^{-1}$), and that CPS 4 seems to be tracing only part of the backside hemisphere of an expanding bubble, indicates that the source of this shell lies foreground to the cloud.
In summary, either CPS 4 is driven by a yet undetermined  young star close to the front of the Perseus cloud or it is
 a shell completely unrelated (and foreground) to the Perseus cloud driven by HD 22197.

\subsection{CPS 5}
\label{cps5sec}
\begin{figure}
\epsscale{1.0}
\plotone{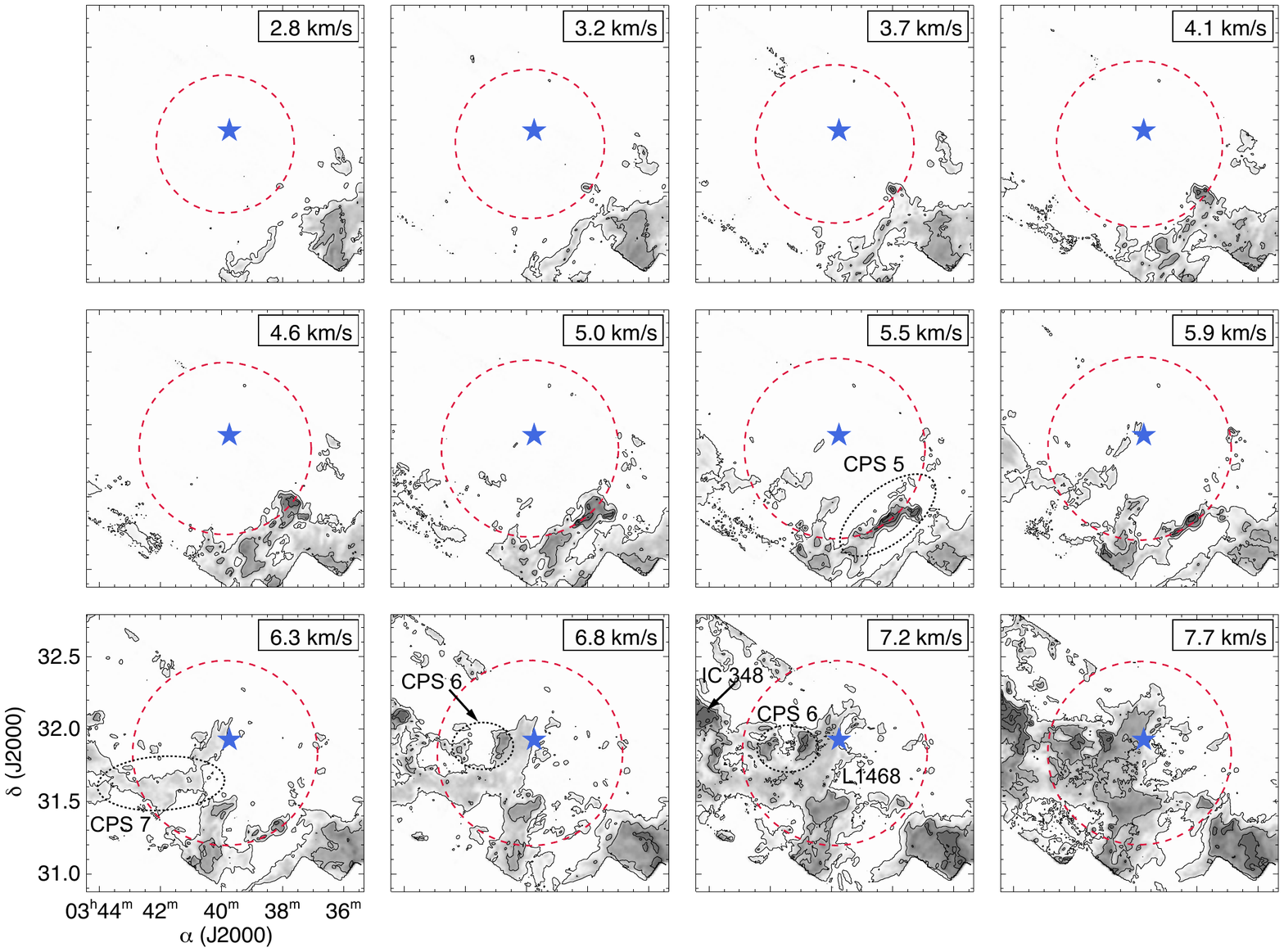}
\caption{
Channel maps of $^{12}$CO emission near CPS 5. 
 The number on the upper right corner of each panel indicates the central $V_{LSR}$  of the channel map. 
  Starting contour and contour steps are 0.6 and 2.4 K, respectively.
 Dashed circles shows the expected extent, 
at different radial velocities, of an expanding bubble with a radius, $V_{\mathrm{exp}}$ and central LSR velocity of 38.3\arcmin, 6~km~s$^{-1}$ and
6.8~km~s$^{-1}$, respectively (using the model discussed in \S~\ref{shellid}).
  The filled star symbol shows the position of the candidate driving source HD 278942.
The location of  the  main arc associated with CPS 5 is shown in the channel map at $V_{LSR} = 5.5$km~s$^{-1}$. The locations
of CO emission associated with the IC 348 cluster, the L1468 cloud, CPS 6 and CPS 7 are also indicated.
\label{cps5fig1}}
\end{figure}

\begin{figure}
\epsscale{1.0}
\plotone{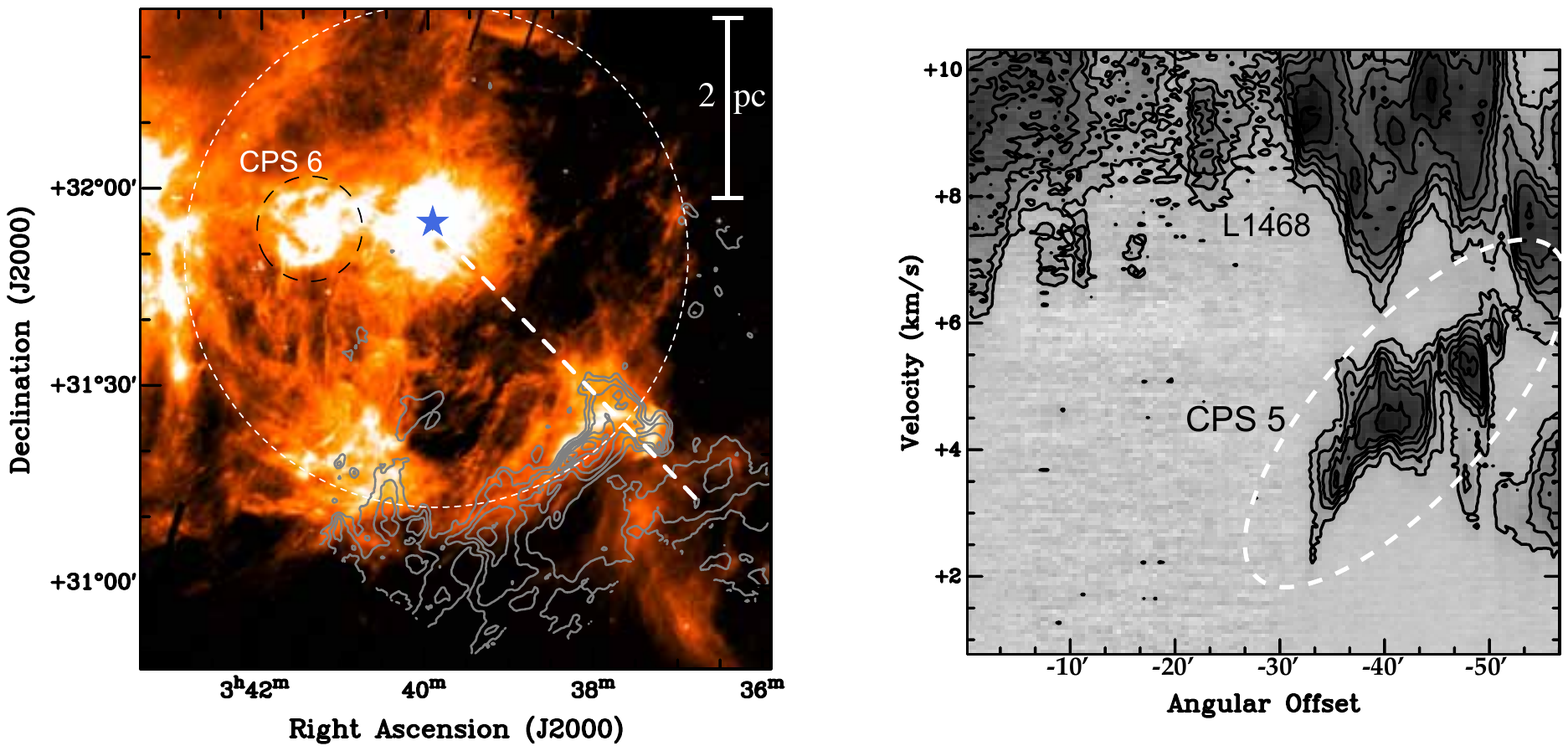}
\caption{Integrated intensity map and  $p-v$ diagram of $^{12}$CO emission near CPS 5.
The left panel shows the $^{12}$CO(1-0) integrated intensity contours (for  $3.9 < V_{LSR} < 6.2$~km~s$^{-1}$) depicting CPS 5, overlaid 
on the  MIPS 24 $\micron$ \/ map. Starting contour and contour steps are both 2 K km s$^{-1}$.
 The white (black) dashed circle shows the approximate extent of the circular IR nebulosity associated with CPS 5 (CPS 6). 
 The filled star symbol shows the position of the candidate driving source HD 278942.
The right panel shows the $p-v$ diagram along the cut shown by the diagonal white dashed line in the integrated intensity map. 
Positions southwest of the center of CPS 5 are shown as negative offsets. 
The approximate extent of the CO emission associated with CPS 5 in the $p-v$ diagram is shown as a dashed white oval. The CO emission associated 
with the L1468 cloud is also labeled in the $p-v$ diagram.
\label{cps5fig2}}
\end{figure}

\begin{figure}
\epsscale{1.0}
\plotone{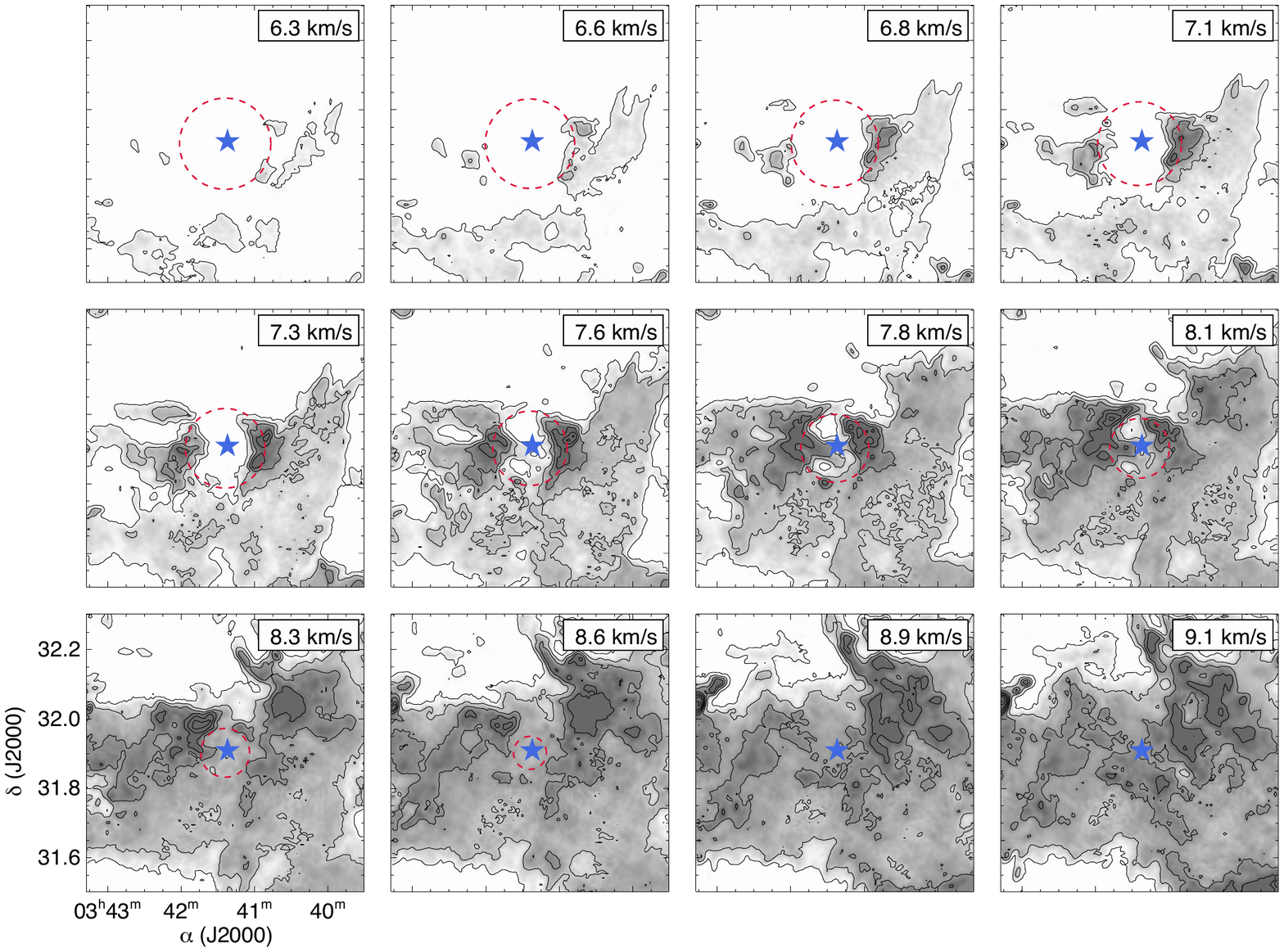}
\caption{ Channel maps of $^{12}$CO emission near CPS 6. 
 The number on the upper right corner of each panel indicates the central $V_{LSR}$  of the channel map. 
  Starting contour and contour steps are both 1.5 K.
 Dashed circles shows the expected extent, 
at different radial velocities, of an expanding bubble with a radius, $V_{\mathrm{exp}}$ and central LSR velocity of 8\arcmin, 3~km~s$^{-1}$ and
5.8~km~s$^{-1}$, respectively (using the model discussed in \S~\ref{shellid}).
The filled star symbol shows the position of the candidate driving source  IRAS  03382+3145.
\label{cps6fig1}}
\end{figure}

The infrared nebulosity and shell associated with CPS 5 is clearly evident in all {\it Spitzer}  IRAC and MIPS bands, as well as in IRAS images of the region 
\citep{Ridge+06b, Jorgensen+06, Rebull+07}. This prominent circular structure, with a radius of about $38\arcmin$ ($\sim 2.8$ pc), has been previously reported by several studies
(see Bally et al.~2007, and references therein) and all agree that HD 278942, at the center of the ring, is the star powering CPS 5. There is some disagreement on the spectral type of HD 278942,  but most estimates indicate that it is a late O or early B star 
(see Bally et al.~2007 for more discussion).
 The study by \citet{Ridge+06b} indicates that the shell is mostly behind the Perseus cloud, and 
 that  it is interacting with parts of the molecular cloud. In the molecular line maps, this interaction is most clearly seen in  the western part of the shell 
 (see Figures~\ref{cps5fig1} and \ref{cps5fig2}).

  We first noticed the molecular gas shell when, searching for outflows in the region, we detected high velocity $^{12}$CO in the southwestern edge of the infrared shell using the 3D visualization 
  technique described in \citet{Arce+10}.  
  CO emission associated with CPS 5 is observed over a range of velocities,  from $V_{LSR} \sim 2$ to 7~km~s$^{-1}$,
  but the arc-like CO structure is most clearly seen  at $V_{LSR} = 5.5$~km~s$^{-1}$, where it shows a morphology very similar to the IR nebulosity 
 (Figures~\ref{cps5fig1} and \ref{cps5fig2}).
The peak CO emission in this arc 
 ($T_{mb} \sim 22$ K) is significantly brighter than the surrounding emission, by about a factor of two. Assuming the emission is optically thick, we estimate the CO excitation temperature
  in this region to be about 25 K (using Equation 1). The increase in excitation temperature in this region is more likely due to the interaction between the wind from HD 278942 and the cloud.
At $V_{LSR} > 6$~km~s$^{-1}$ CO emission from CPS 7, CPS 6 and the L1468 cloud start appearing in the region, and at $V_{LSR} > 6.5$~km~s$^{-1}$ there is little or no CO emission associated with CPS 5 (see Figures~\ref{cps5fig1}). 
 
The $p-v$ diagram for CPS 5 shows a clear velocity gradient in the CO emission. In an expanding bubble, this velocity structure is most likely to arise from an arc that is part of the front hemisphere (consistent with HD 278942 being background to the cloud).
The limited extension of the CO emission (i.e., an arc that is much less than half the circumference of the ring traced by the IR emission) makes it hard to constrain the shell's expansion velocity. 
We use $V_{exp} = 6$~km~s$^{-1}$ and a central velocity of $V_{LSR} = 6.8$~km~s$^{-1}$ for CPS 5 as it results in a reasonable fit of the expanding bubble model to the channel maps (Figure~\ref{cps5fig1}) and it is consistent with the limited $p-v$ diagram of this source. 
We note that the expansion velocity of this source could easily be higher given that we do not to trace CO emission associated with this shell for angular 
offsets between 0 and 30\arcmin \/ in the $p-v$ diagram shown in Figure~\ref{cps5fig2}.
The mass region for this source is constrained to the part of the CPS 5 annulus that lies in the right ascension range between $03^h36^m$ and $03^h40^m$, and declination from $30\arcdeg54\arcmin$ \/ to $32\arcdeg08\arcmin$, in order to avoid emission from the L1468 cloud. 


 
 
\subsection{CPS 6}
\label{cps6sec}
\begin{figure}
\epsscale{1.0}
\plotone{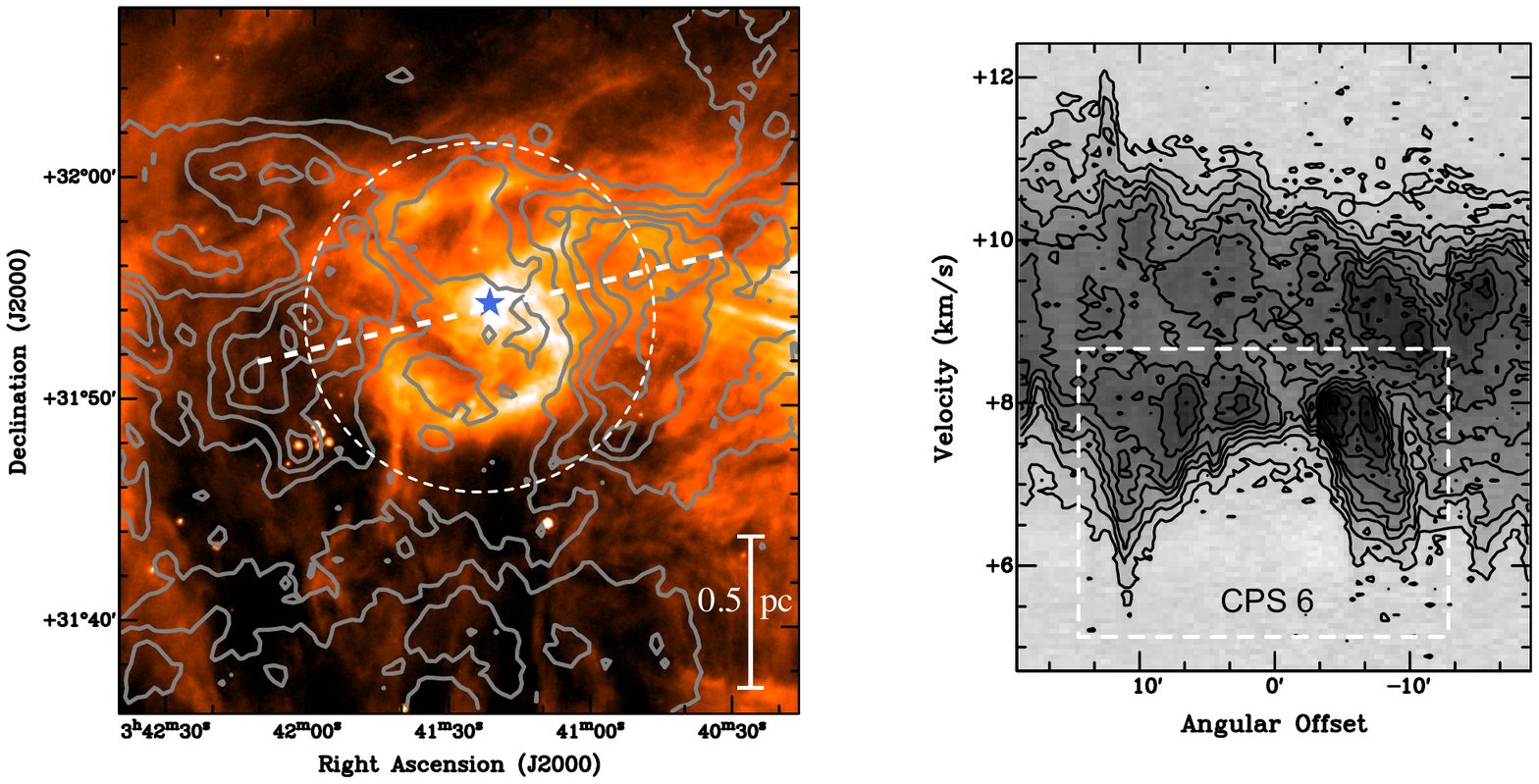}
\caption{Integrated intensity map and  $p-v$ diagram of $^{12}$CO emission near CPS 6.
The left panel shows the $^{12}$CO(1-0) integrated intensity contours (for  $6.48 < V_{LSR} < 8.06$~km~s$^{-1}$) depicting CPS 6, overlaid 
on the  MIPS 24 $\micron$ \/ map. Starting contour and contour steps are 1 and 1.5 K km s$^{-1}$, respectively.
 The white dashed circle shows the approximate extent of the CO circular structure associated with CPS 6. 
 The filled star symbol shows the position of the candidate driving source  IRAS  03382+3145.
The right panel shows the $p-v$ diagram along the cut shown by the diagonal white dashed line in the integrated intensity map. 
Positions southeast (northwest) of the center of CPS 6 are shown as positive (negative) offsets. 
The approximate extent of the CO emission associated with CPS 6 in the $p-v$ diagram is shown as a dashed white rectangle. 
\label{cps6fig2}}
\end{figure}

About $18\arcmin$ due east of HD 278942 lies IRAS 03382+3145, an infrared source that is located at the center of a ring of about 
$5\arcmin$ in radius,  
seen in {\it Spitzer}  MIPS maps \citep{Rebull+07}. 
At near- to mid-IR wavelengths, the MSX (8.1 and 12.1 $\micron$ bands) and {\it Spitzer}  IRAC images exhibit  bright nebulous emission coincident with the interior of the ring-like structure seen in the MIPS images \citep{Kramer+03, Jorgensen+06}.
We first noticed the molecular gas shell when we detected high velocity $^{12}$CO in the region 
and further inspection revealed that the channel maps show a clear ring-like morphology coincident with the 
IR nebulosity, with a velocity structure  consistent with that of an expanding bubble.
 The CO emission associated with CPS 6 is most clearly seen  
from  $V_{LSR} \sim 7$ to 8~km~s$^{-1}$, but emission associated with this shell is seen over a wider range of velocities 
 where the ring structure is observed to slightly diminishes in size 
as the LSR velocity increases (see Figure~\ref{cps6fig1}). The $^{12}$CO integrated intensity (for $ 6.5 < V_{LSR} < 8.1$~km~s$^{-1}$) shows a clear ring structure with a radius 
of  about $8\arcmin$ ($\sim 0.6$ pc), slightly larger than (and enclosing) the ring seen in the {\it Spitzer}  MIPS 24\micron \/ image (see Figure~\ref{cps6fig2}).
The $p-v$ diagram shows an inverse "U" shape with a minimum LSR velocity at  $\sim 5.5$~km~s$^{-1}$ and a maximum LSR velocity of  
 $\sim 8.5$~km~s$^{-1}$, where the shell emission blends with the cloud emission (Figures~\ref{cps6fig2}).
This velocity structure is consistent with that expected of the backside hemisphere of an expanding bubble 
that has a central velocity offset (towards the blue) with respect to the cloud central velocity (similar to the scenario presented in  Figure~\ref{shellmodel}d).
From the channel maps and $p-v$ diagram we estimate  $V_{exp} = 3$~km~s$^{-1}$ and the central velocity to be at $V_{LSR}$ = 5.8 km~s$^{-1}$.
%

The source IRAS 03382+3145, 
right at the center of CPS 6, is the best candidate driver of the surrounding expanding shell. There is not much information on this source, nor there are any 
estimates of its spectral type. In the c2d catalog, the point source  SSTc2d J034122.1+315434, detected in all 2MASS and IRAC bands, is 
the closest source to the position (and lies within the position error ellipse) of IRAS 03382+3145, making it the most likely counterpart to the IRAS source.
It  is classified  as  a ``star'' in the c2d catalog 
(i.e, with NIR SED consistent with a stellar photosphere), and has an $\alpha = -2.65$.  
 Yet, its clear association to the Perseus cloud
 imply that this is not likely a typical main-sequence star and it is very probable that this is a young pre-main-sequence star.
 There is no reliable MIR detection in the MIPS bands  
 because the source is confused with the bright nebulosity in this region \citep{Rebull+07}
 
\subsection{CPS 7}
\label{cps7sec}
\begin{figure}
\epsscale{1.0}
\plotone{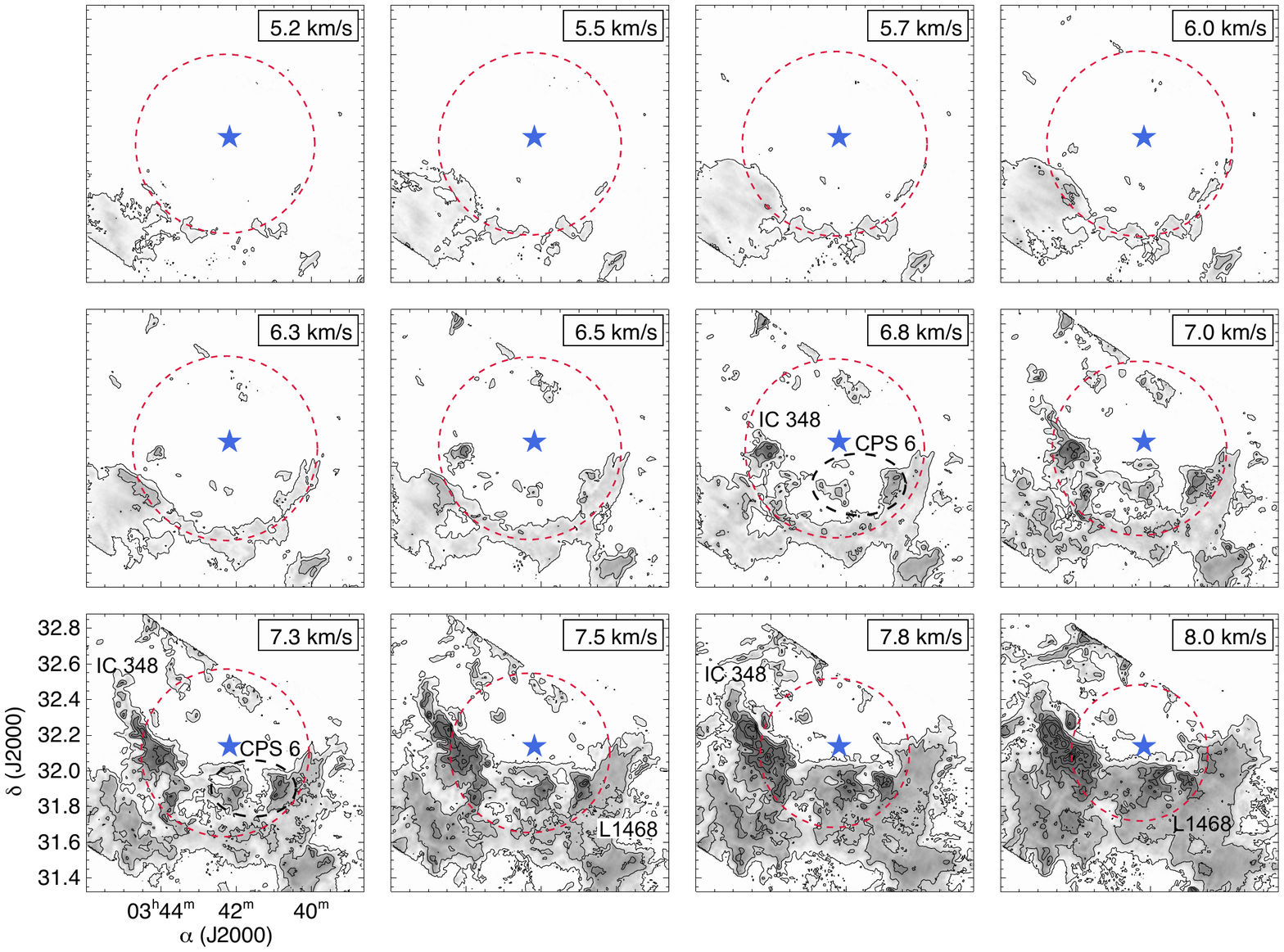}
\caption{ Channel maps of $^{12}$CO emission near CPS 7. 
 The number on the upper right corner of each panel indicates the central $V_{LSR}$  of the channel map. 
  Starting contour and contour steps are 0.75 and 2 K, respectively.
 Dashed circles shows the expected extent, 
at different radial velocities, of an expanding bubble with a radius, $V_{\mathrm{exp}}$ and central LSR velocity of 31.1\arcmin, 3~km~s$^{-1}$ and
6.0~km~s$^{-1}$, respectively (using the model discussed in \S~\ref{shellid}).
The filled star symbol shows the position of the candidate driving source  IRAS 03390+3158.
The location
of CO emission associated with the IC 348 cluster, the L1468 cloud, and CPS 6 is also indicated.
\label{cps7fig1}}
\end{figure}

\begin{figure}
\epsscale{1.0}
\plotone{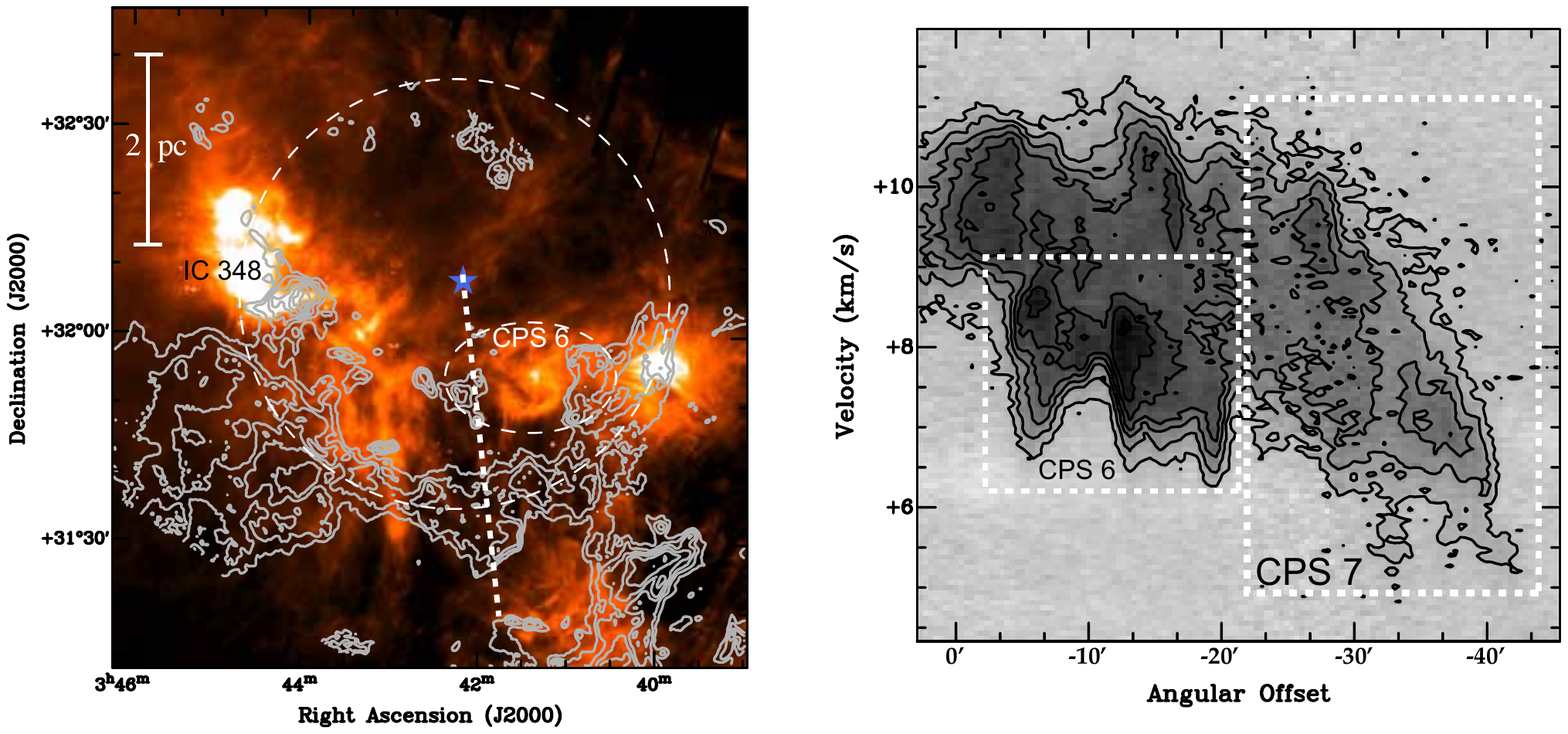}
\caption{Integrated intensity map and  $p-v$ diagram of $^{12}$CO emission near CPS 7.
The left panel shows the $^{12}$CO(1-0) integrated intensity contours (for  $5.52 < V_{LSR} < 7.11$~km~s$^{-1}$) depicting CPS 7, overlaid 
on the  MIPS 24 $\micron$ \/ map. Starting contour and contour steps are both 1  K km s$^{-1}$.
 The white dashed circle shows the size and position of the circle we visually fitted to the CPS 7 CO emission. The white dashed oval
 indicates the position of CPS 6.  
 The filled star symbol shows the position of the candidate driving source  IRAS 03390+3158.
 The position of the IR nebulosity associated with IC 348 is also labeled. 
The right panel shows the $p-v$ diagram along the cut shown by the thick white dashed line in the integrated intensity map. 
Positions southwest of the center of CPS 7 are shown as  negative offsets. 
The approximate extent of the CO emission associated with CPS 7 (CPS 6) in the $p-v$ diagram is shown as a large (small) dashed white rectangle. 
\label{cps7fig2}}
\end{figure}

CPS 7 was first noticed when, using our 3D visualization technique used to search for high-velocity features in the Perseus cloud, 
we detected a few positions that show blue-shifted CO emission, at velocities significantly  different from 
the cloud velocity, along the southern edge of the Perseus molecular cloud (about $20\arcmin$ southeast of the IC 348 cluster).
Examining the channel maps we noticed a wide arc of about $70\arcmin$ in extent, tracing approximately $3/8$ of a circle with a radius of approximately  $30\arcmin$ centered around  
$03^{h}42^{m}18^{s}$, $32\arcdeg06\arcmin10\arcsec$ (J2000), that is clearly separate from the rest of the cloud emission (see Figures~\ref{cps7fig1} and \ref{cps7fig2}).
 We detect faint clumpy emission associated with CPS 7 down to  $V_{LSR} \sim 5$~km~s$^{-1}$, but the arc structure is best seen at  $V_{LSR}$ between 6 and 7~km~s$^{-1}$. 
 At $V_{LSR} \geq 6.8$~km~s$^{-1}$ emission associated with the IC 348 cluster and CPS 6 overlaps with the eastern and western ends of CPS 7, respectively. Additional emission
associated with the L1468 cloud at $V_{LSR} \geq 7.5$~km~s$^{-1}$ confuses the region, making it impossible to isolate  
 the structure associated with CPS 7 at these velocities (Figures~\ref{cps7fig1}).
 The $p-v$ diagram in Figures~\ref{cps7fig2} shows a clear velocity gradient in the CO emission  associated with CPS 7 (i.e., the emission at angular offset from -22\arcmin \/ to -40\arcmin).
This is consistent with the velocity structure expected from part of the backside hemisphere of an expanding bubble.
 From the channel maps and $p-v$ diagram we estimate an expansion velocity of 3~km~s$^{-1}$  and the central velocity to be at $V_{LSR}$ = 6.0 km~s$^{-1}$.
 Although there is no clear IR counterpart to the CPS 7 structure observed in CO, it does not necessarily mean that it does not exist. This region of the cloud is filled with the bright nebulous IR emission associated with CPS 5 and CPS 6 to the west and the IC 348 cluster (and CPS 8) to the east. Thus any faint IR emission associated with CPS 7 may be confused with the bright emission from these nearby shells. 
The mass region is constrained to the part of the CPS 7 annulus that lies west of $03^h44^m$, in order to avoid emission associated with IC 348.


The most promising candidate driving source for CPS 7 is IRAS 03390+3158, which lies approximately $2.6\arcmin$ from the estimated center of the shell.
This source, which by chance is coincident with the IR nebulosity associated with the CPS 5 dust ring (from HD 278942),
 is an optical variable star known as V633 Per (with an observed variation in $V$ from 13.0 to 13.5 mag), and it is listed in the 2MASS 
 and c2d point source catalogs.  SSTc2d J034210.9+320817, as it is known in the c2d point source catalog, is only detected in IRAC bands 1 and 3 as the other two IRAC bands did not cover the area.  The source is also confidently detected at  $12\micron$ with IRAS and $24\micron$ in the MIPS1 band. Detection of this source at $60\micron$ and 
 $70\micron$ with IRAS and {\it Spitzer} , respectively, is probably hampered by contamination from the  surrounding nebulous emission from the dust ring associated with CPS 5. 
 The IRAS point source catalog list a detection at $100\micron$, but the flux at this wavelength is probably contaminated by the surrounding nebulosity. 
 From the location of this star  in a J-H vs.~H-K plot, it is apparent that its IR colors are consistent with a
  reddened K5 to M0 main-sequence star or reddened classical Be star \citep{LA92}. It is highly unlikely that this star would be a late type main-sequence star, as its $B$ and $V$ apparent magnitudes (15.7 and
  13.0 to 13.5 mag, respectively) and color ($B-V = 2.2$ to 2.7 mag) imply that it would be very close to the Sun (11 to 32 pc) and have an extremely unlikely 
   high extinction  ($A_V = 2.5$ to 4.0 mag) for a nearby star. It, seems more plausible that this is a young late B or early A star with an extinction of about 6 to 8 mag, residing in or closely to the  Perseus cloud at a distance of about 250 pc. 
  
\subsection{CPS 8}
\label{cps8sec}
\begin{figure}
\epsscale{1.0}
\plotone{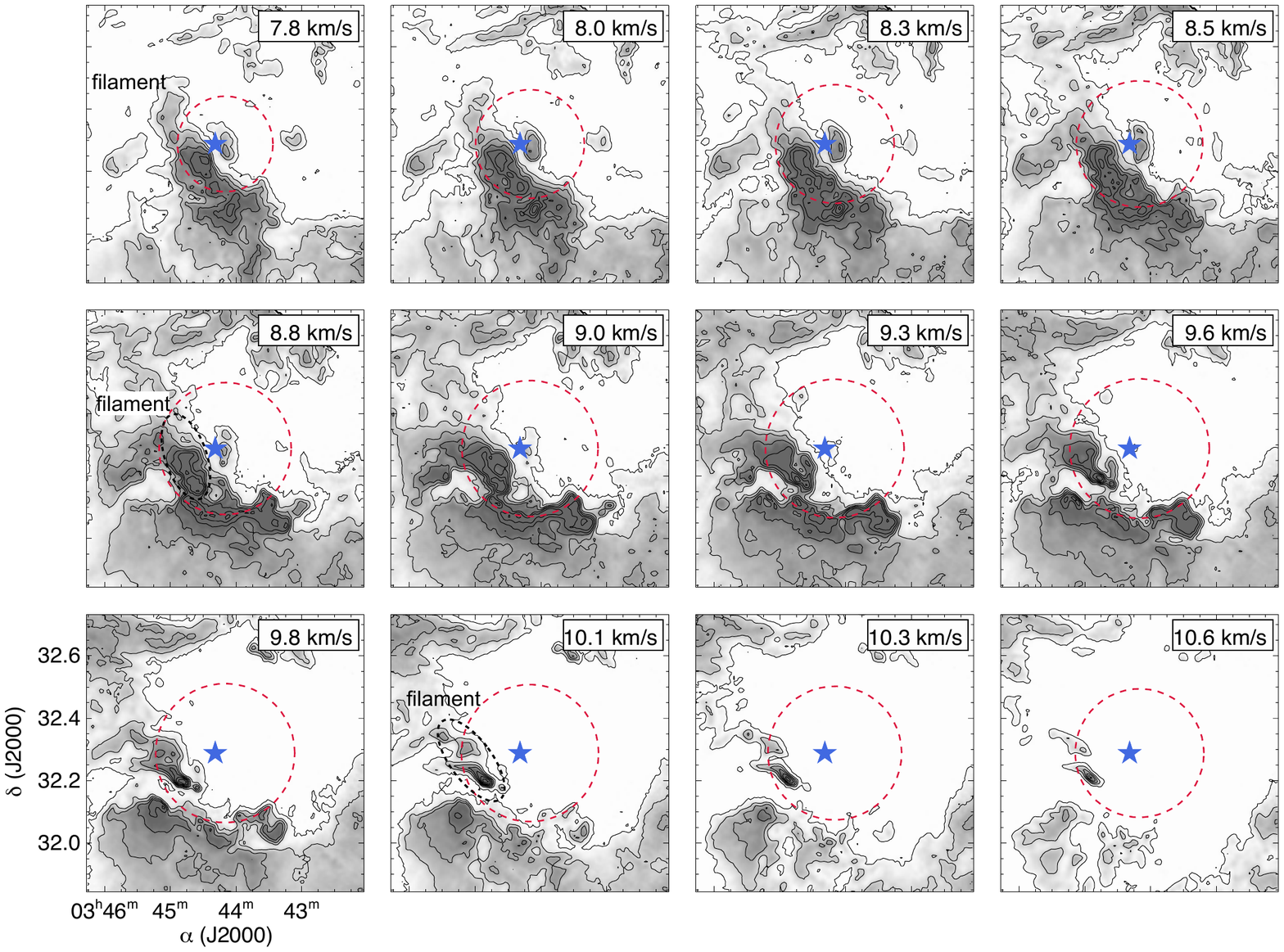}
\caption{ Channel maps of $^{12}$CO emission near CPS 8. 
 The number on the upper right corner of each panel indicates the central $V_{LSR}$  of the channel map. 
  Starting contour and contour steps are 1 and 2 K, respectively.
 Dashed circles shows the expected extent, 
at different radial velocities, of an expanding bubble with a radius, $V_{\mathrm{exp}}$ and central LSR velocity of 13.4\arcmin, 2.5~km~s$^{-1}$ and
9.6~km~s$^{-1}$, respectively (using the model discussed in \S~\ref{shellid}).
The filled star symbol shows the position of the candidate driving source $o$ Per.
\label{cps8fig1}}
\end{figure}

\begin{figure}
\epsscale{1.0}
\plotone{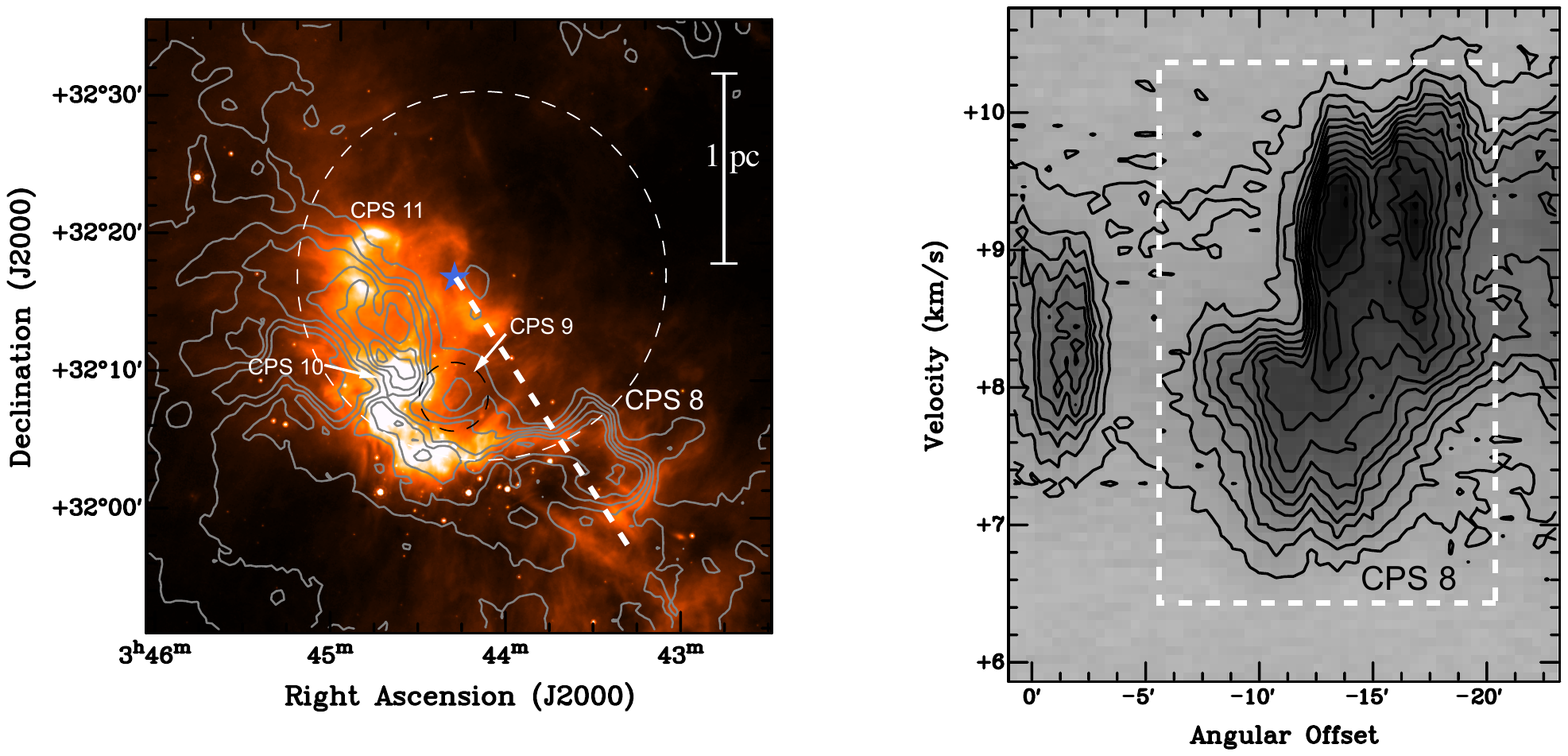}
\caption{Integrated intensity map and  $p-v$ diagram of $^{12}$CO emission near CPS 8.
The left panel shows the $^{12}$CO(1-0) integrated intensity contours (for  $9.27< V_{LSR} < 9.65$~km~s$^{-1}$) depicting CPS 8, overlaid 
on the  MIPS 24 $\micron$ \/ map. Starting contour and contour steps are 3 and 2 K km s$^{-1}$, respectively.
 The white dashed circle shows the size and position of the circle we visually fitted to the CPS 8 CO emission.  
The location of CO emission associated with CPS 9 is shown with a black dashed circle. 
The filled star symbol shows the position of the candidate driving source $o$ Per.
The positions of the IR nebulosity associated with CPS 10 and CPS 11 are also shown.
The right panel shows the $p-v$ diagram along the cut shown by the diagonal thick white dashed line in the integrated intensity map. 
Positions southwest of the center of CPS 8 are shown as  negative offsets. 
The approximate extent of the CO emission associated with CPS 8 in the $p-v$ diagram is shown as a dashed white rectangle. 
\label{cps8fig2}}
\end{figure}

\begin{figure}
\epsscale{1.0}
\plotone{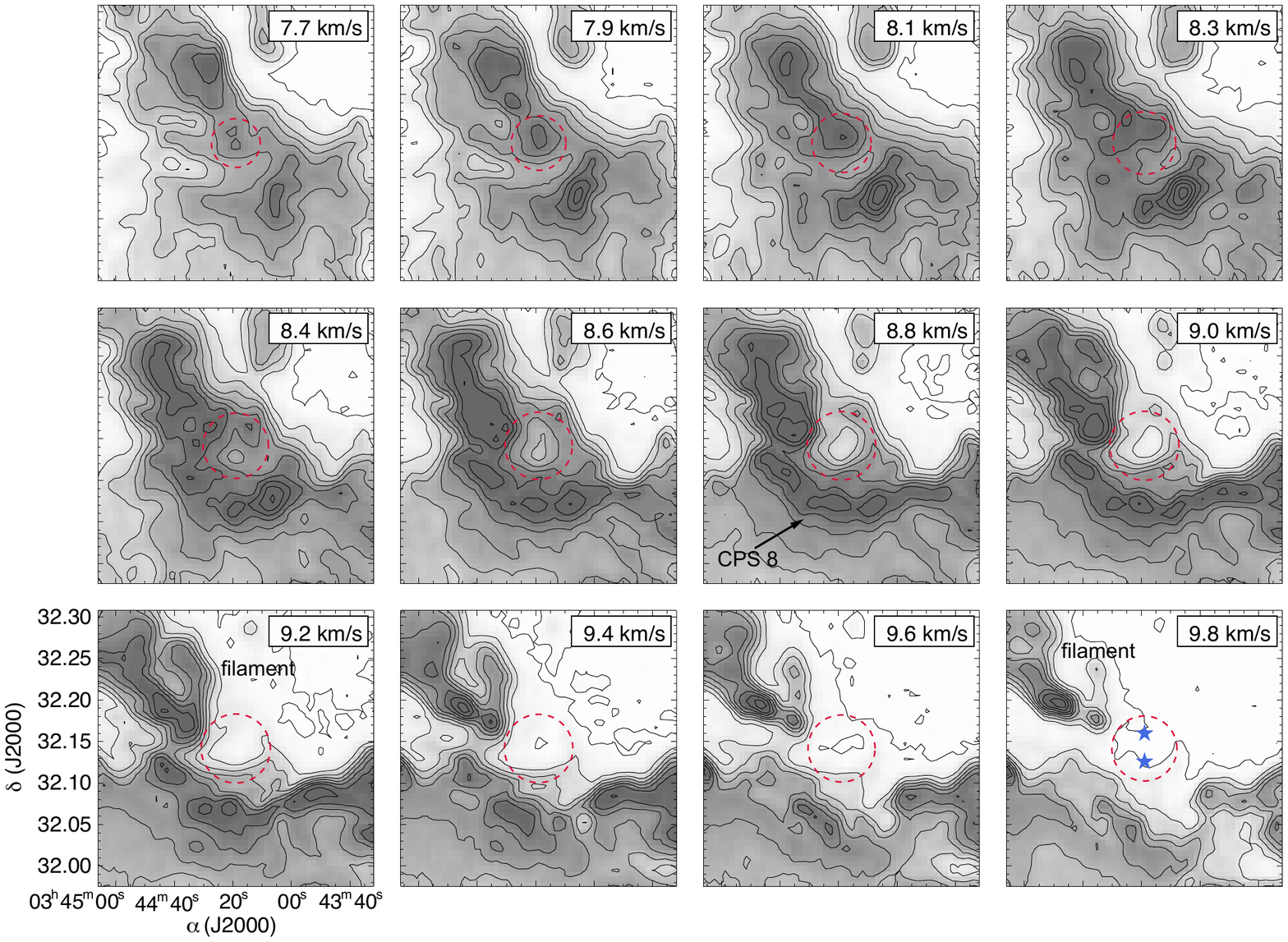}
\caption{ Channel maps of $^{12}$CO emission near CPS 9. 
 The number on the upper right corner of each panel indicates the central $V_{LSR}$  of the channel map. 
  Starting contour and contour steps are 0.8 and 1.6 K, respectively.
 Dashed circles shows the expected extent, 
at different radial velocities, of an expanding bubble with a radius, $V_{\mathrm{exp}}$ and central LSR velocity of 2.5\arcmin, 2~km~s$^{-1}$ and
9.1~km~s$^{-1}$, respectively (using the model discussed in \S~\ref{shellid}).
The filled star symbols in the channel map at $V_{LSR} = 9.8$~km~s$^{-1}$
show the position of the two candidate  sources: V* 695 Per (to the south) and IC 348 LRL  30 (to the north).
\label{cps9fig1}}
\end{figure}

CPS 8 is associated with the bright IR nebulosity observed in the {\it Spitzer}  images presented by \citet{Muench+07} that lies in the southern part of the IC 348 cluster. 
We detected this shell when we examined the CO channel maps and noted  that, 
south of the central part of the main IC 348 cluster, the  emission at the northern edge of the molecular cloud
 has a curved (arc-like) morphology that is significantly brighter than the surrounding CO  emission and is coincident with the bright border of the IR nebulosity
 (see Figures~\ref{cps8fig1} and \ref{cps8fig2}).
 The bright curved wall of molecular gas emission, which traces an arc that is less than half a circle, changes position with LSR velocity, consistent with the kinematics of an expanding shell (see Figure~\ref{cps8fig1}). Molecular gas emission associated with this structure is seen from $V_{LSR} \sim 8$~km~s$^{-1}$ 
up to approximately  10~km~s$^{-1}$, but the structure is most clearly discerned at  $V_{LSR} \sim 9.3$~km~s$^{-1}$ where it shows a peak 
$T_{mb}$ of about 30 K (which corresponds to $T_{ex} \sim 34$~K , assuming the CO is optically thick and using Equation 1).
This is consistent with the gas being heated by the same external source responsible for the nebulosity.
At all velocities where CO emission associated with CPS 8 is detected, but specially at  $V_{LSR} \leq 8.8$~km~s$^{-1}$,
there is also emission associated with a filament that extends northwest of (and it is unrelated to) CPS 8.
The $p-v$ diagram shows a rough velocity gradient in the arc-like CO emission (i.e., increasing $V_{LSR}$ with increasing distance from the center, for angular offsets 
from 10\arcmin \/ to 20\arcmin).
Assuming an expanding bubble is responsible for this gradient, it is then most likely that the CO arises from  part of its front hemisphere. 
Similar to CPS 5 and 7 (where the CO shell is limited to less than half a circle) it is hard to constrain the shell's expansion velocity and we can only obtain a lower limit. A reasonable estimate, 
using the $p-v$ diagram and fitting the expanding bubble model to the channel maps, is an expansion velocity of $\sim 2.5$~km~s$^{-1}$ and a central velocity 
at  $V_{LSR} = 9.6$~km~s$^{-1}$. 
The mass region in CPS 8 is limited to the shell area south of $32\arcdeg11\arcmin$, in order to avoid emission from the filament in the region.

A circle with a radius of $\sim 13\arcmin$ centered on $03^{h}44^{m}10^{s}$, $32\arcdeg17\arcmin20\arcsec$ (J2000) provides a reasonable  visual fit to the  curved wall of high-intensity CO associated with CPS 8
Within $3\arcmin$ of this circle's center the only known early type star capable of illuminating the bright IR nebula is the B1III star $o$ Per (at $\sim 1.9\arcmin$ from the center of CPS 8). It has been typically assumed that this star is not a member of the IC 348 cluster nor of the nearby Per OB association, mainly due to its peculiar proper motions as compared to the rest of the association \citep{Fredrick56, deZeeuw+99}. 
However, the estimated distance to this star by \citet{Cernis93}, 
about $280 \pm 50$ pc, is consistent with the average distance to the cluster of $300 \pm 15$~pc 
obtained from different methods \citep{Herbst08}, and it is quite possible that $o$ Per is sufficiently close (within $\sim 1$ pc) to the cluster  to be responsible for CPS 8. 
This bright source is observed in the optical and the infrared, and it is classified as a YSO candidate in the c2d point source catalog, where it is named
SSTc2d J034419.1+321718.

Evidence that $o$ Per might be responsible for CPS 8 also comes  from the fact that extended 8~\micron \/ emission (in the {\it Spitzer} IRAC map of the region) is coincident with the 
CO and 24~\micron \/ arc-like structures \citep{Jorgensen+06}.
It is generally thought that the 8~\micron \/ emission is dominated by PAHs that are heated by far-UV photons and  $o$ Per is the only source in the region that seems to be capable of producing enough UV photons to heat the entire arc associated with CPS 8. 
The candidate drivers  of CPS 10 and 11 are able to heat their surroundings and form  small rings that are seen in the IR, but these only extend to no more than 2\arcmin \/ from their sources' position ---too small to be responsible for the structure associated with CPS 8. Further indirect evidence that $o$ Per is responsible for CPS 8
comes from the study of \citet{Sun+08}, who derive the far-UV (FUV) field at different positions in the IC 348 region by a fitting a PDR model to the observed intensity ratios, using lines of CO, $^{13}$CO and [CI]. The derived FUV flux at positions coincident with CPS 8 is a factor of 3 to 7 higher than the FUV flux expected to arise solely from HD 281159 (the source of CPS 10). These results imply that an additional source of FUV photons is needed to account for the high FUV flux in the region, and  $o$ Per seems to be the most likely source.

\subsection{CPS 9}
\label{cps9sec}
This relatively small shell, seen only in the molecular line maps, lies approximately $5\arcmin$ north of the eastern part of CPS 8 (see Figure~\ref{cps8fig2}). It was first noticed when we examined in detail the CO channel maps in the IC 348 region.  
The blueshifted (foreground) cap of the expanding bubble is observed at $V_{LSR} \sim 7.7 - 8.1$~km~s$^{-1}$. At larger LSR velocities the CO shows a ring-like structure that increases slightly with radius, and reaches a diameter of approximately $5\arcmin$ at its largest extent (see Figure~\ref{cps9fig1}). We detect CO associated with CPS 9 all the way up to $V_{LSR} \sim 9.9$~km~s$^{-1}$. However, at $V_{LSR} > 9 $~km~s$^{-1}$ only an arc-like structure is observed in the southern part of the shell. It appears that CPS 9 has blown-out of the 
low-density (northern) edge of the cloud. 
The CO structure associated with CPS 9 has a slight elongation in the southwest-northeast direction at $V_{LSR} = 8.6 - 9.0$~km~s$^{-1}$. This morphology could be due to a 
lower expansion rate towards the region with higher density in the southwestern part of the shell. 
We  need higher angular resolution data to confirm this.
We find no clear IR counterpart to the CPS 9 structure observed in CO. However, this  does not necessarily mean that it does not exist as  this region of the cloud is filled with the bright nebulous IR emission associated with CPS 8 and 10, and the  CPS 9 IR emission (it it exists) could be confused with other IR nebulosity in the region. 

\begin{figure}
\epsscale{1.0}
\plotone{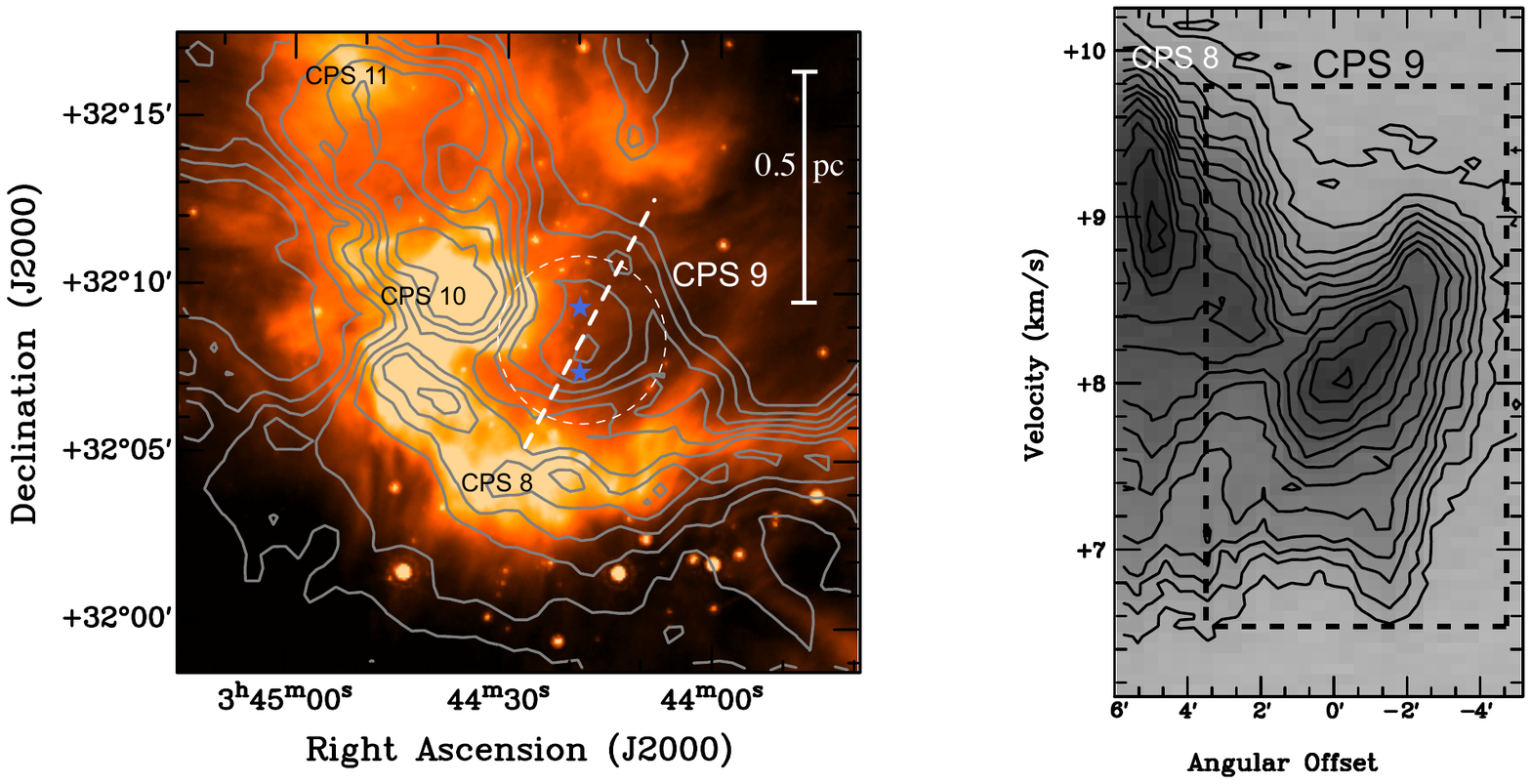}
\caption{Integrated intensity map and  $p-v$ diagram of $^{12}$CO emission near CPS 9.
The left panel shows the $^{12}$CO(1-0) integrated intensity contours (for  $8.57< V_{LSR} < 9.21$~km~s$^{-1}$) depicting CPS 9, overlaid 
on the  MIPS 24 $\micron$ \/ map. Starting contour and contour steps are 2 and 1 K km s$^{-1}$, respectively.
 The white dashed circle shows the extent of the CO emission associated with CPS 9.  
The location of CO emission associated with CPS 9 is shown with a black dashed circle. 
The filled star symbols show the position of the two candidate  sources: V* 695 Per (to the south) and IC 348 LRL  30 (to the north).
The positions of the IR nebulosity associated with CPS 8, CPS 10 and CPS 11 are also shown.
The right panel shows the $p-v$ diagram along the cut shown by the diagonal thick white dashed line in the integrated intensity map. 
Positions southwest (northeast) of the center of CPS 9 are shown as  positive (negative) offsets. 
The approximate extent of the CO emission associated with CPS 9 in the $p-v$ diagram is shown as a dashed (black) rectangle. The CO emission
associated with CPS 8 is also indicated in the $p-v$ diagram.  
\label{cps9fig2}}
\end{figure}

The $p-v$ diagram in Figure~\ref{cps9fig2} shows an asymmetric ``V-like'' structure
with a minimum LSR velocity at  $\sim 7.5$~km~s$^{-1}$ and different maxima LSR velocity at the two ends:  
 about 9.0~km~s$^{-1}$ at the northwestern (negative angular offset) end and approximately 9.5~km~s$^{-1}$  at the southwestern (positve angular offset) end.
This asymmetry is due to the fact that there is 
little or no CO emission in the northern edge of CPS 9 at $V_{LSR}  > 9 $~km~s$^{-1}$, as indicated above, and possibly also due to emission associated to CPS 8 in the southern part of CPS 9.
 The velocity structure in the $p-v$ diagram is consistent with that of the front hemisphere of an expanding bubble (similar to that shown in Figure~\ref{shellmodel}c).  From the channel maps and $p-v$ diagram we estimate an expansion velocity of 2~km~s$^{-1}$  and the central velocity to be at $V_{LSR}$ = 9.1 km~s$^{-1}$. 
The mass region for CPS 9 is constrained not to include regions included in CPS 8 or CPS 10.

A search in SIMBAD for known stars within $65\arcsec$ of the center of CPS 9 yields 35 sources. Two possible candidates are V* 695 Per and IC 348 LRL 30. The former  is 
an M3.75 star that is an optical variable  with detected X-ray emission \citep{PZ01} and is classified as a YSO candidate  in the c2d catalog (where it is named SSTc2d J034419.2+320735).  
IC 348 LRL 30 is a F0 star that has been classified as a YSO candidate in the c2d catalog (where it is named SSTc2d J034419.1+320931) and has been also detected in X-ray (CXOPZ J034419.1+320931, Preibisch \& Zinnecker 2001). These two sources are plotted in  Figure~\ref{cps9fig2}, where it can be seen that none of these lie at the exact center, V* 695 Per is
$\sim 55\arcsec$ to the south of the center, while IC 348 LRL 30 lies $\sim 60\arcsec$ to the north. Out of these two sources, IC 348 LRL 30 seems the most likely as it the star with the 
earliest spectral classification. Another F type star in the region is IC 348 LRL 28 (at about $60\arcsec$ from the center), but it appears to be a foreground star \citep{Luhman+03} so we discard it as a candidate source.

 \subsection{CPS 10}
 \label{cps10sec}
\begin{figure}
\epsscale{1.0}
\plotone{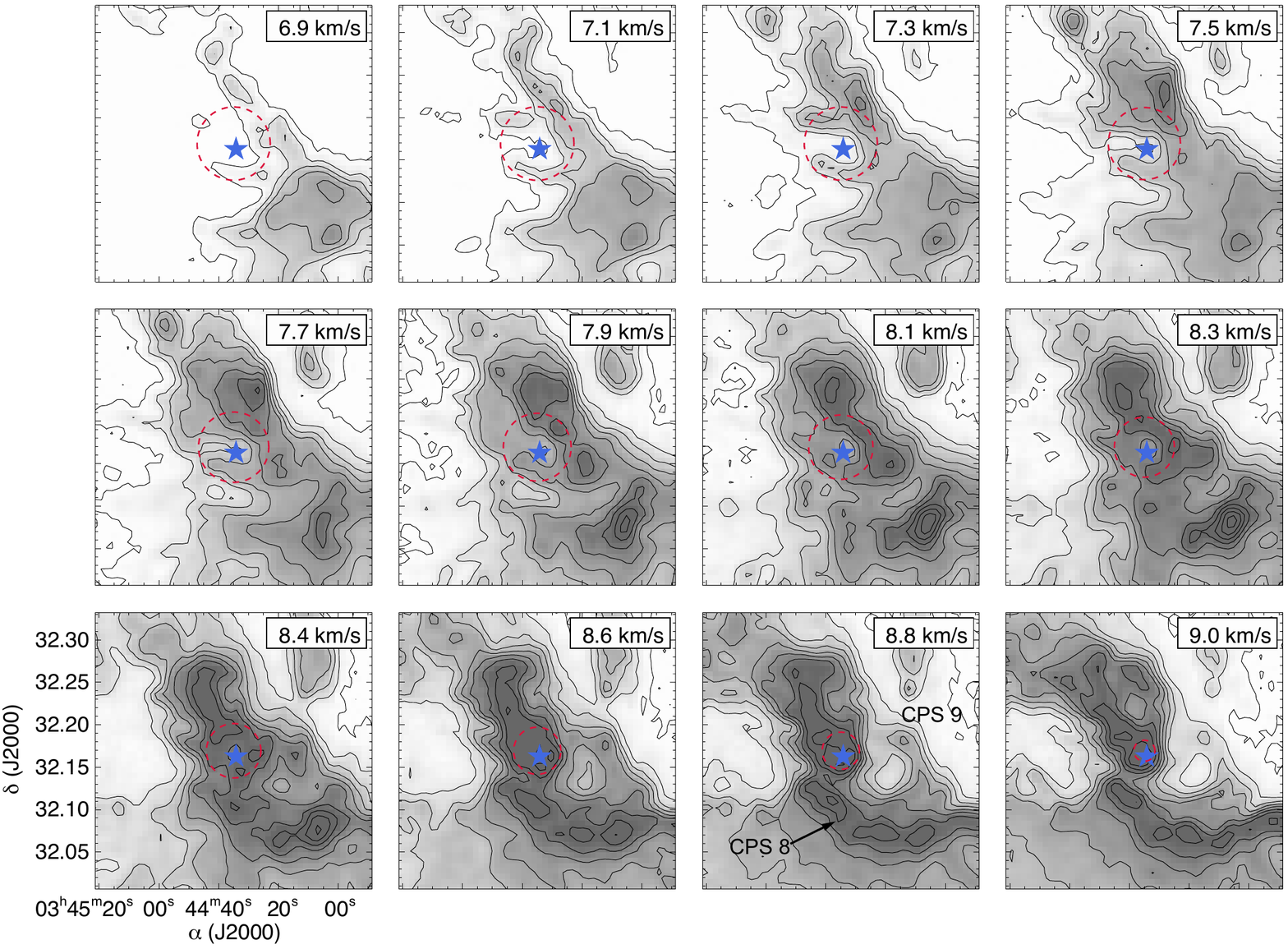}
\caption{ Channel maps of $^{12}$CO emission near CPS 10. 
 The number on the upper right corner of each panel indicates the central $V_{LSR}$  of the channel map. 
  Starting contour and contour steps are 0.8 and 1.6 K, respectively.
 Dashed circles shows the expected extent, 
at different radial velocities, of an expanding bubble with a radius, $V_{\mathrm{exp}}$ and central LSR velocity of 2.6\arcmin, 2~km~s$^{-1}$ and
7.1~km~s$^{-1}$, respectively (using the model discussed in \S~\ref{shellid}).
The filled star symbol shows the position of the candidate driving source HD 281159.
The locations of structures associated with CPS 8 and CPS 9 are also labeled.
\label{cps10fig1}}
\end{figure}

\begin{figure}
\epsscale{1.0}
\plotone{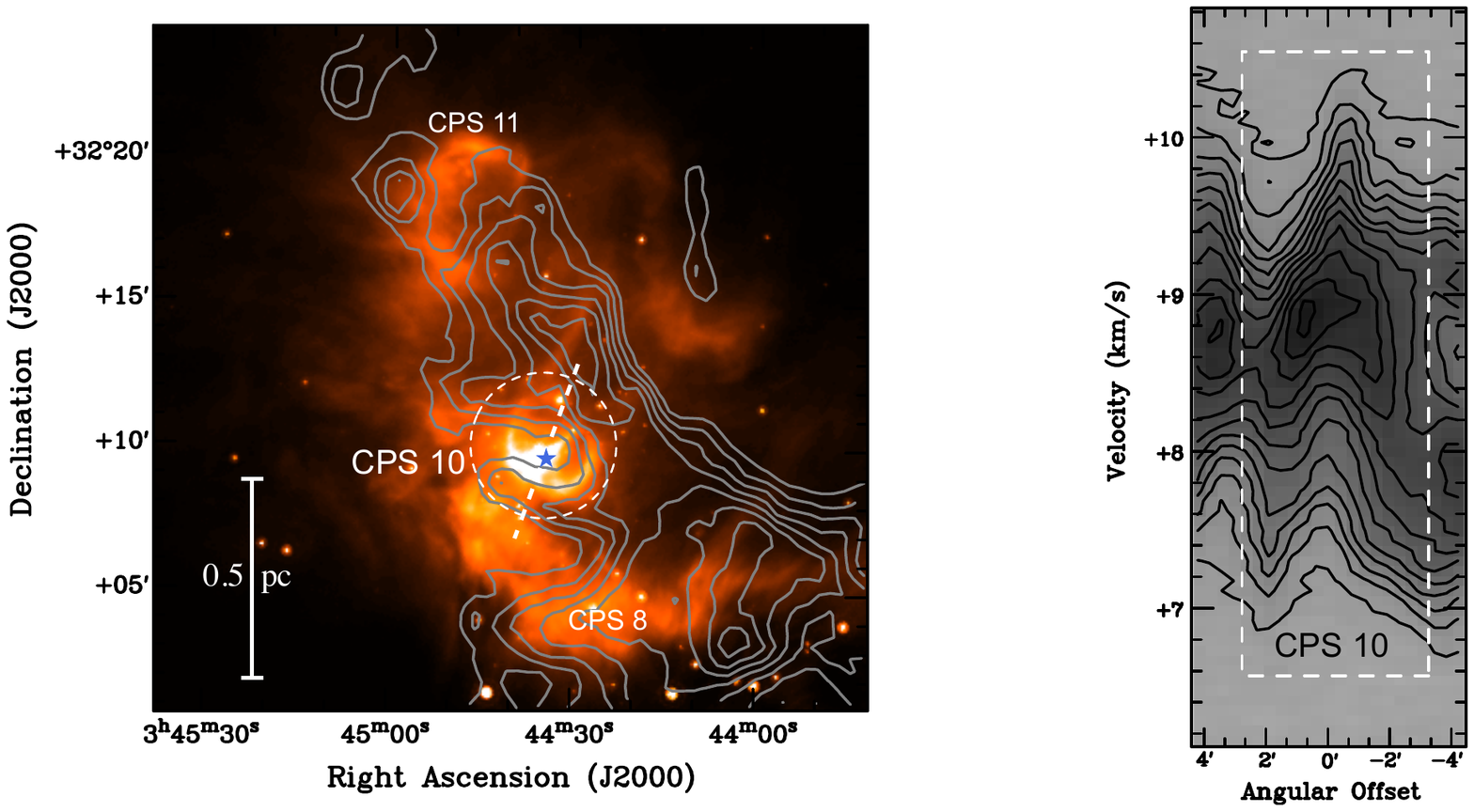}
\caption{Integrated intensity map and  $p-v$ diagram of $^{12}$CO emission near CPS 10.
The left panel shows the $^{12}$CO(1-0) integrated intensity contours (for  $6.8 < V_{LSR} < 7.8$~km~s$^{-1}$ depicting CPS 10, overlaid 
on the  MIPS 24 $\micron$ \/ map. Starting contour and contour steps are 2.4 and 1 K km s$^{-1}$, respectively.
 The white dashed circle shows the extent of the CO emission associated with CPS 10.  
The filled star symbol shows the position of the  candidate driving source HD 281159.
The positions of the IR nebulosity associated with CPS 8 and CPS 11 are also shown.
The right panel shows the $p-v$ diagram along the cut shown by the diagonal thick white dashed line in the integrated intensity map. 
Positions southwest (northeast) of the center of CPS 10 are shown as  positive (negative) offsets. 
The approximate extent of the CO emission associated with CPS 10 in the $p-v$ diagram is shown as a dashed (white) rectangle. 
\label{cps10fig2}}
\end{figure}

\begin{figure}
\epsscale{1.0}
\plotone{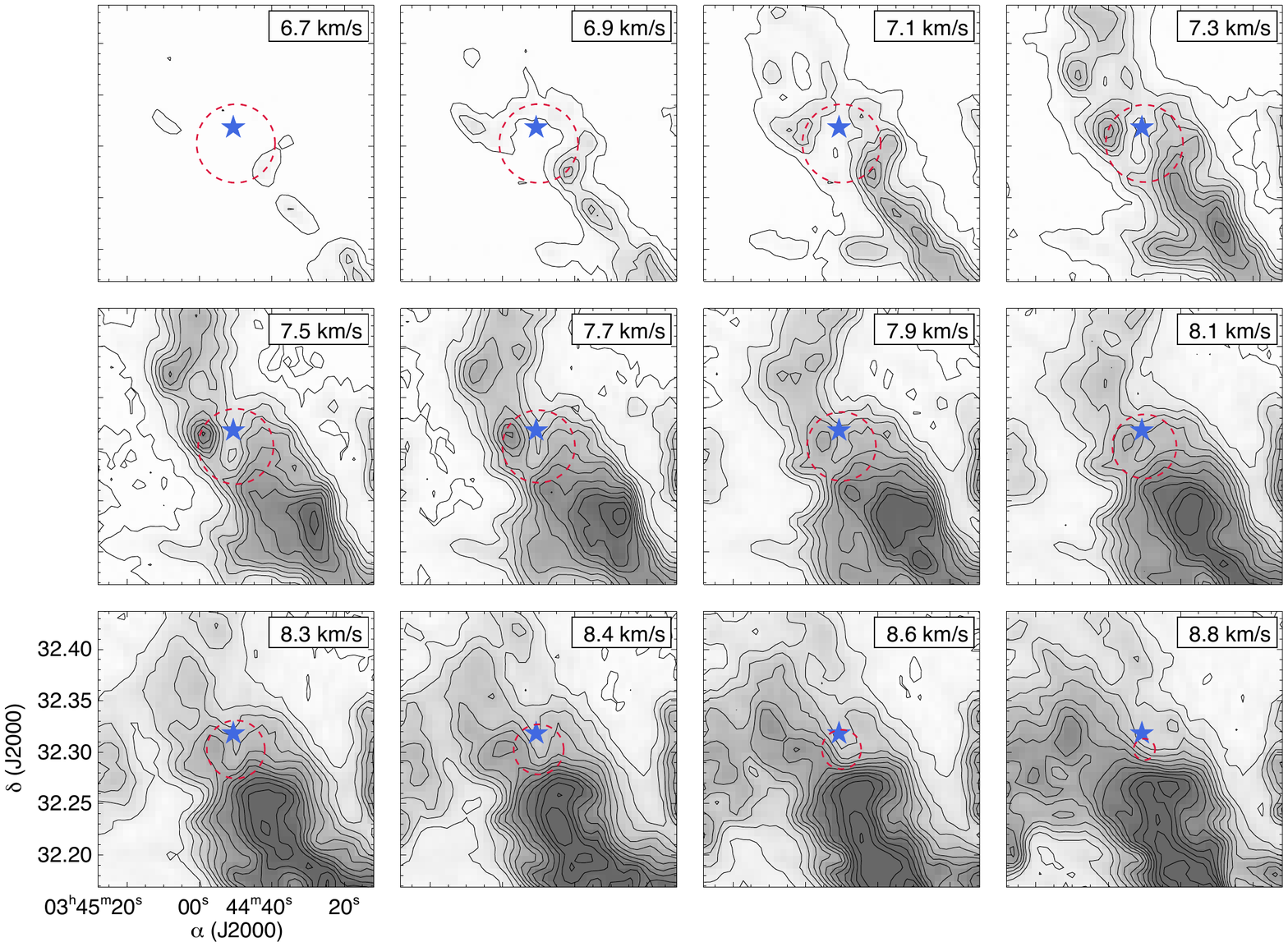}
\caption{ Channel maps of $^{12}$CO emission near CPS 11. 
 The number on the upper right corner of each panel indicates the central $V_{LSR}$  of the channel map. 
  Starting contour and contour steps are 0.5 and 1.0 K, respectively.
 Dashed circles shows the expected extent, 
at different radial velocities, of an expanding bubble with a radius, $V_{\mathrm{exp}}$ and central LSR velocity of 2.3\arcmin, 2~km~s$^{-1}$ and
6.9~km~s$^{-1}$, respectively (using the model discussed in \S~\ref{shellid}).
The filled star symbol shows the position of the candidate driving source IC 348 LRL 3.
\label{cps11fig1}}
\end{figure}

CPS 10 is associated with the well known source HD 281159 in IC 348. A circular structure of about $3\arcmin$ in diameter around this source is observed in all {\it Spitzer}  IRAC bands and in the  MIPS $24\micron$ \/ and $70\micron$ \/ images 
\citep{Jorgensen+06, Rebull+07}. The existence of this IR nebulosity prompted our search for a similar structure in the molecular line maps, and we found that 
the CO emission surrounding this source has a ring-like morphology with part of the eastern region missing (see Figures~\ref{cps10fig1} and \ref{cps10fig2}) . 
The presumed back (redshifted) cap of the bubble is observed at  $V_{LSR} \sim 9$~km~s$^{-1}$ and 
the size and shape of the CO structure changes slightly with decreasing LSR velocity (Figure~\ref{cps10fig1}). 
The $p-v$ diagram shows an upside down ``V'' structure with  minimum LSR velocity at approximately 7~km~s$^{-1}$ (Figure~\ref{cps10fig2}).
 This velocity structure is consistent with that of the back hemisphere of an expanding bubble.
 From the channel maps and $p-v$ diagram we estimate an expansion velocity of 2~km~s$^{-1}$  and the central velocity to be at $V_{LSR}$ = 7.1 km~s$^{-1}$.  
The driving source (HD 281159) is also known as 
SSTc2d J034434.2+320946 in the c2d catalog, where it is classified as a YSO candidate.

\subsection{CPS 11}
 \label{cps11sec}
\begin{figure}
\epsscale{1.0}
\plotone{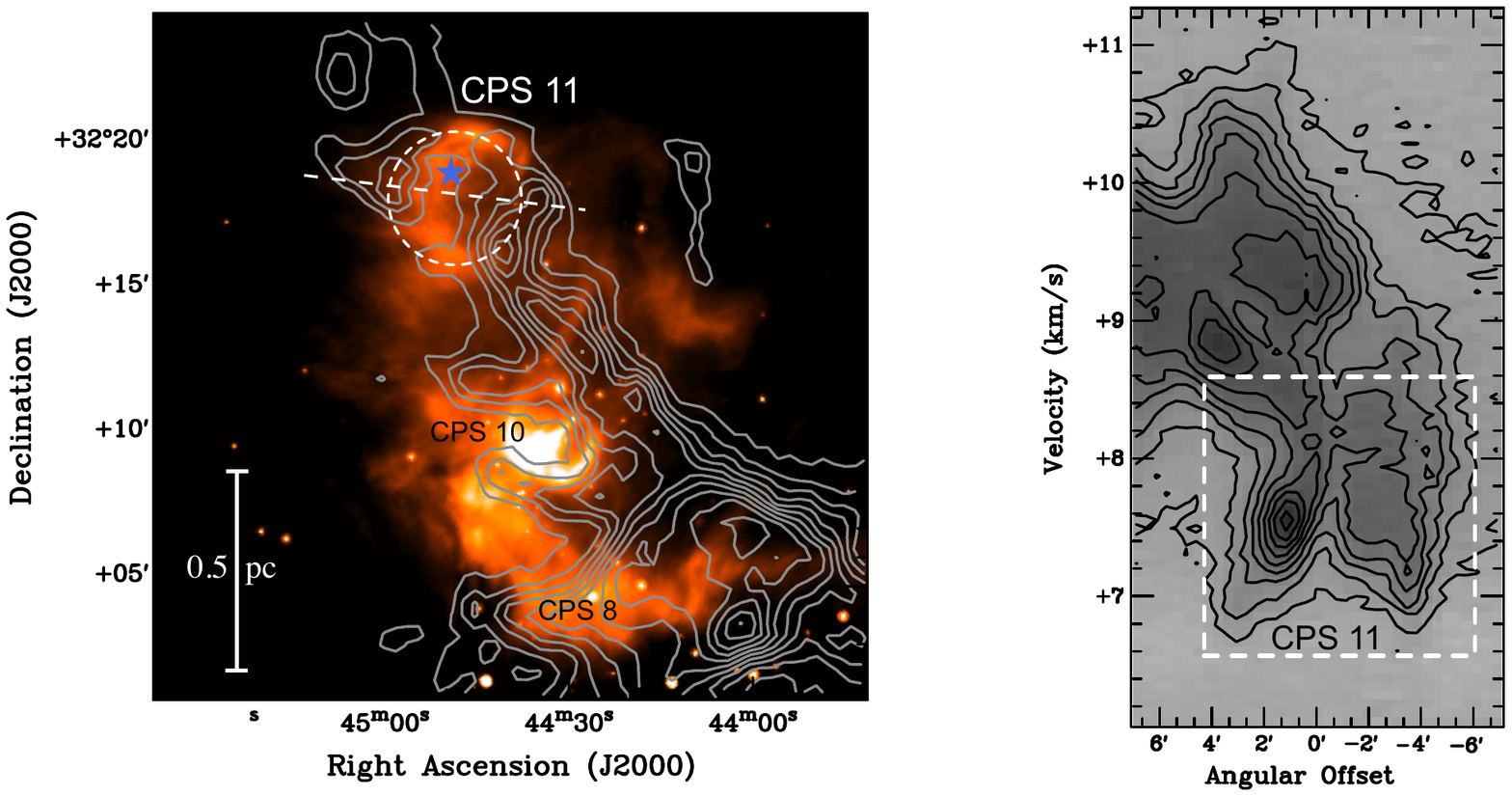}
\caption{Integrated intensity map and  $p-v$ diagram of $^{12}$CO emission near CPS 11.
The left panel shows the $^{12}$CO(1-0) integrated intensity contours (for  $6.7 < V_{LSR} < 7.4$~km~s$^{-1}$ depicting CPS 11, overlaid 
on the  MIPS 24 $\micron$ \/ map. Starting contour and contour steps are 0.6 and 0.5 K km s$^{-1}$, respectively.
 The white dashed circle shows the extent of the CO emission associated with CPS 11.  
The filled star symbol shows the position of the  candidate driving source IC 348 LRL 3.
The positions of the IR nebulosity associated with CPS 8 and CPS 10 are also shown.
The right panel shows the $p-v$ diagram along the cut shown by the diagonal thick white dashed line in the integrated intensity map. 
Positions northeast (southwest) of the center of CPS 11 are shown as  positive (negative) offsets. 
The approximate extent of the CO emission associated with CPS 11 in the $p-v$ diagram is shown as a dashed (white) rectangle. 
\label{cps11fig2}}
\end{figure}

\begin{figure}
\epsscale{1.0}
\plotone{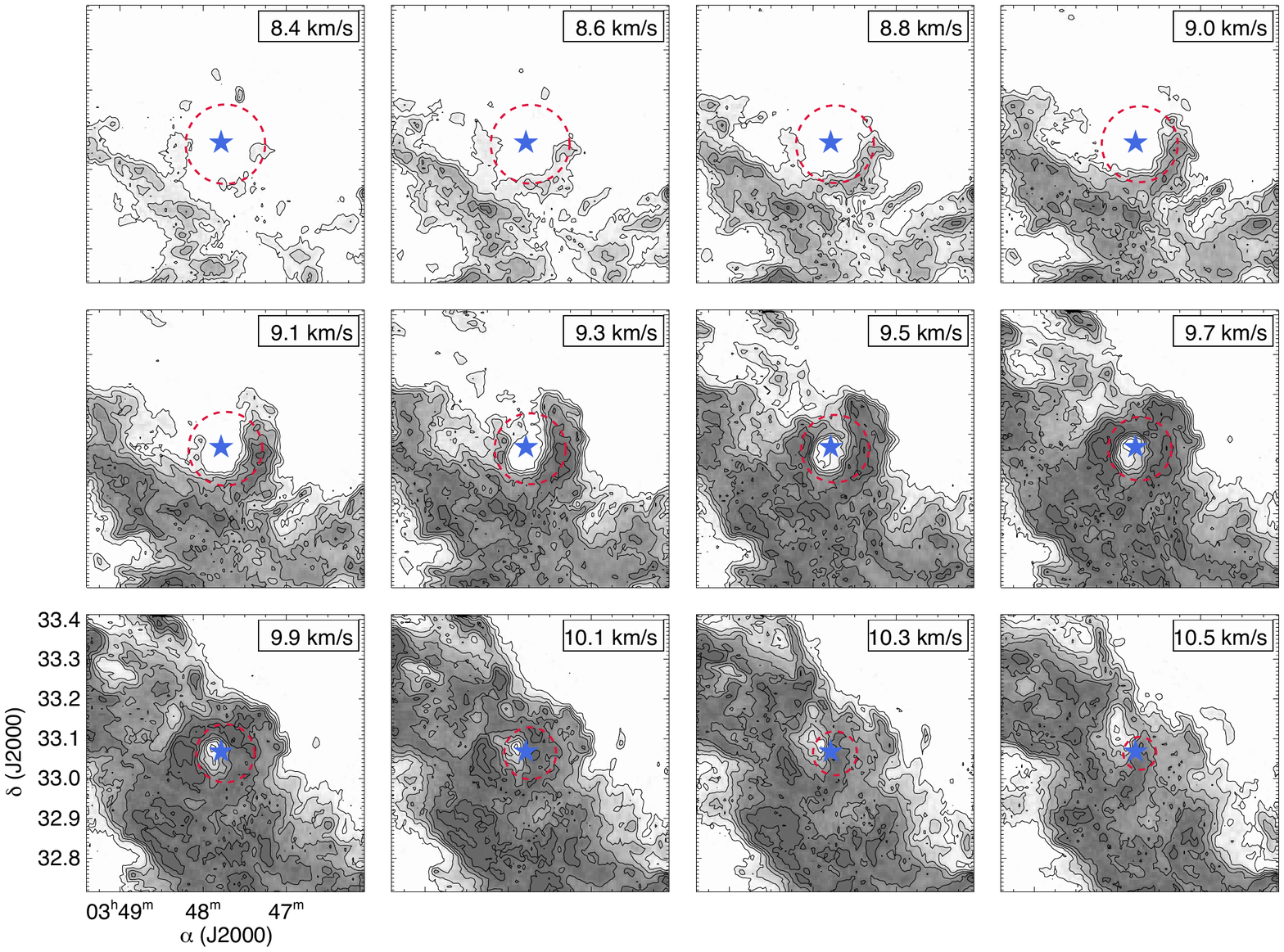}
\caption{ Channel maps of $^{12}$CO emission near CPS 12. 
 The number on the upper right corner of each panel indicates the central $V_{LSR}$  of the channel map. 
  Starting contour and contour steps are 0.9 and 0.9 K, respectively.
 Dashed circles shows the expected extent, 
at different radial velocities, of an expanding bubble with a radius, $V_{\mathrm{exp}}$ and central LSR velocity of 6\arcmin, 2.5~km~s$^{-1}$ and
8.2~km~s$^{-1}$, respectively (using the model discussed in \S~\ref{shellid}).
The filled star symbol shows the position of the candidate driving source IRAS 03446+3254.
\label{cps12fig1}}
\end{figure}

\begin{figure}
\epsscale{1.0}
\plotone{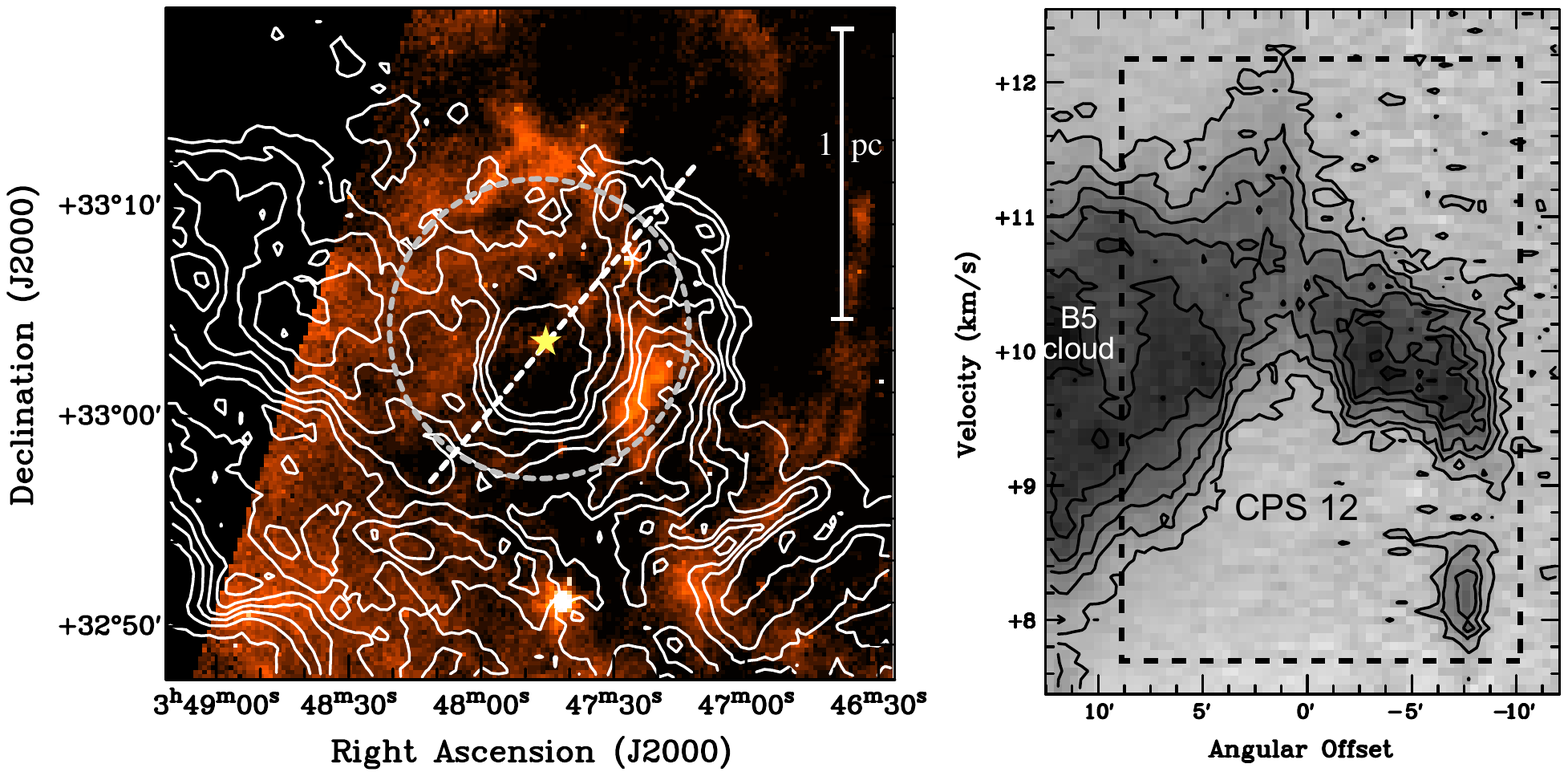}
\caption{Integrated intensity map and  $p-v$ diagram of $^{12}$CO emission near CPS 12.
The left panel shows the $^{12}$CO(1-0) integrated intensity contours (for  $8.4 < V_{LSR} < 9.7$~km~s$^{-1}$ depicting CPS 12, overlaid 
on the  MIPS 24 $\micron$ \/ map. Starting contour and contour steps are both 0.8 K km s$^{-1}$.
 The grey dashed circle shows the approximate extent of the CO emission associated with CPS 12.  
The filled star symbol shows the position of the  candidate  driving source IRAS 03446+3254.
The right panel shows the $p-v$ diagram along the cut shown by the diagonal thick white dashed line in the integrated intensity map. 
Positions northwest (southeast) of the center of CPS 12 are shown as  positive (negative) offsets. 
The approximate extent of the CO emission associated with CPS 12 in the $p-v$ diagram is shown as a dashed (black) rectangle. 
\label{cps12fig2}}
\end{figure}

In the northeastern part of the IC 348 cluster there is a bowl-shaped nebulosity with a diameter of about $2\arcmin$ clearly detected   
 in all {\it Spitzer}  IRAC bands and in the MIPS $24\micron$ \/ and $70\micron$ \/ images \citep{Muench+07}.  Similar to CPS 10, the circular nebulosity observed with {\it Spitzer}  motivated our search for a shell structure in the molecular maps in this region.  
 Our CO observations barely resolve the structure associated with CPS 11, yet a ring-like structure is discernible at LSR velocities near  7.3~km~s$^{-1}$, and emission associated with this structure is seen from $V_{LSR} \sim 6.7$~km~s$^{-1}$  to about 8.5~km~s$^{-1}$  (see Figure~\ref{cps11fig1}). The CO ring structure almost fully surrounds the circular IR nebulosity
 (see Figure~\ref{cps11fig2}).
 Similar to CPS 10, the $p-v$ diagram of CPS 11 shows an upside down ``V'' structure with  minimum and maximum LSR velocity 
 at approximately 6.8 and 8.5~km~s$^{-1}$, respectively (Figure~\ref{cps11fig2}).
 This velocity structure is consistent with that of the back hemisphere of an expanding bubble.
 From the channel maps and $p-v$ diagram we estimate an expansion velocity of 2~km~s$^{-1}$  and the central velocity to be at $V_{LSR}$ = 6.9 km~s$^{-1}$.  

 
 Inside the nebulosity there are two relatively bright A stars, and it is the closest to the center of the nebula (and the one that lies $\sim 1\arcmin$ from the center of CSP 11), 
 that is the most likely driving source; the star named Dust-Bowl in SIMBAD. This is an A0 star that is also known as IC 348 LRL 3, 2MASS J03445064+3219067, and SSTc2d J034450.7+321906, and is a known X-ray   emitter \citep[CXOPZ J034450.7+321904,][]{PZ01}.

\subsection{CPS 12}
 \label{cps12sec}
 CPS 12 lies in the northeastern part of the B5 dark cloud, at the easternmost edge of the Perseus molecular cloud complex. The ring structure, with a radius of about $6\arcmin$ is clearly seen in both $^{12}$CO and $^{13}$CO maps as well as in the MIPS ($160\micron$) map. The emission in the MIPS $24\micron$ \/ and $70\micron$ \/ maps  associated with this structure is very faint and the ring-like structure is barely discernible at these shorter wavelengths.  The ring is best seen in $^{12}$CO and $^{13}$CO at LSR velocities between 
 9.5 and 9.9~km~s$^{-1}$, but molecular gas emission associated with this structure is seen from $V_{LSR} \sim 8$  to 10~km~s$^{-1}$ (see Figure~\ref{cps12fig1}). The radius of the structure decreases slightly as the LSR velocity increases, and it appears that in CPS 12 we are mostly observing the back side of an expanding  bubble. 
 This is confirmed by the $p-v$ diagram in Figure~\ref{cps12fig2} , which shows an upside down ``V'' structure (similar to that shown in Figure~\ref{shellmodel}d).  
 Using the channel maps and $p-v$ diagram, we estimate an expansion velocity of 2.5~km~s$^{-1}$  and a central velocity at $V_{LSR}$ = 8.2 km~s$^{-1}$. 
 
 This circular cavity in the molecular cloud has been reported before by 
 several authors  \citep{Goldsmith+86, Langer+89, Yu+99} and all agree that the star at the center, IRAS 03446+3254, is the source responsible for this structure.    This source, also known as SSTc2d J034747.1+330404,
 is believed to be a young stellar object  \citep{Goldsmith+86, Myers+87} and the presence of a disk is deduced from IR excess and 
  {\it Spitzer}  IRS spectra that show 10 and 20 $\micron$ silicate lines presumably from dust in a flattened circumstellar structure
  \citep{Kessler-Silacci+06}. 
    Moreover, the detection of [NeII] in the IRS spectrum of this source is believed to arise from a disk heated by X-ray radiation from the central star \citep{Lahuis+07}. 
 This source, however,  is not classified as a YSO candidate in the c2d catalog, as this area was not covered by the c2d IRAC maps.


\section{Analysis \& Discussion}
\label{analysis}

\subsection{Mass, Momentum, Energy of Shells}
\label{mass}

To obtain the mass of the shells, we utilized a method that  uses the  $^{12}$CO(1-0) to $^{13}$CO(1-0)  ratio to correct for 
 the opacity in the $^{12}$CO(1-0)  line, as a function of velocity. 
In the general, the line opacity increases the closer to the 
central cloud velocity. Using an optically thick line without properly correcting for its velocity-dependent opacity will result in an underestimation of the mass, momentum, and kinetic energy. 

A brief description of the method is discussed below, but for more detail see \citet{Arce+10} and \citet{AG01}. 
The region used to obtain the mass for each CPS (i.e., the shell mass region) is an annulus with the same radius as that of the CPS and a thickness slightly wider than the thickness of the shell\footnote{We include an annulus that is slightly wider than the shell's thickness as not all emission associated with the shell is included within the ring defined by the profile's FWHM (see Figure~\ref{shellprofiles}).
The annulus is widen by increasing (and decreasing) the outer (and inner) radii  
by three pixels (i.e., 70\arcsec) in the CO maps, always avoiding including emission from unrelated structures.  
When needed, we use different ring radii for different velocities, 
where the radius and thickness for each velocity range is defined with the same procedure as described above 
(using the azimuthally averaged profile of the integrated intensity map over that particular velocity range).}.
 For each shell we calculate average spectra of $^{12}$CO(1-0) and $^{13}$CO(1-0) over the defined shell mass region
in order to estimate the ratio of  $^{12}$CO (1-0) to  $^{13}$CO (1-0), hereafter denoted $R_{12/13}$,  as a function of velocity. 
The line ratios obtained for each shell mass region were then  fitted with a second order
polynomial. To calculate the shell mass at a given position,
we directly use the  $^{13}$CO emission
(given by the main beam corrected antenna temperature of the line, $T_{mb}^{13}$)
 whenever is detected.
At large velocities  away from the central cloud velocity, where the $^{13}$CO emission was not reliably detected, we use the  $^{12}$CO (1-0) emission ($T_{mb}^{12}$) and the fit to  $R_{12/13}(v)$ to estimate the value of $T_{mb}^{13}$
at the given velocity and position, using the simple equation
$T_{mb}^{13} = T_{mb}^{12}/ R_{12/13}(v)$.
 We then obtain a value of the $^{13}$CO opacity ($\tau_{13}$) using equation 1 of \citet{AG01}, a value of the $^{13}$CO column density ($N_{13}$), 
 and subsequently the mass, using
equations 3, and 4 of \citet{AG01}. We only use spectral data that is greater than or equal to twice the rms noise of the spectrum.
To obtain the total molecular hydrogen column density we used  the region-dependent values of  [H$_2$]/[$^{13}$CO] obtained for
the  Perseus molecular cloud by \citet{Pineda+08} and
 a ratio of   $^{12}$CO to $^{13}$CO of 62 from \citet{LP93}.

 We estimate the excitation temperature of the cloud gas, $T_{ex}$, for each shell region assuming that the $^{12}$CO(1-0) line core is optically thick. We  measured the peak $T_{mb}^{12}$ of each spectrum in the shell, and use Equation~1 
  to obtain a distribution of  the ambient cloud $T_{ex}$ values in the region of interest. 
  Most shells have $T_{ex}$ from about 15 to 30 K, and a few exhibit excitation
  temperatures of more than 30 K. We assume a temperature of 25 K for all shells, and caution that this may be a lower limit as
  it is very probable (and expected)
   that the actual gas temperature is significantly higher than the value of $T_{ex}$ derived from the CO(1-0) data.  
 Studies of collimated outflows that observe different transitions of the same molecule show that temperatures of the entrained molecular
  gas are higher than the  excitation temperature of the ambient gas and is not uncommon for the outflowing gas to reach temperatures of 50 to 100~K \citep[e.g.,][]{HT01}.
 The study by \citet{BW10} indicates that in the shells around high mass star forming regions they observed with the CO(3-2) line, the gas is most likely to have a
 kinetic temperature of 40 to 60 K ---higher than the excitation temperature obtained from the peak CO(3-2) emission. The uncertainties in gas temperature,   gas opacity, and shell mass region result in an uncertainty in the shell mass estimate of about a factor of two.

\begin{figure}
\epsscale{0.9}
\plotone{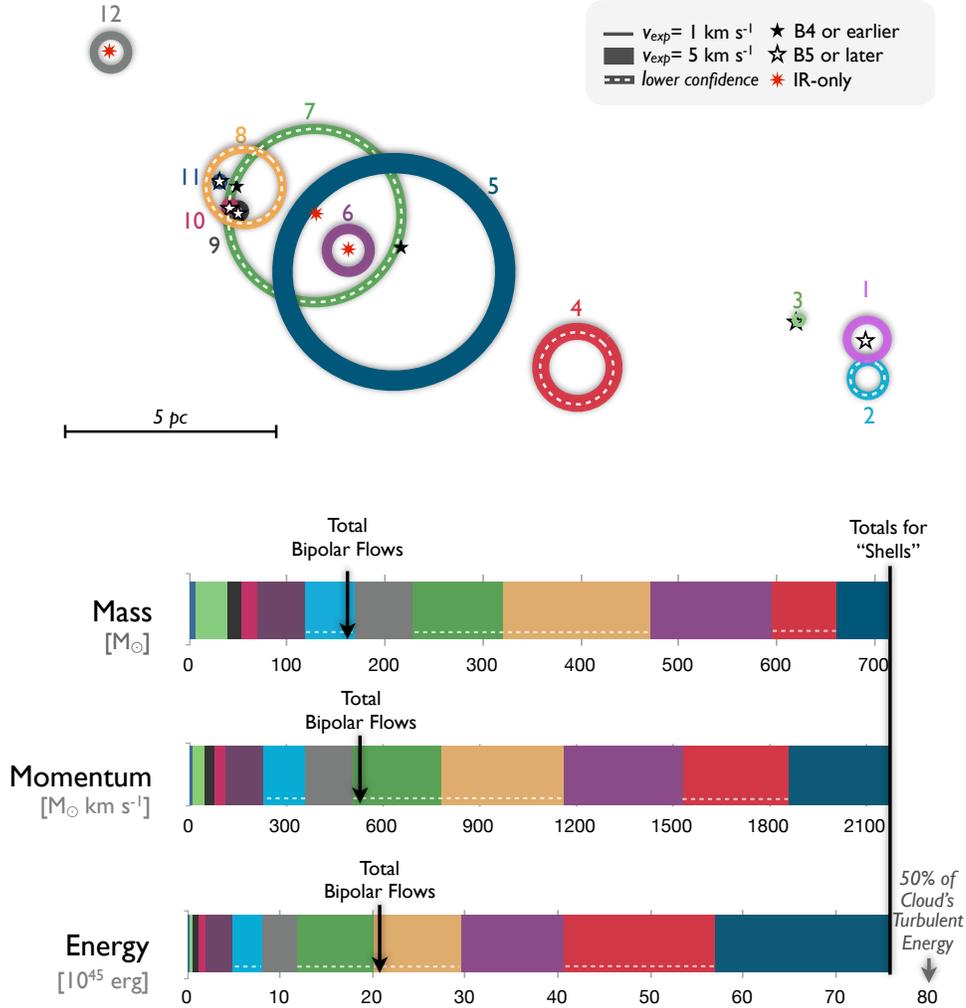}
\vspace{-1in}
\caption{Schematic picture of size, mass, momentum and energy distribution of shells in Perseus. The radius of each circle (and position) are proportional to the radius (and location) of the shell in the cloud, while the ring thickness is proportional to the expansion velocity (see legend on upper right corner). Shells with a confidence level of 3 or less (from Table~\ref{shellgrade}) are indicated by a dashed white line. Candidate powering sources with a B5 spectral type or later are shown as white stars symbols, while those with earlier spectral type (i.e., high-mass stars) are shown as black (filled) star symbols.  Candidate sources with no known spectral type (but known $\alpha$) are shown as red stars. The relative mass, momentum and kinetic energy of the shells are shown in the three horizontal bars (where the colors indicate the value for each shell). The total outflow mass, momentum and kinetic energy of the molecular outflows in Perseus  (from Arce et al.~2010) are shown for comparson.
\label{cartoonfig}}
\end{figure}

Using our estimate of the shell mass and expansion velocity (see above) we can then obtain the shell momentum and kinetic energy. 
 These are given by  $P_{\mathrm{shell}} =  M_{\mathrm{shell}} V_{\mathrm{exp}}$ and  $E_{\mathrm{shell}} = 0.5 M_{\mathrm{shell}} V_{\mathrm{exp}}^2$, respectively, and are shown for each shell in Table~\ref{shellsprop}. Figure~\ref{cartoonfig} shows a graphic comparison of 
 the relative mass, momentum and energy from each CPS. 
 Note, that for most shells $P_{\mathrm{shell}}$ and $E_{\mathrm{shell}}$ are lower limits as $V_{\mathrm{exp}}$ is the minimum possible  expansion velocity, as discussed above.

\subsection{Shell Shapes and Cloud Geometry}
\label{geometry}
The distribution of molecular gas associated with the shells  can be used to constrain the geometry of the expanding bubble and to some extent the thickness of the host cloud \citep{BW10}. As stated above, although all shells in Perseus show evidence of expansion, most are only observed over a limited range of velocities. For some of the shells only the front (blueshifted)  hemisphere is seen (CPS 1, 2, and 9), while in others only the back (redshifted) hemisphere is seen (CPS 6, 10, 11, and 12).  Only one source  shows blue- and red-shifted emission at the center of the circular shell (CPS 3), consistent with having front and back caps (although the $p-v$ diagram of this shell is not fully compatible with this picture, see \S~\ref{cps3sec} for more detail). The four largest shells (CPS 4, 5, 7, and 8) show neither front or back cap and the CO shells only traces arcs, not complete circles. We expect the winds that drive the observed shells to be spherical (see below), and we only detect the shells in CO where the winds are entraining the cloud; we assume that parts of the shells not detected in CO (including caps) are regions where there is no molecular gas. 
The sources of these four shells are observed at the edge of the cloud and in some cases the sources are thought to be in front or behind the cloud. 
It is therefore not surprising that for these 
 large shells the CO does not fully trace circles as the bubbles driven by these sources are not immersed in the cloud.
 
 The other eight shells presented in this work are seen in projection to be well within the cloud boundaries and would be expected to be detectable in CO at all velocities if they were fully surrounded by molecular gas material. However, as indicated above most of the shells show only one hemisphere (or part of it). 
 This indicates that these shells  lie closer to the  cloud's front (shells that only show the redshifted or back hemisphere) or back (shells that only show the blueshifted or front hemisphere).
This could happen if either: 1) most shell sources were at the front or back edge of a thick ($\sim 1$ pc) cloud; 
or 2) the clouds are thin along the line of sight (comparable to the shell radius) and a slight offset in the position of the bubble's center with respect to 
the cloud's center  results in one  hemisphere (the one that is observed in CO) entraining most of the gas while the other entrains very little or no gas
(and is not seen in CO). The former scenario seems unlikely as it would require that most sources be in the front or back of the cloud.  
The shell sources are thought to be young ($\sim 1 - 2$ Myr, see below) and given the low velocity of young stars with respect to their cloud (see above), 
they could not have moved much more than 0.3 pc from their birth site.   Hence, a more likely scenario is that the parent cloud is relatively thin along the line of sight 
(as suggested by Beaumont \& Williams 2010).

 The medium-size shells (CPS 1, 2, 6, and 12) have a radii ranging from about 0.4 to 0.5 pc, and show only one hemisphere. This suggests that the cloud thickness (along the line of sight) is not likely to be much more than approximately 0.5 pc in the regions close to these shells. In the B5 cloud (the host cloud of CPS 12), the major and minor axes on the plane of the sky  are about 3.5 and 1.5 pc. From the discussion above, we estimate the cloud thickness to be about  15\% to 30\% of the axes on the plane of the sky, consistent with the aspect ratios found by \citet{BW10} in a study of bubbles in high-mass star forming regions.  
 One would expect that many of the small bubbles in Perseus, traced by the shells with radii between 0.1 and 0.2 pc, would be fully surrounded by molecular gas. Yet, all the small shells in IC 348  (CPS 9, 10, and 11) only show the front or the back hemisphere. This is probably because these sources lie on a relative narrow filament (0.35 pc wide on the plane of the sky) that juts out from the northern part of the cloud. From the size of the shells, it appears that the thickness (along the line of sight) of this filament is not much more than 0.2 pc ---consistent with the fact that we expect the filament's thickness to be similar to its width (on the plane of the sky). The only shell that shows evidence of possibly being a bubble fully immersed in the cloud is CPS 3. This shell has a radius of about 0.15 pc and lies close to the central part of the cloud associated with NGC~1333, where we would expect the cloud to be more than 0.3 pc thick.

\subsection{Origin of Shells}
\label{origins}
Cavities of different sizes and shapes have been found in many clouds forming low- and high-mass stars. However, the shells we report here are different from those previously reported in the literature.  
Shells surrounding young or evolved high-mass stars stars are commonly traced by atomic (HI) maps (Chu 2008; Cappa et al.~2003, and refereces therein), molecular line maps \citep{BW10}, and observations of the dust IR emission 
\citep{Churchwell+06, Churchwell+07}. Most of these have radii of more than 1 pc and 
 it is clear that they are produced by stellar winds from one or more high-mass stars inside the shell.
In nearby regions where low- and intermediate-mass stars are forming, 
 cavities associated with active and fossil outflows are common. 
One example is NGC~1333, where the approximately twenty cavities that have been found
are substantially smaller (with sizes 0.1 to 0.2 pc), and most have lower expansion velocities than  the shells reported here \citep{KS00, Quillen+05}.
 Many of the cavities in NGC~1333 cannot be associated with a central source, and it has been suggested that these are fossil outflow lobes ---similar to the highly elongated cavities found in the Circinus cloud by Bally et al.~(1999). 
 On the other hand the shells reported here have a circular morphology and are much larger than typical  outflow-driven cavities, yet are (in general) smaller than the bubbles associated with high-mass stars. Moreover, most (ten out of the twelve) of the shell candidate sources in Perseus are low and intermediate mass stars (see below), very different from the well-known bubbles in our galaxy which are mostly driven by high-mass stars that produce HII regions \citep[e.g.,][]{Deharveng+10}.
 


The difference between the shells in Perseus and those commonly observed near high-mass stars and the outflow cavities that have been reported in other regions of active star formation 
warrants a discussion on the possible nature of the COMPLETE Perseus Shells. 
The  fact that the majority of the shells  coincide with an IR nebulosity of a similar shape, the existence of a convincing candidate driving source (in most cases),
and the velocity structure of the shells traced by the CO line emission all strongly suggest that they are formed by stellar winds that drive expanding bubbles in the Perseus molecular cloud.
The location of the candidate sources, either inside or at the edge of the cloud, as well as their optical-IR SED indicate that the sources are likely to be relatively young (pre-main sequence) stars. Many of the candidate sources show evidence of youth: evidence of a circumstellar disk,  X-ray emission, or optical variability. Except in two cases 
($o$ Per and HD 278942), all other candidate sources with a known spectral type are B5 or later. Even for candidate sources without spectral classification we can safely deduce from their colors and luminosity that they are low or intermediate-mass stars. 

Apart from the H$\alpha$ emission observed inside CPS 5 driven by HD 278942 \citep{Ridge+06b}  
there is no evidence of warm or hot ( $ > 10^{4}$ K) gas associated with any of the other shells  reported here. 
If these expanding shells are driven by stellar winds, then we can safely assume that they are in the momentum conserving (snowplow) phase, as it is typically assumed for winds from low-mass stars \citep[see][]{NS80,Levreault83,McKee89}. 
We can then obtain   a rough estimate of the wind mass loss rate ( $\dot{m}_{\mathrm{w}}$) required to produce the observed shells using:  
\begin{equation} 
 \dot{m}_{\mathrm{w}} = \frac{P_{\mathrm{shell}}}{v_{\mathrm{w}} \tau_{\mathrm{w}} }
 \end{equation}
where   
$P_{\mathrm{shell}}$ is the total shell momentum, the wind velocity is $v_{\mathrm{w}}$, and  $\tau_{\mathrm{w}}$ is the wind timescale (the amount of 
time the wind has been active). 
We obtain the total shell momentum from the data (see Table~\ref{shellsprop}) and make reasonable assumptions on the values of 
$v_{\mathrm{w}}$ and  $\tau_{\mathrm{w}}$ (see below)
to obtain an estimate of $ \dot{m}_{\mathrm{w}}$. 
The wind velocity is typically assumed to be close to the escape velocity of the star, which is about $1 - 4 \times 10^2$ km s$^{-1}$ for low- and intermediate-mass stars \citep{LC99}. Here we assume $v_{\mathrm{w}} = 200$ km s$^{-1}$.
The slope of the SED between 2 and 20 $\micron$ is generally used 
to estimate the evolutionary stage of a pre-main sequence star \citep{Evans+09}.
For most of the candidate sources we obtain estimates of $\alpha$ from the c2d point source catalog (see Table~\ref{sourceprop}).               
For the most part, the candidate sources appear to be in the Class II or Class III stage, which implies that these have ages of about 1 to 3 Myr \citep{Evans+09}. Assuming that the wind has been active for most of the lifetime of the star, then we can assume that  $\tau_{\mathrm{w}} \sim 1$ Myr.
We use equation 2 to obtain a rough estimate of the wind mass loss rate for each source, which  ranges from $10^{-8}$ to $10^{-6}$ M$_{\sun}$~yr$^{-1}$
(see Table~\ref{sourceprop}).
We note that by using this formula, and assuming that the winds are spherical, we only obtain a lower limit of the wind mass loss rate needed to
 drive the observed shell with $P_{\mathrm{shell}}$. This is because we
 only detect  CO shells where the winds are interacting with the molecular gas and for many of the shells discussed here 
 the full bubble is not detected in CO (see \S~\ref{geometry}).



We corroborate our estimate of the wind mass loss rates by comparing the measured radius of the shells with the expected radius of 
a wind-driven bubble in the momentum-conserving snowplow phase. This is given by:
\begin{equation}
R_{\mathrm{b}} =\left( \frac{3  \dot{m}_{\mathrm{w}} v_{\mathrm{w}} \tau_{\mathrm{w}}^2}{2 \pi \rho_{\mathrm{o}}}\right)^{1/4}  
\end{equation}
 where $\rho_{\mathrm{o}}$ is the density of the interstellar medium surrounding the bubble and $R_{\mathrm{b}}$ is the bubble radius \citep{LC99}. 
 Using Equation 2, we can then express the equation above as:
\begin{equation}
R_{\mathrm{b}}  =   0.5~\mathrm{pc} \left(\frac{P_{\mathrm{shell}}}{100~\mathrm{M_{\sun}~km~s^{-1}}} \right)^{1/4} 
 \left(\frac{\tau_{\mathrm{w}}}{10^6~\mathrm{yr}} \right)^{1/4} \left(\frac{5 \times 10^{4}~\mathrm{cm^{-3}}}{n_{\mathrm{o}}} \right)^{1/4} 
\end{equation}
where $n_{\mathrm{o}}$ is the number density of molecular hydrogen in the surrounding molecular cloud.
Using the  value of $P_{\mathrm{shell}}$  for each CPS (from Table~\ref{shellsprop}) and the same 
value of the cloud density ($n_{\mathrm{o}} = 5 \times 10^{4}$ cm$^{-3}$) and wind timescale ($\tau_{\mathrm{w}} = 1$  Myr) 
in Equation 3, the resulting bubble radius ($R_{\mathrm{b}}$) is within a factor of two for  nine  of the twelve shells in our sample.
For the remaining three shells, in one of them (CPS 3) the estimated $R_{\mathrm{b}}$ from Equation 4 is much larger than the observed CPS radius
of 0.14 pc, while for two of them (CPS 5 and 7) the  estimated $R_{\mathrm{b}}$ is much smaller than the observed CPS radius of $\sim 2.5$ pc. 
CPS 3 is in the central region of the protostellar cluster NGC 1333, and 
 its driver is probably younger (with $ \tau_{\mathrm{w}} < 10^6$ yr) and likely surrounded by higher density gas ($n_{\mathrm{o}} >  5 \times 10^4$~cm$^{-3}$) 
compared to the rest of the shell driving sources in our sample (which will result in a smaller bubble radius). 
The two large shells are driven by sources that appear to be outside the cloud (or  at the cloud's edge) 
where $n_o$ is expected to be significantly less than $5 \times 10^4$~cm$^{-3}$ (which will result in a larger bubble radius). 
From our discussion above we can safely assume that, for the most part, 
the shells observed in the Perseus molecular cloud are 
driven by spherical winds with  mass loss rates between $\sim 10^{-7} - 10^{-6}$~M$_{\sun}$~yr$^{-1}$.


\subsubsection{What kind of wind drives the shells?}
\label{windkindsec}

Stars drive winds with a range of mass loss rates that are produced by different mechanisms. Our results can constrain which of these mechanisms are most likely for the sources 
discussed here. In addition to having the approximate mass loss rate given above, the
winds that produced the shells in the Perseus molecular cloud should be poorly collimated (i.e., spherical or quasi-spherical) 
winds in order to produce the 
circular structures observed in the molecular line maps. 
Although young embedded protostars can power winds with mass loss rates ranging from $10^{-9}$ to $10^{-6}$ M$_{\sun}$~yr$^{-1}$  \citep{Lizano+88,Hartigan+94,Bally+06}, consistent 
with our estimate of $\dot{m}_{\mathrm{w}}$, these winds are typically bi-polar and collimated \citep[e.g.,][]{AS05, Stojimirovic+06}
and it is highly unlikely that they will produce the circular morphologies of the CO shells observed in our maps. As protostars evolve the outflow opening angle increases 
\citep{AS06} and young stars in the pre-main sequence (Class II) stage may drive outflows with very wide opening angles that could resemble spherical winds (see below).
The radial winds driven by radiation pressure (commonly observed in high-mass stars) 
or those driven by thermal pressure (i.e., coronal winds) could in principle produce circular shells. 
However, these two mechanisms cannot provide a mass loss rate that is high enough to explain the observed shells in Perseus, as 
radiation-driven winds from low-mass main sequence stars have  
 mass loss rates in the order of $10^{-12}$ to $10^{-10}$ M$_{\sun}$~yr$^{-1}$ \citep{LC99}, and  
coronal winds produce mass loss rates of only up to $10^{-9}$ M$_{\sun}$~yr$^{-1}$ \citep{DeCampli81}.

The inferred properties of the stellar wind responsible for the CO shells in Perseus are more 
consistent with those of  
winds detected in pre-main sequence (T Tauri and Herbig Ae/Be) stars. These winds, 
thought to be accretion-driven, are known to have mass loss rates of up to a few times $10^{-7}$ M$_{\sun}$~yr$^{-1}$ and are believed 
to be spherical or to have very wide opening angles \citep{Nisini+95,Edwards+06}. Recent observations by \citet{Edwards+06} of the He I $\lambda10830$
profile of a sample of accreting T Tauri stars indicate that in many sources this line exhibits a P Cygni profile similar to that expected
from stars with spherical stellar winds. 
Further modeling of the line shapes by \citet{Kwan+06} suggests that about half of the stars power 
wide-angle polar (stellar) winds, and  the other half have line shapes that are well fitted by a 
disk wind emerging at a constant angle relative to the disk surface \citep[see also][]{Kurosawa+11}. 
These results are consistent with a picture where both wide-angle polar winds and disk winds are 
present at the same time in T Tauri stars, resulting in a quasi-spherical wind \citep[see Figure 3 in][]{Edwards09}, and the wind that will dominate the observed line shape 
depends on the inclination angle of the system with respect to the observer 
\citep{Kwan+06,Edwards09}.
These stellar and disk winds are accretion-driven  winds launched through magnetohydrodynamic  
processes in the star-disk system \citep{Ferreira+06}.

We note that although the proposed scenario in which the observed Perseus shells are  
formed by accretion driven winds is the most compatible with our results, there might be some  problems with this picture.
 First,  a combination of very wide angle polar winds and disk winds does not 
 produce a fully spherical wind. In principle these winds would  unlikely be able to produce fully circular shells. 
 However, if the direction of the ejection axis of the wide angle wind changes on timescales much shorter than $\tau_{\mathrm{w}}$ (for example, due to precession of the disk, see
Terquem et al.~1999), it will allow the wind to interact with the cloud  over a wider range of angles, and could result in a circular CO shell around the source. 
The other problem arises when we try to estimate the required reservoir of material  in the disk in order to drive winds with 
 $\dot{m}_{\mathrm{w}}   \sim 10^{-7} - 10^{-6}$~M$_{\sun}$~yr$^{-1}$. If these winds are active for $\tau_{\mathrm{w}} \sim 10^{6}$ yr this means that a total of 
 0.1 to 1 M$_{\sun}$ of material will be launched  throughout the lifetime of these winds.  Models of accretion-driven winds indicate that 
 the ratio of wind mass loss  rate ($\dot{m}_{\mathrm{w}}$)  to disk accretion rate ($\dot{m}_{\mathrm{acc}}$)  is 0.1 to 0.3 \citep{KP00, Shu+00, Ferreira+06}.
 Assuming 
 $\dot{m}_{\mathrm{w}} / \dot{m}_{\mathrm{acc}} \sim 0.3$, then disks would need to be more massive than what they are typically assumed to be (between 0.3 and 3~M$_{\sun}$) 
 at the start of the wind phase  in order to  
 drive winds with high mass loss rates for about 1 Myr.  Recent 
numerical simulations  of embedded sub-solar protostars with disks by \citet{Vorobyov10} indicate that at the end of various runs (representing an age of 0.7 Myr) 
the disk to star mass ratio could by as high as 0.5. 
If this result scales to the formation of higher mass stars, it might then be possible for  G, A, and B5 stars to have disk masses of about 0.5, 1 and 
3~M$_{\sun}$, respectively, at the start of their pre-main sequence phase.  
 
Our results indicate that out of all the low- and intermediate- mass young stars in the Perseus molecular cloud only a small fraction drive winds with exceptionally high mass loss rates 
($\gtrsim 5 \times 10^{-7}$  M$_{\sun}$~yr$^{-1}$).
  Out of the 12 shells discussed here, only six (CPS 1, 2, 4, 6, 7, 12) come from presumed low- or intermediate-mass stars   and  require 
  $\dot{m}_{\mathrm{w}}  \gtrsim 5 \times 10^{-7}$  M$_{\sun}$~yr$^{-1}$ to drive the observed shells (see Table~\ref{sourceprop}).
  We might be missing a few additional shells due to incomplete coverage of the cloud or the relative low angular resolution of our CO maps. 
 We have no means to estimate how complete is our shell sample. Yet, it seems very unlikely that we would be missing more than 50\% of the existing shells produced by winds 
with  $\dot{m} > 5 \times 10^{-7}$  M$_{\sun}$~yr$^{-1}$. 
 A rough estimate of the total number of young stars in Perseus can be obtained using the c2d catalog of YSO candidates for this cloud \citep{Evans+09}. 
This catalog shows nearly 400 YSO candidates, but it only lists young stars with a detectable infrared excess, and the total number of pre-main sequence stars may be slightly underestimated.  
Taking into consideration pre-main sequence stars with no
IR excess could increase the number of young stars by 10 to 20\%  \citep{Evans+09}. 
From our results (and assuming our shell sample is no more than 50\% incomplete), 
we then estimate that roughly $1 - 3$\% of the low-mass young stars in Perseus currently drive spherical winds with high mass loss rates greater than
 $\sim 5 \times 10^{-7}$  M$_{\sun}$~yr$^{-1}$. 
 As stated above pre-main sequence stars power winds with a wide range of mass loss rates. Even though most have $\dot{m}$ less than $10^{-8}$ M$_{\sun}$~yr$^{-1}$
\citep{Calvet97},
winds with up to $\dot{m} \sim 3 \times 10^{-7}$  M$_{\sun}$~yr$^{-1}$ have been detected \citep{Edwards+06}, and 
it is not implausible that a very small fraction of pre-main sequence stars could have winds with mass loss rates as high as  $\sim10^{-6}$ M$_{\sun}$~yr$^{-1}$. 
Moreover, most of the T Tauri stars for which  there exists an estimate of the wind mass loss rate are M or K  stars \citep[e.g.,][]{Edwards+06}. It is conceivable that
slightly more massive (F, A or late B) pre-main sequence stars could power winds that have, on average, higher mass loss rates than those of T Tauri stars 
 \citep{Nisini+95}.

 Observations of the inner two-thirds of the galactic plane by the {\it Spitzer}  GLIMPSE survey using IRAC images revealed approximately 600 ring structures which  the authors suggest delineate the walls of bubbles produced by winds from high-mass stars
 \citep{Churchwell+06,Churchwell+07}. A recent study by \citet{Deharveng+10}, using a sub-sample of the GLIMPSE shells catalogue, showed that about 90\% of the shells coincide with ionized gas from HII regions, traced by radio continuum emission. It is possible that the other 10\% are formed by 
  stars that are too cool to form detectable HII regions \citep{Churchwell+06}. In light of the results shown here,  
   it could very well be that some of the bubbles with no HII regions are powered by
   late B or A stars that are members of the $\sim 1 - 3\%$ group of young low- and intermediate-mass stars that have exceptionally high mass loss rates.


\subsection{Impact of winds on the cloud}
\label{impact}
Winds from protostars will inject energy and momentum into the cloud and may help disrupt their surroundings or drive turbulence in the gas \citep{NL07}.
 The existence of the shells indicates that stellar winds are having a significant impact on the cloud. 
 The most obvious effect of the winds is to clear the gas surrounding the young star  and pile it up in a ring structure, thereby changing the cloud's density distribution. 
If enough gas is accelerated to velocities above the cloud's escape speed, 
the winds could significantly disrupt the region. However, from our results it appears that the impact from winds to the integrity of the Perseus molecular cloud complex will be minimal.
 Using a total mass of $7 \times 10^3$~M$_{\sun}$ for the Perseus cloud complex and an effective radius of 7 pc (from the approximate geometrical mean of the cloud extent), we calculate the gravitational binding energy of the Perseus complex to be $\sim 6 \times 10^{47}$ erg.   The total kinetic energy of all expanding shells in the Perseus cloud is $\sim 7.6 \times 10^{46}$, or approximately 13\% of the estimated binding energy. 
 Clearly the observed shells do not have the energy to potentially unbind the entire Perseus molecular cloud complex. However, some of the shells are seen at the edge of the cloud and have expanding velocities similar to (or more than) the escape velocity of the region   \citep[see][]{Arce+10}. A fraction of the gas can therefore break from the cloud's gravitational pull, and potentially disrupt the immediate region surrounding (within $\sim 1$ pc)  the powering young star.   
 
The total turbulent energy ($E_{turb}$) in the Perseus molecular cloud complex is $1.6 \times 10^{47}$~erg \citep{Arce+10}. The total kinetic energy of the shells is approximately one
half of $E_{turb}$ (see Figure~\ref{cartoonfig}), 
but given the uncertainty in the shell mass and expansion velocity estimates the total energy in shells may be comparable to the total turbulence energy in Perseus. This suggests that stellar winds could potentially help drive the turbulence in the molecular cloud complex. 

One way to assess whether the winds that produce the observed shells can 
sustain the turbulence  in the molecular cloud is to compare the wind energy injection rate into the cloud ($\dot{E}_{\mathrm{w}}$) with the turbulent energy dissipation rate ($L_{turb}$). We obtain an  approximate value of $\dot{E}_{\mathrm{w}}$ by dividing the total kinetic energy of all shells by the wind timescale ($\tau_{\mathrm{w}} \sim 10^6$ yr), and obtain $\dot{E}_{\mathrm{w}} \sim 2 \times 10^{33}$ erg s$^{-1}$. If we only consider shells with a confidence level of 4 or more (see Table~ \ref{shellgrade}), the 
 wind energy injection rate would be about half of the original estimate.
Another way to estimate the wind energy injection rate is using Equation 3.7 from \citet{McKee89}:
\begin{equation}
 \dot{E}_{w} = \frac{1}{2}(\dot{M}_{\mathrm{w}} v_{\mathrm{w}})v_{\mathrm{rms}}
 \end{equation}
  Here, the total mass loss rate from all the winds in the cloud is given by $\dot{M}_{\mathrm{w}}$. The rms velocity of the turbulent motions
($v_{\mathrm{rms}}$), is $\sqrt{3}$ times the velocity dispersion.  
Using the estimate of the wind mass loss rate for each of the sources, shown in Table~\ref{sourceprop}, we 
obtain $\dot{M}_{\mathrm{w}} \sim 1 \times 10^{-5}$ M$_{\sun}$~yr$^{-1}$. The average velocity width (FWHM) in Perseus is about 2~km~s$^{-1}$ \citep{Arce+10}, which gives $v_{\mathrm{rms}} \sim 1.5$~km~s$^{-1}$. Assuming $v_{\mathrm{w}} \sim 2 \times 10^{2}$~km~s$^{-1}$ (see above), we obtain that $\dot{E}_{\mathrm{w}} \sim 10^{33}$ erg s$^{-1}$, consistent with our previous estimate. If we only consider shells with confidence level 4 or higher, we would obtain $\dot{E}_{\mathrm{w}} \sim 5 \times 10^{32}$ erg s$^{-1}$ (still consistent, within a factor of two, with the previous estimate).
 The turbulent energy dissipation rate is given by $L_{turb} = E_{turb}/t_{diss}$, where the turbulence dissipation timescale is $t_{diss} = \eta t_{ff}$, which is given in terms of the free-fall time ($t_{ff}$). Estimates of the value of $\eta$ from numerical simulations of clouds range between $\sim 1$ and 10 \citep{McKee89,MacLow99}. 
Assuming a gas density averaged over the entire cloud complex of $\sim 10^3$ cm$^{-3}$ and 
$\eta = 5$, we obtain $t_{diss} \sim 5 \times 10^{6}$ yr, which results in a turbulent energy dissipation rate of approximately $10^{33}$ erg s$^{-1}$. 
The crude estimates of the wind energy input rate and turbulent energy dissipation rate indicate that these two are approximately the same (within a factor of two or less), and our results reveal that powerful  spherical winds  from young stars have the potential  for driving turbulence  in the Perseus molecular cloud complex. This conclusion remains even if we were to only consider the energy input of shells with a confidence level of 4 or more.

In Perseus, the detected shells are spread throughout most of the cloud complex (see Figure~\ref{all13co}), yet only a small fraction of the cloud ($10$\%, by mass) is currently being swept-up in shells. This, however, does not mean that they cannot help drive the turbulence in the entire cloud complex.  Clouds are known to be magnetized, and it is possible that  winds and shells interact  with the cloud's magnetic field.  In fact, \citet{Ridge+06b} argue that high polarization vectors (from measurements of the polarization of background starlight) found to be aligned with the warm dust emission from the shell associated with HD 278942 (i.e., CPS 5) may be the result of the magnetic field in Perseus  being swept-up by the shell.
The interaction between winds and magnetic fields may generate large amplitude Alfv\'en waves that travel throughout the cloud \citep{NL07,Wang+10}.
 These MHD waves could transform the shell motions into turbulent motions and help drive turbulence  in the cloud near and far from the expanding shell.

In their study, \citet{Arce+10} indicate that even though outflows have enough power to drive the turbulence in localized regions  
of active star formation (with sizes of 1 to 4 pc), 
they lack the necessary energy to feed the observed turbulence in the entire
Perseus complex. The results shown here indicate that, from a pure energy-budget perspective, 
powerful spherical winds from young stars can supply the energy needed to maintain 
 turbulence on a global cloud scale in Perseus.  
External sources of turbulence are probably important during cloud formation if, for example, clouds are formed through the compression of large-scale atomic gas streamers in the interstellar medium \citep{Ballesteros+99}. However, once star formation takes place in the cloud, outflows and bubbles can provide enough energy input to sustain the cloud turbulence.
The radius of the shells observed in Perseus  range from $\sim 0.1$ to about 2.8 pc.  
 Consequently, shells would be expected to drive turbulence at a range of scales (similar to outflows) and a distinct turbulence driving  length would be 
 hard to detect \citep[see][]{Arce11}.

  


\section{Summary \& Conclusions}
\label{summary}

We detect twelve shells in the COMPLETE  molecular  line maps of the Perseus  molecular cloud complex, a star forming region typically assumed to harbor low- and intermediate-mass stars.
 The shells, observed in $^{12}$CO(1-0) and $^{13}$CO(1-0), have circular or arc-like morphologies and range in radius from about 0.1 to 2.8 pc. Most of them are coincident with IR nebulosity that have similar shape to the CO shells, and for many we can assign a candidate source that lies near the center of the circular structure. 
Their velocity-dependent structure is consistent with the shells being produced by expanding bubbles. All these indicate that the COMPLETE Perseus Shells most likely trace the 
molecular gas that has been entrained by spherical (or very wide angle) winds from young stars.

Two out of the twelve shells seem to be powered by known high-mass stars, while the rest appear to be driven by low- or intermediate-mass pre-main sequence stars.  
One of the high-mass stars is HD 278942, a star that is thought to be behind (and possibly associated with) the Perseus cloud complex  and close enough that its stellar winds interact with the molecular gas \citep{Ridge+06b}. The other is $o$ Per, a star close to IC 348 that has been typically assumed not to be associated with the stellar cluster in  Perseus. Here we propose that this early B star 
is the most likely source of the relatively powerful wind that interacts with the northern part of the Perseus cloud and that it lies within 1 pc of IC 348.
 

Recent evidence suggests that high-mass stars might be impacting the molecular gas in other well-known regions that have been 
typically thought of being predominantly low-mass star forming clouds.
In Ophiuchus, a ring of about 2 pc in radius and approximately centered at the position of the B star $\rho$ Oph is 
discernible in the dust temperature maps,  and appears to be interacting with the nearby 
molecular cloud \citep{Schnee+05}. In Taurus, a cavity of about 3 pc in size is outlined by young stars and condensations that look like cometary globules \citep{Goldsmith+08}. Although there is not much information about the shell-like structure in Taurus and no candidate source has been found, it has the makings of a shell formed by winds from a high-mass star.
It thus seem that, contrary to the conventional assumption, low-mass star forming regions are not free from the impact of winds from high-mass stars. 

Based on corroborating evidence from kinematic signatures, IR nebulosities, and the identification of candidate driving sources, we propose that the shells reported here are likely formed
by momentum-conserving wind-cloud interactions. 
 The observed size and momentum of these shells imply stellar mass loss rates of 
 about $10^{-8}$ to $10^{-6}$ M$_{\sun}$~yr$^{-1}$. At least eight    
require winds with a high mass loss rate of $\dot{m}_w \gtrsim 5 \times10^{-7}$ M$_{\sun}$~yr$^{-1}$, where two of these are the winds the from high-mass stars mentioned above. 
This  implies that at least six  stars  in Perseus drive winds that have an unusually high $\dot{m}_w$ (by at least a factor of three) compared to the range of mass loss rates measured
for other pre-main sequence stars. We claim this is not unfeasible as no more than 
3\% of the total young stars in Perseus would be required to have such high $\dot{m}_w$. In addition, 
most of the estimates of wind mass loss rates have been done for low-mass (K and M) T Tauri stars and it is plausible that
slightly more massive (F, A or late B) pre-main sequence stars could power winds that have, on average, higher mass loss rates than those of T Tauri stars. Although the exact nature of these powerful (quasi-spherical) winds is not known, we argue that they are likely accretion driven winds.

We estimate  the total kinetic energy of all the shells   
to be approximately one half of the total turbulence energy in the Perseus molecular cloud complex, 
which suggests that stellar winds could help drive the turbulence in the molecular cloud complex. Furthermore, comparison of our estimate of the total energy input rate of all shells and the 
turbulence energy dissipation rate in Perseus indicate they are similar. Thus, the spherical winds that produce the COMPLETE Perseus shells have the 
 potential  for driving turbulence  in the molecular cloud complex. In Perseus, a combination of outflows in regions of active star formation \citep{Arce+10} and shells from powerful 
 winds distributed throughout the cloud provides more than enough power to maintain the turbulence throughout the cloud. 
 In principle there is no need to invoke large-scale external drivers (e.g.,  cloud-cloud collisions,  galactic rotation, etc.) to maintain the turbulence 
  in Perseus, at least while the cloud is actively forming stars.  
 
 More observations of entire molecular cloud complexes using CO line maps (to trace the kinematics of the gas) and IR continuum maps (to trace the warm and cold dust) with good spatial resolution 
 ($\sim 0.05$ pc or less) are needed to study the frequency of shells in low-mass star forming regions and their impact on the cloud. Further studies should examine whether this is a  wide-spread phenomenon throughout the Galaxy or the Perseus molecular cloud complex is just an exceptional case. In addition, large-scale maps  of clouds will help investigate the frequency at which winds from high-mass stars impact nearby low-mass star forming regions.
Once available, it will be useful to use the molecular line data from the JCMT Legacy Survey of the Gould Belt \citep{Ward-Thompson+07} and the IR continuum data from the Herschel Gould Belt Survey \citep{Andre+Saraceno05} to search for shells and study their impact on their parent clouds.

\acknowledgments

We would like to thank the rest of the COMPLETE team, and especially Mark Heyer, for their help with the acquisition and reduction of the data. 
We also like to thank Chris McKee and Chris Maztner for their discussion on winds, and Kelle Cruz for her help on colors of very low-mass stars. Comments and suggestions from the
anonymous referee helped improve the paper. 
 Our gratitude goes to the Initiative in Innovative Computing at Harvard for their support of the Astronomical Medicine Project. HGA was partially funded by NSF awards AST-0401568 and AST-0845619 while conducting this study. The COMPLETE Survey of Star Forming Regions is supported by  NSF grant No.~AST-0407172.  HGA is very grateful to the Universidad de Chile and University of Illinois, Urbana-Champaign astronomy departments
 for their kind hospitality while much of this work was  conducted.





{\it Facilities:} \facility{FCRAO}.

\newpage

\begin{deluxetable}{ccccccc}
\tabletypesize{\scriptsize}
\tablecaption{COMPLETE Shells in Perseus
\label{shellstab}}
\tablewidth{0pt}
\tablehead{
\colhead{Shell} & \colhead{Cloud} & \colhead{Center} & \colhead{CO Radius} & \colhead{CO Thickness} & \colhead{IR Radius} & \colhead{Candidate}     \\
\colhead{Name} & \colhead{Region} & \colhead{($\alpha_{2000}$, $\delta_{2000}$)} & \colhead{(arcmin/pc)} &  \colhead{(arcmin/pc)} & \colhead{(arcmin/pc)} &
\colhead{Source} 
}
\startdata                                                                                                                 
  CPS 1  & West of NGC 1333 &  03 27 29, 31 16 50  &   6.0 / 0.44  &  3.0 / 0.22 &  ---  & multiple\\    
  CPS 2  & West of NGC 1333 &  03 27 36, 31 04 50  &   5.9 / 0.44 &  4.7 / 0.34  &  ---  & multiple\\              
  CPS 3  & NGC 1333              &  03 29 26, 31 25 10  &    1.9 / 0.14 & 1.5 / 0.11  &  1.0 / 0.07  & BD+30 549\\         
  CPS 4  & East of B1              &  03 35 25, 31 10 00  &   16.8 / 1.22  & 9.4 / 0.68  & 11.7 / 0.85 &  multiple\\        
  CPS 5  & Btw.~B1 + IC348    &  03 39 53, 31 50 10  &   38.3 / 2.79  & 9.1 / 0.66    & 36.4 /2.65 &  HD 278942\\   
  CPS 6  & L1468                    &  03 41 24, 31 54 10   &  8.0 / 0.58 &  5.1 / 0.37 &  5.5 / 0.40  & IRAS  03382+3145\\   
  CPS 7  & IC348-L1468         &  03 42 18, 32 06 10  &  31.1 / 2.26 & 7.0 / 0.51 &     --- & IRAS 03390+3158\\
  CPS 8  &  IC348                   & 03 44 10, 32 17 20    &   13.4 / 0.97 & 4.8 / 0.35  &  13.0 / 0.95   & omi Per\\            
  CPS 9  &  IC348                   & 03 44 19, 32 08 30    &   2.5 / 0.18 & 1.4 / 0.10   &    ---   & IC 348 LRL  30\\
 CPS 10  &  IC348                   & 03 44 35, 32 10 10    &   2.8 / 0.20  & 2.5 / 0.18    &   1.5 / 0.11       & HD 281159\\   
 CPS 11 &  IC348                    & 03 44 50, 32 18 10    &   2.3 / 0.17 & 1.5 / 0.11   &  1.5 / 0.11 &  IC 348 LRL 3\\ 
 CPS 12 &  B5                         & 03 47 44, 33 03 50    &   6.0 / 0.44 & 5.0 / 0.36  &   5.5 / 0.40  &    IRAS03446+3254\\ 

\enddata

\end{deluxetable}

\begin{deluxetable}{crcrr}
\tabletypesize{\scriptsize}
\tablecaption{Properties of Shells 
\label{shellsprop}}
\tablewidth{0pt}
\tablehead{
\colhead{Shell} & \colhead{Mass} & \colhead{$V_{\mathrm{exp}}$} & \colhead{Momentum} & \colhead{Kinetic Energy}        \\
\colhead{Name} & \colhead{[M$_{\sun}$]} & \colhead{[km s$^{-1}$]} & \colhead{[M$_{\sun}$ km s$^{-1}$]} &  \colhead{[$10^{45}$ erg]}
}
\startdata
 CPS 1  &  48 & 2.5  & 120 &  3.0\\
 CPS 2  & 51 & 2.5 & 127 &  3.2\\
 CPS 3  & 33 & 1.2 &  37 &  0.4\\  
 CPS 4  & 66 & 5 & 330 &  16.4\\
 CPS 5  & 53 & 6 & 315  & 18.8 \\
 CPS 6  & 124 & 3 & 371 & 11.1 \\
 CPS 7  & 92 & 3 &  277 &  8.3\\
 CPS 8  & 151 & 2.5 & 377 &  9.4\\
 CPS 9  &  14 & 2 & 29  & 0.6 \\
CPS 10  & 17  & 2  &  34 &  0.7\\
CPS 11  &  6 &  2 &  11 &  0.2\\ 
CPS 12  & 59  & 2.5  & 149 & 3.7\\  

\enddata
\end{deluxetable}

\begin{deluxetable}{cccccccl}
\tabletypesize{\scriptsize}
\tablecaption{Properties of Candidate Sources 
\label{sourceprop}}
\tablewidth{0pt}
\tablehead{
\colhead{Shell} & \colhead{Candidate} &  \colhead{Position}  & \colhead{Spectral} & \colhead{c2d} & \colhead{$\alpha$} & \colhead{$\dot{m}_w$\tablenotemark{a}} & \colhead{Comments} \\
\colhead{Name} & \colhead{Source} & \colhead{($\alpha_{2000}$, $\delta_{2000}$)} & \colhead{Type} & \colhead{Classification} &   \colhead{} &  \colhead{[$10^{-7}$~M$_{\sun}$~yr$^{-1}$]}  &  \colhead{}                            
}
\startdata
 CPS 1   &   BD+30 543\tablenotemark{b}  & 03 27 40.0, 31 15 40 &   F2  &  star+dust(MP1)  & -2.47 & 6.0 & --- \\
 CPS 2   &  multiple                     &    ---  &  ---  & --- & ---& 6.4 & see~\S~\ref{cps2sec}\\ 
 CPS 3   &   BD+30 549               & 03 29 19.8, 31 24 57 &   B8  &  YSOc\_star+dust(MP1) & -2.33 & 1.9 & X-ray source  \\
 CPS 4   &   multiple                     & --- &    ---  &  ---  & --- & 15.0 & see~\S\ref{cps4sec} \\
 CPS 5   &   HD 278942               & 03 39 44.7, 31 55 33 & O9.5-B3  & star & -2.81  & 15.8 &--- \\
 CPS 6   &   IRAS  03382+3145   &  03 41 22.1, 31 54 34 & ---  &  star & -2.65  & 18.6 & ---  \\
 CPS 7   &   IRAS 03390+3158    &  03 42 10.9, 32 08 17 &  ---  &  star  & -2.67 & 13.9 & variable star \\
 CPS 8   &    $o$ Per                    & 03 44 19.1, 32 17 18 & B1 & YSOc & -2.66 & 18.9 &  variable star, X-ray source \\
 CPS 9    &  IC 348 LRL  30       &  03 44 19.1, 32 09 31    & F0 & YSOc\_star+dust(MP1) & -2.37 & 1.5 & X-ray source\\ 
                                          
 CPS 10  &   HD 281159              & 03 44 34.2, 32 09 46 & B5 & YSOc\_star+dust(MP1) & -2.03 & 1.7 & ---\\
 CPS 11  &  IC 348 LRL 3            & 03 44 50.7, 32 19 06 & A0  & star & -2.65 & 0.6 & X-ray source\\
 CPS 12  &  IRAS 03446+3254    & 03 47 47.1, 33 04 04 & ---  & star+dust(MP1) & -0.14 &  7.5 & evidence of disk from IRS data\\

\enddata
\tablenotetext{a}{Estimate of minimum wind mass loss rate needed to drive the observed CO shells, using Equation 2, the value of $P_{\mathrm{shell}}$ for each shell given in Table~\ref{shellsprop}, $\tau_{\mathrm{w}} = 1$~Myr, and $v_{\mathrm{w}} = 200$~km~s$^{-1}$.}
\tablenotetext{b}{ We assume that the c2d source SSTc2d J032740.5+311540 is the IR counterpart of BD+30 543.}

\end{deluxetable}

\begin{deluxetable}{ccccccc}
\tabletypesize{\scriptsize}
\tablecaption{CPS Confidence Level Grading
\label{shellgrade}}
\tablewidth{0pt}
\tablehead{
\colhead{CPS} & \colhead{Velocity} & \colhead{IR} & \colhead{Circular} & \colhead{$p-v$} & \colhead{Candidate}  & \colhead{Confidence}      \\
\colhead{Number} & \colhead{Structure?\tablenotemark{a}} & \colhead{Nebulosity?\tablenotemark{b}} & \colhead{Structure?\tablenotemark{c}} & \colhead{Diagram?\tablenotemark{d}}
 & \colhead{Source?\tablenotemark{e}}  & \colhead{Level} 
}
\startdata
1  & Y &   N  & Y &     Y &      Y  &  4\\
2  & Y  &  N  &  Y  &     Y  &    N   &    3\\
3  & Y  & Y   &   Y  &     Y  &   Y   &    5\\
4  & Y  & Y   &   Y  &     N  &   N   &    3\\  
5  & Y  &  Y  &   Y  &     N   &  Y   &    4\\
6  & Y  &  Y  &   Y  &     Y  &   Y   &    5\\
7  & Y  &  N  &   N  &     N  &   Y  &     2\\
8  & Y  &  Y  &   N  &     N  &   Y  &     3\\
9  & Y  &  N  &   Y  &     Y  &   Y  &     4\\
10  & Y &  Y &   Y  &     Y  &   Y  &     5\\
11  & Y &  Y &   Y  &     Y  &   Y   &    5\\
12  & Y &  Y  &   Y  &  Y  &     Y  &  5\\    
\enddata
\tablenotetext{a}{Does the shell's radius change with velocity as expected for an expanding bubble?}
\tablenotetext{b}{Is there IR nebulosity associated with this shell?}
\tablenotetext{c}{Does the CO or IR emission associated with this shell show a circular morphology?}
\tablenotetext{d}{Is the $p-v$  diagram consistent with that expected for an expanding bubble?}  
\tablenotetext{e}{Do we find a suitable candidate source?}
\end{deluxetable}

\clearpage

\newpage

\newpage

\newpage

\newpage

\clearpage

\newpage
  
\clearpage

\newpage

\newpage

\newpage

\newpage

\newpage

\newpage

\newpage

\newpage

\clearpage

\newpage

\newpage

\newpage

\newpage

\clearpage

\newpage

\clearpage

\newpage

\newpage

\newpage

\newpage

\end{document}